\documentclass[reqno,11pt]{amsart}
\usepackage{amscd}
\usepackage{slashed}
\usepackage{amssymb}
\usepackage{ushort}
\usepackage{mathtools}
\usepackage{esint}
\usepackage{enumitem}
\usepackage{epsfig}
\usepackage{pst-grad} 
\usepackage{pst-plot} 
\usepackage[mathscr]{eucal}
\numberwithin{equation}{section}
\allowdisplaybreaks[1]

\title[Linear Stability of the Non-Extreme Kerr Black Hole]{Linear Stability of the \\ Non-Extreme Kerr Black Hole}

\author[F.\ Finster]{Felix Finster}
\thanks{F.F.\ is supported in part by the Deutsche Forschungsgemeinschaft.}
\address{Fakult\"at f\"ur Mathematik \\ Universit\"at Regensburg \\ D-93040 Regensburg \\ Germany}
\email{Felix.Finster@mathematik.uni-regensburg.de}

\author[J.\ Smoller]{Joel Smoller \\ \\ August 2016}
\thanks{J.S.\ is supported in part by the National Science Foundation,
Grant No.\ DMS-1105189.}
\address{Mathematics Department \\ The University of Michigan \\ Ann Arbor, MI 48109, USA}
\email{smoller@umich.edu}

\newtheorem{Def}{Def.}[section]
\newtheorem{Thm}[Def]{Theorem}
\newtheorem{Prp}[Def]{Proposition}
\newtheorem{Lemma}[Def]{Lemma}

\newtheorem{Corollary}[Def]{Corollary}

\newcommand{\Thanks}{\vspace*{.5em} \noindent \thanks}
\newcommand{\Proof}{\begin{proof}}
\newcommand{\QED}{\end{proof} \noindent}

\newcommand{\lbra}{\langle}
\newcommand{\lket}{\rangle}
\newcommand{\bra}{<\!\!}
\newcommand{\ket}{\!\!>}
\newcommand{\C}{\mathbb{C}}
\newcommand{\R}{\mathbb{R}}
\newcommand{\1}{\mbox{\rm 1 \hspace{-1.05 em} 1}}
\newcommand{\Z}{\mathbb{Z}}

\newcommand{\N}{\mathbb{N}}

\newcommand{\beq}{\begin{equation}}
\newcommand{\eeq}{\end{equation}}
\newcommand{\umin}{u_{\mbox{\tiny{\rm{min}}}}}
\newcommand{\vmin}{v_{\mbox{\tiny{\rm{min}}}}}
\newcommand{\umax}{{u_{\mbox{\tiny{\rm{max}}}}}}
\newcommand{\uflip}{{u_{\mbox{\tiny{\rm{flip}}}}}}

\newcommand{\WKB}{{\text{\tiny{\rm{WKB}}}}}

\newcommand{\A}{\mathcal{A}}
\renewcommand{\O}{\mathscr{O}}

\renewcommand{\H}{\mathscr{H}}
\newcommand{\la}{\langle}
\newcommand{\ra}{\rangle}

\newcommand{\ul}{u_\ell}
\newcommand{\ur}{u_r}
\newcommand{\Const}{\mathscr{C}}
\newcommand{\const}{\mathfrak{c}}
\newcommand{\D}{\mathscr{D}}

\DeclareMathOperator{\re}{Re}
\DeclareMathOperator{\im}{Im}

\begin{document}
\maketitle

\begin{abstract}
It is proven that for smooth initial data with compact support outside the event horizon,
the solution of every azimuthal mode of the Teukolsky equation for general spin
decays pointwise in time.
\end{abstract}

\tableofcontents

\section{Introduction}
In this paper we prove linear stability of the non-extreme Kerr black hole under perturbations by
gravitational and electromagnetic waves. More precisely, we consider the initial value problem for
the Teukolsky equation of general spin~$s \in \{ 0, \frac{1}{2}, 1, \frac{3}{2}, 2, \ldots\}$
for smooth initial data with compact support outside the event horizon.
Thus, rewriting the equation as a first-order system in time, we analyze the solution for
initial data~$\Psi_0 \in C^\infty_0((r_1, \infty) \times S^2, \C^2)$
(where~$r_1$ is the event horizon, and the two components of~$\Psi_0$ describe the
Teukolsky wave function and its first time derivative at time zero; for details see Section~\ref{secprelim}).
We decompose the initial data into a Fourier series of azimuthal modes,
\beq \label{fouriermode}
\Psi_0(r, \vartheta, \varphi) = \sum_{k \in \Z} e^{-i k \varphi}\: \Psi_0^{(k)}(r, \vartheta) \:.
\eeq
Since the Kerr geometry is axisymmetric, the Teukolsky equation decouples into separate
equations for each mode. Therefore, the solution of the Cauchy problem with initial data~$\Psi_0$
is obtained by solving the Cauchy problem for each mode and taking the sum of the resulting solutions.
With this in mind, we here restrict attention to the Cauchy problem for a single mode, i.e.
\beq \label{initmode}
\Psi(0, r, \vartheta, \varphi) = e^{-i k \varphi}\: \Psi_0^{(k)}(r, \vartheta) \in C^\infty_0 \big(
(r_1, \infty) \times S^2, \C^2 \big) \:.
\eeq
We derive an integral representation of the solution which involves the fundamental solutions
of the ODEs arising in the separation of variables. Moreover, we prove the following pointwise decay result:
\begin{Thm} Consider a non-extreme Kerr black hole of mass~$M$ and angular momentum~$aM$
with~$M^2>a^2>0$. Then for any~$s \geq 0$, the solution~$\Psi$ of the Teukolsky equation
with initial data of the form~\eqref{initmode} decays to zero in~$L^\infty_\text{\rm{loc}}((r_1, \infty) \times S^2)$.
\end{Thm} \noindent
This theorem establishes in the dynamical setting that the non-extreme Kerr black hole
is linearly stable.

In general terms, the problem of linear stability of black holes can be stated mathematically as the question
whether solutions of massless linear wave equations in the Kerr geometry decay in time.
The different types of equations are characterized systematically in the Newman-Penrose
formalism by their spin, taking the possible values~$s=0, \frac{1}{2}, 1, \frac{3}{2}, 2, \ldots$.
From the physical point of view, the most interesting cases are $s=1$ (Maxwell field)
and~$s=2$ (gravitational waves). The case~$s=0$ of scalar waves is a major mathematical
simplification.
The black hole stability problem has a long history and has been studied by many authors.
For brevity, we only mention a few recent results and refer for the broader context to~\cite{bull}
and the references therein. Despite many results for scalar waves in the Kerr geometry
(see for example~\cite{weq, wdecay, tataru+tohaneanu, tataru2, metcalfe+tataru+tohaneanu,
dafermos+rodnianski+rothman}) and for higher spin waves in spherically symmetric space-times
(see for example~\cite{schdecay, sterbenz+tataru, andersson-blue2, dafermos-holzegel-rodnianski}),
only few results are known for higher spin waves in the Kerr geometry.
There are results for the Dirac field~\cite{tkerr} and for the Maxwell
field~\cite{metcalfe+tataru+tohaneanu2, andersson-blue1, andersson-baeckdahl-blue},
all of which use the specific structure of the respective equations. Also, we would like to mention recent
nonlinear stability results in the related
Kerr-De Sitter geometry~\cite{hintz+vasy}. A general framework for analyzing the equations of arbitrary spin
in the Kerr geometry goes back to Teukolsky~\cite{teukolsky}, who
showed that the massless equations of any spin can be rewritten as a single wave equation for
a complex scalar field~$\phi$, referred to as the Teukolsky equation.
If~$s \neq 0$, the coefficients of the Teukolsky equation are {\em{complex}}.
The Teukolsky equation has the remarkable property that it
can be separated into a coupled system of a radial and an angular ODE (for details see for example
the textbook~\cite{chandra}).
The only known stability result in the Teukolsky framework was obtained by
Whiting~\cite{whiting}, who proved that the Teukolsky equation does not admit solutions which
decay both at spatial infinity and at the event horizon and increase exponentially in time.
This so-called {\em{mode stability}} result is also a key ingredient to our analysis of the long-time dynamics
of solutions of the Cauchy problem.

We now outline our method of proof (a more detailed overview is given in the survey article~\cite{CMSA}).
We first bring the Teukolsky equation into Hamiltonian form 
by employing the ansatz
\[ \Psi = \begin{pmatrix} \Phi \\ i \partial_t \Phi \end{pmatrix} \]
and writing the equation as
\[ i \partial_t \Psi = H \Psi \:, \]
where~$H$ is a second-order spatial differential operator. Using suitable PDE estimates,
we show that the resolvent~$R_\omega:= (H - \omega)^{-1}$
exists if~$\omega$ lies outside a strip enclosing the real axis (see Lemma~\ref{lemmaresex}).
We then derive an integral representation for the solution of the Cauchy problem
which involves a Cauchy-type contour integral over the resolvent (see Theorems~\ref{thmcomplete}
and~\ref{thmcauchy0}).
Next, we decompose the resolvent on the contour into an infinite sum of angular
modes (see Theorem~\ref{thmsepres}). These angular modes
arise from our previous paper~\cite{tspectral} where we derive a spectral decomposition of the
angular operator into invariant subspaces. After employing this angular mode decomposition,
Whiting's mode stability~\cite{whiting} makes it possible to move the contour integrals
of the separated resolvent onto the real axis.

At this point two major problems remain: to show that the separated resolvents have no
poles on the real axis, and to control the infinite sum of angular modes uniformly in time.
In order to resolve these problems, we write the radial part of the separated Teukolsky equation in
Sturm-Liouville form
\[ \Big( -\frac{d^2}{du^2} + V \Big) X = 0 \:. \]
By a careful analysis of the potential~$V$ and of the solutions of this ODE,
we show that we can approximate~$X$
in different regions by WKB, Airy and parabolic cylinder functions, with rigorous
error estimates. Here we rely crucially on our previous work on special functions~\cite{special}
and on the ODE estimates developed in~\cite{invariant, tinvariant}.
These results also give rise to corresponding estimates for the separated resolvent
(see Proposition~\ref{prpkernes}).

For smooth initial data with compact support outside the event horizon, we
thus obtain an integral representation of the solution~$\Psi$ of the Cauchy problem
for the Teukolsky equation involving an infinite sum of angular modes
(see Theorem~\ref{thmrep}). We prove that this infinite sum, for large~$n$ is
uniformly small, and that the remaining finite sum decays using the Riemann-Lebesgue lemma
(see Corollary~\ref{cordecay}). This gives the above theorem.

\section{Preliminaries} \label{secprelim}
We consider the Kerr metric in Boyer-Lindquist
coordinates $(t, r, \vartheta, \varphi)$ with $r>0$, $0 <
\vartheta < \pi$, $0 \leq \varphi < 2\pi$ (see for example~\cite{chandra}).
Then the line element takes the form
\begin{align*}
ds^2 &= g_{jk}\:dx^j x^k \\
&= \frac{\Delta}{U} \:(dt - a \:\sin^2 \vartheta \:d\varphi)^2
- U \left( \frac{dr^2}{\Delta} + d\vartheta^2 \right) -
\frac{\sin^2 \vartheta}{U} \:(a \:dt - (r^2+a^2) \:d\varphi)^2 \:,
\end{align*}
where
\beq \label{UDdef}
U(r, \vartheta) = r^2 + a^2 \:\cos^2 \vartheta \qquad \text{and} \qquad
\Delta(r) = r^2 - 2 M r + a^2  \:.
\eeq
Here the parameters~$M$ and~$aM$ denote the mass and the angular momentum of the black hole,
respectively. We shall restrict attention to the {\em{non-extreme case}}
with non-zero angular momentum, i.e.~$M^2 > a^2>0$.
In this case, the function~$\Delta$ has two distinct zeros,
\beq \label{r01def}
r_0 = M - \sqrt{M^2 - a^2} \qquad \text{and} \qquad r_1 = M + \sqrt{M^2 - a^2} \:,
\eeq
corresponding to the Cauchy and the event horizon, respectively.
We shall consider only the region $r>r_1$ {\em{outside the event horizon}}, and thus $\Delta>0$.

Our starting point is the Teukolsky equation in the form given by Whiting~\cite{whiting}
\beq \label{teukolsky}
\begin{split}
&\bigg( \frac{\partial}{\partial r} \Delta \frac{\partial}{\partial r} - \frac{1}{\Delta}
\left\{ (r^2+a^2)\: \frac{\partial}{\partial t} + a\: \frac{\partial}{\partial \varphi} - (r-M) \,s \right\}^2 
- 4 s \: (r+i a \cos \vartheta)\: \frac{\partial}{\partial t} \\
&\qquad + \frac{\partial}{\partial \cos \vartheta} \:\sin^2 \vartheta\: \frac{\partial}{\partial \cos \vartheta} 
+ \frac{1}{\sin^2 \vartheta} \left\{ a \sin^2 \vartheta\: \frac{\partial}{\partial t} 
+ \frac{\partial}{\partial \varphi} + i s \cos \vartheta \right\}^2 \bigg) \phi = 0 \:.
\end{split}
\eeq
We restrict attention to a fixed $\varphi$-mode. Thus for a given~$k \in \Z/2$ we
make the ansatz
\beq \label{ksep}
\phi(t,r,\vartheta, \varphi) = e^{-i k \varphi}\: R(t,r,\vartheta) \:.
\eeq
Moreover, we introduce the Regge-Wheeler coordinate~$u \in \R$ by
\beq \label{RW}
\frac{du}{dr} = \frac{r^2+a^2}{\Delta} \:,\qquad
\frac{\partial}{\partial r} = \frac{r^2+a^2}{\Delta}\: \frac{\partial}{\partial u}
\eeq
and introduce the new function~$\Phi$ by
\beq \label{PhiR}
\Phi(t,u,\vartheta) = \sqrt{r^2+a^2}\: R(t,r,\vartheta) \:.
\eeq
Using the transformation
\begin{align*}
\frac{\partial}{\partial r} \Delta \frac{\partial}{\partial r}
&= \frac{r^2+a^2}{\Delta} \frac{\partial}{\partial u} (r^2+a^2) \frac{\partial}{\partial u} \\
&= \frac{r^2+a^2}{\Delta}\:  \sqrt{r^2+a^2} \left(\frac{\partial^2}{\partial u^2}\: \sqrt{r^2+a^2} 
- \Big( \partial_u^2 \sqrt{r^2+a^2} \Big) \right) ,
\end{align*}
we find that
\[ \frac{\partial}{\partial r} \Delta \frac{\partial}{\partial r} R
=  \frac{1}{\sqrt{r^2+a^2}}\: 
\frac{(r^2+a^2)^2}{\Delta} \left(\frac{\partial^2}{\partial u^2}
- \frac{ \partial_u^2 \sqrt{r^2+a^2} }{\sqrt{r^2+a^2}} \right) \Phi \:. \]
Then the Teukolsky equation takes the form
\beq \label{TPhi}
T \Phi = 0 \:,
\eeq
where
\begin{align*}
T&= \frac{(r^2+a^2)^2}{\Delta} \left(\frac{\partial^2}{\partial u^2}
- \frac{ \partial_u^2 \sqrt{r^2+a^2} }{\sqrt{r^2+a^2}} \right)  \\
&\qquad - \frac{1}{\Delta} \left\{ (r^2+a^2)\: \frac{\partial}{\partial t} - iak - (r-M) \,s \right\}^2 
- 4 s \: (r+i a \cos \vartheta)\: \frac{\partial}{\partial t} \\
&\qquad + \frac{\partial}{\partial \cos \vartheta} \:\sin^2 \vartheta\: \frac{\partial}{\partial \cos \vartheta} 
+ \frac{1}{\sin^2 \vartheta} \left\{ a \sin^2 \vartheta\: \frac{\partial}{\partial t} 
-ik + i s \cos \vartheta \right\}^2\:.
\end{align*}

\section{Hamiltonian Formulation}
In order to write the Teukolsky equation~\eqref{TPhi} in Hamiltonian form, we make the ansatz
\beq \label{twocomp}
\Psi = \begin{pmatrix} \Phi \\ i\partial_{t}\Phi \end{pmatrix} \:.
\eeq
Then the equation takes the form
\beq \label{partH}
i\,\partial_{t}\Psi=H\,\Psi \:,
\eeq
where $H$ is the Hamiltonian
\beq \label{Hamform}
H= \begin{pmatrix} 0 & 1 \\
A & \beta \end{pmatrix} \:,
\eeq
whose matrix entries are the operators
\begin{align}
A &= \frac{r^2+a^2}{\rho} \left(-\frac{\partial^2}{\partial u^2}
+ \frac{ \partial_u^2 \sqrt{r^2+a^2} }{\sqrt{r^2+a^2}} \right) \label{A1} \\
&\qquad +\frac{\Delta}{\rho\: (r^2+a^2)} \left(
-\frac{\partial}{\partial \cos \vartheta} \:\sin^2 \vartheta\: \frac{\partial}{\partial \cos \vartheta} 
+\frac{(-k + s \cos \vartheta)^2}{\sin^2 \vartheta} \right) \label{A2} \\
&\qquad - \frac{\big( a k + i (M-r) s \big)^2}{\rho\: (r^2+a^2)} \label{A3} \\
\beta &= \frac{2}{\rho} \left[ - \Big(a k + i (M-r) s \Big) + \Big(a k - 2 i r s + a s \cos \vartheta
\Big)\: \frac{\Delta}{r^2+a^2} \right] \\
\rho &= r^2+a^2 - a^2 \sin^2 \vartheta\: \frac{\Delta}{r^2+a^2} \:.
\end{align}
As the domain of definition of~$H$ we choose the smooth wave functions
which are compactly supported outside the event horizon. Thus, working with
the Regge-Wheeler coordinate~$u$ throughout (see~\eqref{RW}), we choose
\beq \label{domainH}
\D(H) = C^\infty_0(\R \times S^2, \C^2)\:.
\eeq
We remark that in the limiting case~$a \searrow 0$, the above Hamiltonian reduces to that
in~\cite[Section~4]{schdecay}.
In the case~$s=0$, on the other hand, our Hamiltonian coincides with that in~\cite[eqn~(2.25)]{weq},
except for the factor~$\sqrt{r^2+a^2}$ in the transformation~\eqref{PhiR}.

The next step is to introduce a scalar product. Our starting point is the bilinear form
\beq \label{bil}
\bra \Psi, \tilde{\Psi} \ket = \int_{-\infty}^\infty \frac{\rho}{r^2+a^2}\: du \int_{-1}^1 d\cos \vartheta
\; \la \Psi, \left( \begin{array}{cc} A & 0 \\ 0 & 1 \end{array}
\right) \tilde{\Psi} \ra_{\C^2} \:.
\eeq
This bilinear form has a structure similar to the familiar ``energy scalar product'' used for example
in Minkowski space. In our setting, however, this bilinear form does not have the
symmetry property~$\overline{\bra \Psi_1, \Psi_2 \ket} = \bra \Psi_2, \Psi_1 \ket$
(because the term~\eqref{A3} is complex) and is therefore certainly not positive definite.
Our strategy is to modify~\eqref{bil} in such a way that it becomes
symmetric and positive definite. We first verify the sign of the zero order term in~\eqref{A1}.

\begin{Lemma} Outside the event horizon, the zero order term in~\eqref{A1} is non-negative, i.e.
\[ \frac{ \partial_u^2 \sqrt{r^2+a^2} }{\sqrt{r^2+a^2}} \geq 0 \qquad
\text{for all~$r>r_1$}\:. \]
\end{Lemma}
\Proof By direct computation, one finds that
\[ \frac{ \partial_u^2 \sqrt{r^2+a^2} }{\sqrt{r^2+a^2}}
= \frac{\Delta}{(r^2+a^2)^4} \: f(r) \qquad \text{with} \qquad
f(r) := a^4 - 4 a^2 M r + a^2 r^2 + 2 M r^3 \:. \]
We want to show that the function~$f$ is non-zero outside the event horizon.
To this end, we first note that its derivative
\[ f'(r) = - 4 a^2 M + 2 a^2 r + 6 M r^2 \]
is obviously monotone increasing. Therefore, a direct computation using~\eqref{r01def} gives
\[ f'(r) \geq f'(r_1) = \big(12 M^3 - 8 a^2 M \big) + \big(12 M^2+2a^2 \big) \sqrt{M^2-a^2} \geq 0 \:, \]
where in the last step we used that~$a^2<M^2$. We conclude that~$f$ is monotone increasing.
Therefore,
\[ f(r) \geq f(r_1) = 8 M^2 \big(M^2 - a^2 \big) + \big(8 M^3 - 4 a^2 M \big) \sqrt{M^2-a^2} \geq 0 \:, \]
where we again used~\eqref{r01def} together with the inequality~$a^2<M^2$. This concludes the proof.
\QED
In view of this lemma, the term~\eqref{A1} gives a positive contribution to the
bilinear form~\eqref{bil}. Obviously, the same is true for the term~\eqref{A2}.
In order to get rid of the troublesome complex term~\eqref{A3} we set
\[ \delta = 1 + \frac{\big( a k + i (M-r) s \big)^2}{\rho\: (r^2+a^2)} \]
and introduce the scalar product~$(.,.)$ by
\beq \label{sprod}
( \Psi, \tilde{\Psi} ) = \int_{-\infty}^\infty \frac{\rho}{r^2+a^2}\: du \int_{-1}^1 d\cos \vartheta
\; \lbra \Psi, \left( \begin{array}{cc} A+\delta & 0 \\ 0 & 1 \end{array}
\right) \tilde{\Psi} \lket_{\C^2} \:.
\eeq
Taking the completion of the domain~\eqref{domainH} gives rise to a Hilbert
space~$(\H, (.,.))$. Using that~$A+\delta \geq \1$ and that the weight factor in~\eqref{sprod} written as
\[ \frac{\rho}{r^2+a^2} = 1 - \frac{\Delta\; a^2 \sin^2 \vartheta}{(r^2+a^2)^2} \]
is clearly bounded uniformly from above and below in~$u$ and~$\vartheta$,
the corresponding Hilbert space norm~$\|.\|$ is equivalent
to the Sobolev norm on~$(H^{1,2} \oplus L^2)(\R \times S^2, \C^2)$.

\section{Resolvent Estimates}
Obviously, the Hamiltonian~$H$ is not symmetric on~$(\H, (.,.))$.
But, as is verified by direct computation,
we obtain a symmetric operator by modifying the Hamiltonian to
\[ H_+ = \begin{pmatrix} 0 & 1 \\ A+\delta & \re \beta \end{pmatrix} \:, \]
where we again choose the domain~\eqref{domainH}.
The difference of~$H$ and~$H_+$ is a bounded operator. Namely,
\begin{align*}
\|(H-H_+) \Psi\| &= \bigg\| \begin{pmatrix} 0 & 0 \\ -\delta & i \im \beta \end{pmatrix}
\begin{pmatrix} \Psi_1 \\ \Psi_2 \end{pmatrix} \bigg\| 
= \big\| - \delta \Psi_1 + i \im \beta \, \Psi_2 \big\|_{L^2} \\
&\leq \sup_{\R \times S^2} \big( |\delta| + |\im \beta| \big) \: \Big(  \big\|\Psi_1\|_{L^2} +\|\Psi_2\|_{L^2} \Big)
\leq \sup_{\R \times S^2} \big( |\delta| + |\im \beta| \big) \: \|\Psi\| \:,
\end{align*}
implying that
\beq \label{HmHp}
\|(H-H_+)\| \leq c := \sup_{\R \times S^2} \big( |\delta| + |\im \beta| \big) \:.
\eeq
Now we use a method similar as in~\cite[Lemma~4.1]{weq} to obtain resolvent estimates.
\begin{Lemma} \label{lemmaresex}
For every~$\omega$ with
\beq \label{imlower}
|\im \omega| > c \:,
\eeq
the resolvent~$R_\omega = (H-\omega)^{-1}$ exists and is bounded by
\beq \label{Rwes}
\| R_\omega \| \leq \frac{1}{|\im \omega|-c}\:.
\eeq
\end{Lemma}
\Proof The inequality~\eqref{HmHp} gives rise to the bound
\begin{align*}
\| H - H^* \| &= \big\| (H-H_+) - (H-H_+)^* \big\| \\
&\leq \|H-H_+\| + \|(H-H_+)^*\| = 2\:\|H-H_+\| \leq 2c \:.
\end{align*}
It follows that for every normalized vector $\Psi \in \D(H)$,
\begin{align}
\|(H-\omega) \Psi\| &\geq \big| \big(\Psi, (H-\omega) \Psi \big) \big|
\geq \big|{\mbox{Im}}\, \big(\Psi, (H-\omega) \Psi \big) \big| \nonumber \\
&\geq |{\mbox{Im}}\, \omega|\:\|\Psi\|^2 - \frac{1}{2} \:\big| \big(\Psi, (H-H^*) \Psi \big) \big| \notag \\
&\geq \big( |{\mbox{Im}}\, \omega| - c \big) \:\|\Psi\|^2\:. \label{Hmoes}
\end{align}
It follows that the operator~$(H-\omega)$ is injective.

In order to show that this operator is also surjective, we first note that the estimate~\eqref{Hmoes}
implies that the image of~$(H-\omega)$ is a closed subspace of~$\H$. Therefore, it
suffices to show that the image of the operator~$H-\omega$ is dense in $\H$.
If this were not the case, there would exist a non-zero vector~$\hat{\Psi} \in \H$ such that
\[ \big( (H-\omega)\Psi,\hat{\Psi} \big) = 0 \qquad \text{for all $\Psi \in \D(H)$} \:. \]
In other words, $\hat{\Psi}$ is a weak solution of the adjoint equation
$(H^*-\overline{\omega}) \hat{\Psi}=0$. By the interior regularity theorem for elliptic operators
(cf.~\cite[Section~6.3.1]{evans}), every weak solution
of this equation is a solution in the strong sense.
On the other hand, repeating the estimate~\eqref{Hmoes}
with~$H$ replaced by~$H^*$ and~$\omega$ replaced by~$\overline{\omega}$, we
get the inequality
\[ \big\|(H^*-\overline{\omega}) \hat{\Psi} \big\| \geq \big( |{\mbox{Im}}\, \omega| - c \big) \,\big\| \hat{\Psi} \big\|^2 \:. \]
This is a contradiction.

The above arguments show that the resolvent~$R_\omega$ exists.
Applying the inequality~\eqref{Hmoes} for~$\Psi = R_\omega \Phi$ gives the estimate~\eqref{Rwes}.
This concludes the proof.
\QED

\section{Contour Integrals and Completeness}
\begin{figure}
\psscalebox{1.0 1.0} 
{
\begin{pspicture}(0,-2.792334)(7.735,2.792334)
\rput[bl](6.78,-0.0023339272){\normalsize{$\re \omega$}}
\psarc[linecolor=black, linewidth=0.04, dimen=outer](3.32,-0.19233392){2.28}{30.0}{150.0}
\psarc[linecolor=black, linewidth=0.04, dimen=outer](3.32,-0.19233392){2.28}{-150.0}{-30.0}
\psline[linecolor=black, linewidth=0.04, arrowsize=0.02cm 8.0,arrowlength=1.4,arrowinset=0.4]{->}(0.02,-0.19233392)(6.82,-0.19233392)
\psline[linecolor=black, linewidth=0.04, arrowsize=0.02cm 8.0,arrowlength=1.4,arrowinset=0.4]{->}(3.32,-2.7923338)(3.32,2.807666)
\psline[linecolor=black, linewidth=0.02](3.32,-0.19233392)(1.92,1.587666)
\rput[bl](2.24,1.2876661){\normalsize{$R$}}
\psline[linecolor=black, linewidth=0.02, linestyle=dashed, dash=0.17638889cm 0.10583334cm](0.0,0.94766605)(6.8,0.94766605)
\psline[linecolor=black, linewidth=0.02, linestyle=dashed, dash=0.17638889cm 0.10583334cm](0.0,-1.3323339)(6.8,-1.3323339)
\rput[bl](3.615,2.512666){\normalsize{$\im \omega$}}
\rput[bl](6.985,0.8526661){\normalsize{$\im \omega=2c$}}
\rput[bl](4.715,1.692666){\normalsize{$C_1$}}
\rput[bl](4.995,-2.1573339){\normalsize{$C_2$}}
\rput[bl](6.995,-1.4473339){\normalsize{$\im \omega=-2c$}}
\end{pspicture}
}
\caption{The contour~$C_R$.}
\label{figcontour}
\end{figure}
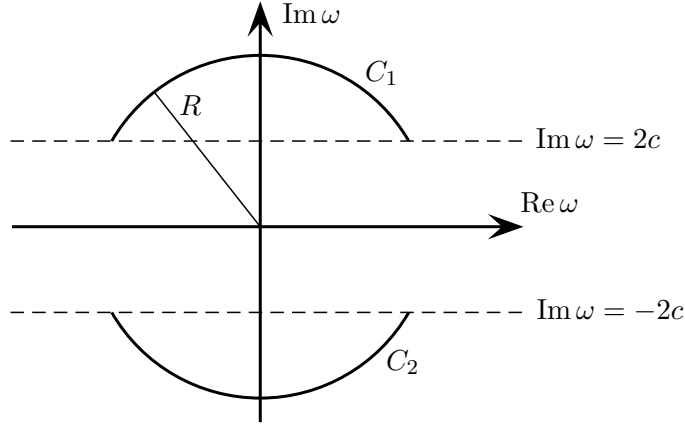

Given~$R>0$, we consider the two contours~$C_1$ and~$C_2$ in the complex
$\omega$-plane defined by
\[ C_1 = \partial B_R(0) \cap \{ {\mbox{Im}}\, \omega > 2c \} \:,\qquad
C_2 = \partial B_R(0) \cap \left\{ {\mbox{Im}}\, \omega <-2c \right\} , \]
both taken with positive orientation (see Figure~\ref{figcontour}).
We set~$C_R = C_1 \cup C_2$.
We can now state the following completeness result.
The proof uses similar methods as in~\cite[Section~7]{schdecay}
and is based on an idea which we learned from A.\ Bachelot~\cite[Proof of Theorem~2.12]{bachelot}.

\begin{Thm} \label{thmcomplete}
For every~$\Psi \in \D(H)$, we have the representation
\beq \label{psirep}
\Psi = -\frac{1}{2 \pi i} \lim_{R \rightarrow \infty}
\int_{C_R} (R_\omega \Psi)\: d\omega\:.
\eeq
\end{Thm}
\Proof Given~$\omega$ satisfying the inequality~\eqref{imlower} and~$\Psi \in \D(\H)$, we know that
\beq \label{RomegaH}
R_\omega \,(H-\omega) \,\Psi = \Psi \:.
\eeq
Solving for~$R_\omega \Psi$ gives the identity
\[ R_\omega \Psi = -\frac{\Psi}{\omega} + \frac{1}{\omega}\: R_\omega (H \Psi) \:. \]
Clearly, this identity also holds for~$\Psi$ replaced by~$H \Psi$. This makes it possible to
iterate the identity to obtain
\beq \label{Rwrel}
R_\omega \Psi = -\frac{\Psi}{\omega} - \frac{1}{\omega^2}\: (H \Psi) +  \frac{1}{\omega^2}\: R_\omega \big(
H^2 \Psi \big) \:.
\eeq
Integrating over the contour~$C_R$ gives the estimate
\begin{align*}
\bigg\| \int_{C_R} \Big( R_\omega \Psi + \frac{\Psi}{\omega} \Big) \,d\omega \bigg\|
\leq \Big( \|H \Psi\| + \big\|R_\omega (H^2 \Psi) \big\| \Big) \int_{C_R} \frac{d|\omega|}{|\omega^2|}\:.
\end{align*}
Using the resolvent estimate~\eqref{Rwes} and noting that the length of the contour
grows only linearly in~$R$, one sees that the right side tends to zero as~$R \rightarrow \infty$.
Hence
\[ \lim_{R \rightarrow \infty} \int_{C_R} (R_\omega \Psi)\: d\omega
= - \Psi \lim_{R \rightarrow \infty} \int_{C_R} \frac{d\omega}{\omega} = -2 \pi i\: \Psi \:, \]
where the last step can be verified by computing the integral explicitly or
by using the estimate
\[ \left| \ointctrclockwise_{\partial B_R(0)} \frac{d\omega}{\omega} -
\int_{C_R}  \frac{d\omega}{\omega} \right| \leq
\frac{12\,c}{R} \xrightarrow{R \rightarrow \infty} 0\:. \]
This concludes the proof.
\QED

The integral representation in Theorem~\ref{thmcomplete} has the disadvantage that the
integrand decays at infinity only like~$1/|\omega|$, making it impossible to work with unbounded
contours (because these would not converge in the Hilbert space).
In order to avoid this problem, we use the method introduced in~\cite[Section~7]{schdecay}
to subtract counter terms which do not change the value of the contour integral.
We thus obtain the following result.

\begin{Thm} \label{thmcauchy0}
Choosing~$C$ as the contour
\[ C = \big\{ \omega \:\big|\: \im \omega = 2c \big\} \cup 
\big\{ \omega \:\big|\: \im \omega = - 2c \big\} \]
with counter-clockwise orientation, the following completeness relation holds for every~$\Psi \in \D(H)$,
\beq \label{cunbound}
\Psi = -\frac{1}{2 \pi i} \int_C \Big( R_\omega \Psi + \frac{\Psi}{\omega+3 i c} \Big)\: d\omega \:.
\eeq
Moreover, the Cauchy problem for the Teukolsky equation with initial data~$\Psi|_{t=0} = \Psi_0 \in \D(H)$
has a unique solution given by
\beq \label{propagator}
\Psi(t) = -\frac{1}{2 \pi i} \int_C e^{-i \omega t}\:
\Big( R_\omega \Psi_0 + \frac{\Psi_0}{\omega+3 i c} \Big)\: d\omega \:.
\eeq
\end{Thm}
\Proof Since the resolvent is holomorphic in the region~$\{|\im \omega|>c\}$
(see~\cite[Section~III.6.1]{kato}), we may
continuously deform the contour~$C_R$ in~\eqref{psirep} for any~$R$.
In particular, we may deform the contours to new contours~$\tilde{C}_R$ which
all lie inside the region~$|\im \omega| < 3c$. Then the function~$1/(\omega+3ic)$,
having its poles outside this region, does not contribute to the contour integral in the
the limit~$R \rightarrow \infty$, i.e.
\beq \label{contint}
\Psi = -\frac{1}{2 \pi i} \lim_{R \rightarrow \infty}
\int_{\tilde{C}_R} \Big( R_\omega \Psi + \frac{\Psi}{\omega+3 i c} \Big)\: d\omega\:.
\eeq
Using~\eqref{Rwrel} and expanding, one finds
\[ R_\omega \Psi = -\frac{\Psi}{\omega} - \frac{H \Psi}{\omega^2}
+ \frac{R_\omega (H^2 \Psi)}{\omega^2}
= -\frac{\Psi}{\omega} + \O \big( \omega^{-2} \big) = -\frac{\Psi}{\omega+3 i c} + \O \big( \omega^{-2} \big)\:. \]
Hence the norm of the integrand in~\eqref{contint} decays quadratically for large~$|\omega|$.
Therefore, in the limit~$R \rightarrow \infty$ the integrals converges to the unbounded
contour integral~\eqref{cunbound}.

In order to prove~\eqref{propagator}, we insert one more counter term (which again does not
change the value of the contour integral) to obtain
\beq \label{complete3}
\Psi_0 = -\frac{1}{2 \pi i} \int_{C} \Big( R_\omega \Psi_0 + \frac{\Psi_0}{\omega+3 i c} 
+ \frac{(H+3ic)\Psi_0}{(\omega+3 i c)^2}\Big)\: d\omega \:.
\eeq
A direct computation using the identity
\[ R_\omega \Psi_0 = -\frac{\Psi_0}{\omega} - \frac{H \Psi_0}{\omega^2} - \frac{H^2 \Psi_0}{\omega^3}
+ \frac{R_\omega (H^3 \Psi_0)}{\omega^3} \]
(obtained again by iterating~\eqref{Rwrel}) shows that the integrand in~\eqref{complete3}
decays even cubically for large~$|\omega|$.
Clearly, the additional counter term can also be inserted in~\eqref{propagator}, so that
the function~$\Psi(t)$ as defined by~\eqref{propagator} takes the form
\beq \label{Psit3}
\Psi(t) = -\frac{1}{2 \pi i} \int_{C} e^{-i \omega t} \:\Big( R_\omega \Psi_0 + \frac{\Psi_0}{\omega+3 i c} 
+ \frac{(H+3ic)\Psi_0}{(\omega+3 i c)^2}\Big)\: d\omega \:.
\eeq
Setting~$t=0$ and using~\eqref{complete3}, we find that~$\Psi(0)=\Psi_0$, showing that
the initial conditions are satisfied. It remains to show that~$\Psi(t)$ satisfies the
Teukolsky equation~$(i \partial_t - H) \Psi(t)=0$. Using that the integrand in~\eqref{Psit3} decays
cubically for large~$\omega$ and that the time derivative generates a factor~$\omega$,
Lebesgue's dominated convergence theorem allows us to interchange the differential operator with the integration.
We thus obtain
\begin{align*}
(i \partial_t - H) \Psi(t) &= \frac{1}{2 \pi i} \int_{C} e^{-i \omega t} \:(H - \omega)
\Big( R_\omega \Psi_0 + \frac{\Psi_0}{\omega+3 i c} 
+ \frac{(H+3ic)\Psi_0}{(\omega+3 i c)^2}\Big)\: d\omega \\
&=\frac{1}{2 \pi i} \int_{C} e^{-i \omega t} \:
\Big( \Psi_0 + \frac{(H-\omega) \Psi_0}{\omega+3 i c} 
+ \frac{(H-\omega) (H+3ic)\Psi_0}{(\omega+3 i c)^2}\Big)\: d\omega = 0 \:,
\end{align*}
because no poles are enclosed by the contour. This shows that~$\Psi(t)$ as given by~\eqref{propagator}
really is a solution of the Cauchy problem. Uniqueness follows immediately because
the Teukolsky equation is hyperbolic.
\QED

We now derive alternative integral representations for the solution of the Cauchy problem
which will be useful for our estimates.
\begin{Corollary} For any integer~$p \geq 1$, the solution of the Cauchy problem for the Teukolsky equation
with initial data~$\Psi|_{t=0} = \Psi_0 \in \D(H)$ has the representation
\beq \label{propagatorp}
\Psi(t) = -\frac{1}{2 \pi i} \int_C e^{-i \omega t}\: \frac{1}{(\omega + 3 i c)^p} \;
\Big( R_\omega \,\big(H + 3 i c \big)^p \,\Psi_0  \Big)\: d\omega \:.
\eeq
\end{Corollary}
\Proof Rewriting~\eqref{RomegaH} as
\[ R_\omega \Psi = -\frac{1}{\omega + 3 i c}\: \Psi + \frac{1}{\omega + 3 i c}\: R_\omega\, (H+3 i c)\, \Psi \:, \]
we can iterate similar to~\eqref{Rwrel} to obtain
\begin{align}
R_\omega \Psi &= -\frac{1}{\omega + 3 i c}\: \Psi -\frac{(H+3 i c)\, \Psi}{(\omega + 3 i c)^2}
- \cdots -\frac{(H+3 i c)^{p-1}}{(\omega + 3 i c)^p}\: \Psi \label{Rrel1} \\
&\quad\: +\frac{1}{(\omega + 3 i c)^p}\: R_\omega\, (H+3 i c)^p\, \Psi \:. \label{Rrel2}
\end{align}
Using this identity in~\eqref{propagator}, the first summand in~\eqref{Rrel1} cancels.
For all the other summands in~\eqref{Rrel1}, one can compute the integral in~\eqref{propagator} with
residues to obtain zero. Therefore, only the summand~\eqref{Rrel2} remains, giving the result.
\QED

For negative times, the integral representation of the solution of the Cauchy problem
can be further simplified. This is the representation which we will use in the remainder
of this paper.
\begin{Corollary} \label{corprop}
For negative times, the solution of the Cauchy problem 
for the Teukolsky equation with initial data~$\Psi|_{t=0} = \Psi_0 \in \D(H)$
has the integral representation
\[ \Psi(t) = -\frac{1}{2 \pi i} \int_{\R - 2 i c} e^{-i \omega t}\:
\Big( R_\omega \Psi_0 + \frac{\Psi_0}{\omega+3 i c} \Big)\: d\omega \qquad \text{(if~$t<0$)}\:. \]
Moreover, for any integer~$p \geq 1$,
\beq \label{propagatorp2}
\Psi(t) = -\frac{1}{2 \pi i} \int_{\R - 2 i c} e^{-i \omega t}\: \frac{1}{(\omega + 3 i c)^p} \;
\Big( R_\omega \,\big(H + 3 i c \big)^p \,\Psi_0  \Big)\: d\omega \:.
\eeq
\end{Corollary}
\Proof Starting from~\eqref{propagator} and~\eqref{propagatorp}, we take the limit where the upper part of the
contour~$\R+2i c$ is deformed towards~$\im \omega \rightarrow \infty$,
making use of the fact that the factor~$e^{-i \omega t}$ decays exponentially in this limit.
\QED

\section{Separation of Variables and Jost Solutions}
Our methods rely on the fact that the Teukolsky equation is completely separable into
a coupled system of a radial and an angular ODE. We now recall this procedure.
Let~$\omega \in \C$. We make the separation ansatz
\[ \Phi(t,u, \vartheta) = e^{-i \omega t} \,X(u)\, Y(\vartheta) \:. \]
This gives rise to the coupled system of ODEs
\beq \label{coupled}
{\mathcal{R}}_\omega X(u) = -\lambda X(u) \:,\qquad
\A_\omega Y(\vartheta) = \lambda Y(\vartheta) \:,
\eeq
where the radial operator~${\mathcal{R}}_\omega$ and
the angular operator~$\A_\omega$ are given by
\begin{align}
{\mathcal{R}}_\omega &= -\frac{(r^2+a^2)^2}{\Delta} \left(\frac{\partial^2}{\partial u^2}
- \frac{ \partial_u^2 \sqrt{r^2+a^2}}{\sqrt{r^2+a^2}} \right) \notag \\
&\qquad + \frac{1}{\Delta} \Big( -i \omega \,(r^2+a^2) - iak - (r-M) \,s \Big)^2 - 4 i s r \omega + 4 k\, a \omega \label{Rop} \\
\A_\omega &= 4 s a \omega \cos \vartheta - 4 k\, a \omega \notag \\
&\qquad - \frac{\partial}{\partial \cos \vartheta} \:\sin^2 \vartheta\: \frac{\partial}{\partial \cos \vartheta} 
- \frac{1}{\sin^2 \vartheta} \Big( -i a \omega \sin^2 \vartheta -ik + i s \cos \vartheta \Big)^2 \notag \\
&= - \frac{\partial}{\partial \cos \vartheta} \:\sin^2 \vartheta\: \frac{\partial}{\partial \cos \vartheta} 
- \frac{1}{\sin^2 \vartheta} \Big( i a \omega \sin^2 \vartheta -ik + i s \cos \vartheta \Big)^2 \notag \\
&= - \frac{\partial}{\partial \cos \vartheta} \:\sin^2 \vartheta\: \frac{\partial}{\partial \cos \vartheta} 
+\frac{1}{\sin^2 \vartheta} \Big( -a \omega \sin^2 \vartheta +k - s \cos \vartheta \Big)^2 \:. \label{Aop}
\end{align}
The operator~$\A_\omega$ coincides with the angular
Teukolsky operator as studied in~\cite{tinvariant, tspectral} (with the aspherical parameter~$\Omega
=-a \omega$).
Note that in order to get this agreement, we added and subtracted the constant~$4 k a \omega$.

The radial equation can be written as the Sturm-Liouville equation
\begin{equation} \label{schroedinger}
\left( -\frac{d^2}{du^2} + V \right) X = 0 \:,
\end{equation}
where~$V$ is the potential
\beq \label{Vdef}
\begin{split}
V(u) &= \frac{\lambda\, \Delta}{(r^2+a^2)^2} +  \frac{ \partial_u^2 \sqrt{r^2+a^2}}{\sqrt{r^2+a^2}} \\
&\quad + \frac{4 \omega\, \Delta}{(r^2+a^2)^2}\: \big( ak - i r s\big)
-\Big( \omega + \frac{ak - i (r-M) \,s}{r^2+a^2} \Big)^2 \:.
\end{split}
\eeq
Near spatial infinity, the potential has the asymptotics
\beq \label{VRasy}
V(u) = -\omega^2 - \frac{2is \omega}{u} + \O\left( \frac{\log u}{u^2} \right) \qquad \text{if~$u \rightarrow \infty$}\:.
\eeq
Likewise, near the event horizon, we have the asymptotics
\beq \label{VLasy}
V(u) = -\Omega^2 + \O\big( e^{\gamma u} \big) \qquad \text{if~$u \rightarrow -\infty$}\:,
\eeq
where
\beq \label{Omegadef}
\Omega := \omega + \frac{ak - i (r_1-M) \,s}{r_1^2+a^2} \:, \qquad \gamma := \frac{r_1-r_0}{r_1^2+a^2}
\eeq
(and~$r_0$ and~$r_1$ are again the horizons~\eqref{r01def}).
These asymptotics are the same as in the Schwarzschild geometry, if at the
event horizon we replace~$\omega$ by~$\omega + (ak)/(r_1^2+a^2)$ and~$r_1^2$
by~$r_1^2+a^2$. Therefore, we can proceed exactly as in~\cite[Section~3]{schdecay}
to construct Jost solutions~$\acute{\phi}_\pm$ and~$\grave{\phi}_\pm$ with
prescribed asymptotics at $u \rightarrow \pm \infty$.
Near the event horizon, the following result was established in~\cite[Theorem~3.1]{wdecay}:
\begin{Thm} \label{thm31}
For every~$\omega$ in the domain
\[ D_- := \Big\{ \omega \:\Big|\:
 {\mbox{\rm{Im}}}\, \omega < \frac{(r_1-M) s}{r_1^2+a^2} + \frac{\gamma}{2} \Big\} \]
there is a solution~$\acute{\phi}_-$ of~\eqref{schroedinger}
having the asymptotics
\beq \label{acutephiasy}
\lim_{u \to -\infty} e^{-i \Omega u} \:\acute{\phi}_-(u) = 1 \:,\qquad
\lim_{u \to -\infty} \left(e^{-i \Omega u} \:\acute{\phi}_-(u) \right)' = 0 \:.
\eeq
These solutions can be chosen to form a holomorphic family, in the sense that for every~$u \in \R$,
the function~$\acute{\phi}_-(u)$ is holomorphic in~$\omega \in D_-$.
Similarly, on the domain
\[ D_+ := \Big\{ \omega \:\Big|\:
 {\mbox{\rm{Im}}}\, \omega > \frac{(r_1-M) s}{r_1^2+a^2}- \frac{\gamma}{2} \Big\} \]
there is a holomorphic family of solutions~$\acute{\phi}_+$ of~\eqref{schroedinger}
with the asymptotics
\[ \lim_{u \to -\infty} e^{i \Omega u} \:\acute{\phi}_+(u) = 1 \:,\qquad
\lim_{u \to -\infty} \left(e^{i \Omega u} \:\acute{\phi}_+(u) \right)' = 0 \:. \]
\end{Thm} \noindent
Loosely speaking, the method of proof is to use a Picard-type iteration
on the unbounded interval~$(-\infty, -u_0)$ for sufficiently small and
negative~$u_0$, taking a plane wave as
the starting point. For a general introduction to this method
we refer to the classical text book~\cite{alfaro+regge}.

At infinity, one must keep in mind that the potential is of long range,
making it necessary to modify the plane wave asymptotics by
factors~$u^{\pm s}$. The following result was obtained
in~\cite[Theorem~3.3]{schdecay}:
\begin{Thm} \label{thm34}
On the domain $E_+ := \{ \omega \:|\: \omega \neq 0 {\mbox{ and }} {\mbox{\rm{Im}}}\, \omega \geq 0 \}$,
there is a family of solutions~$\grave{\phi}_+(u)$ of~\eqref{schroedinger},
holomorphic in the interior of~$E_+$, having the asymptotics
\[ \lim_{u \rightarrow \infty} u^s\, e^{-i \omega u}\: \grave{\phi}_+(u) = 1\:,\qquad
\lim_{u \rightarrow \infty} \left( u^s\, e^{-i \omega u}\: \grave{\phi}_+(u)\right)' = 0 \:. \]
Likewise, on the domain $E_- := \{ \omega \:|\: \omega \neq 0 {\mbox{ and }} {\mbox{\rm{Im}}}\, \omega \leq 0 \}$,
there is a family of solutions~$\grave{\phi}_-(u)$ of~\eqref{schroedinger}, holomorphic
in the interior of~$E_-$, with the asymptotics
\beq \label{phiasy}
\lim_{u \rightarrow \infty} u^{-s}\, e^{i \omega u}\: \grave{\phi}_-(u) = 1\:,\qquad
\lim_{u \rightarrow \infty} \left( u^{-s}\, e^{i \omega u}\: \grave{\phi}_-(u)\right)' = 0 \:.
\eeq
\end{Thm}

\section{Separation of the Resolvent}
Our strategy for getting more explicit information on the solution of the
Cauchy problem in Corollary~\ref{corprop} is to express the resolvent
in terms of the Jost solutions. We now explain how this ``separation'' of the
resolvent is ca be derived based on the spectral decomposition of the angular operator
as derived in~\cite{tspectral}.

For any~$\omega \in \R-2ic$, we choose the Jost function as
\[ \acute{\phi} = \acute{\phi}_- \qquad \text{and} \qquad
\grave{\phi} = \grave{\phi}_+ \:. \]
These functions all decay exponentially in their asymptotic ends.
We now apply the results of~\cite{tspectral} on the spectral decomposition of the
angular operator. According to~\cite[Theorem~1.1]{tspectral}, for any~$\omega$
in the strip~$U$ defined by
\beq \label{stripdef}
U := \big\{ \omega \in \C \text{ with } |{\mbox{\rm{Im}}}\, \omega| < 3c \big\} \:,
\eeq
there is a family~$(Q^\omega_n)_{n \in \N \cup \{0\}}$ of idempotent operators on~$L^2(S^2)$
whose images are invariant subspaces of the angular operator~$\A_\omega$
with the following properties:
\begin{itemize}[leftmargin=2em]
\item[(i)] The $Q^\omega_n$ are complete in the sense that
\beq \label{strongcomplete}
\sum_{n=0}^\infty Q^\omega_n = \1
\eeq
with strong convergence of the series.
\item[(ii)] The operators~$Q^\omega_0$ have rank at most~$N$, where~$N$ can be chosen
uniformly in~$\omega$.
The operators~$Q^\omega_1, Q^\omega_2, \ldots$ have rank at most two. \label{pageii}
\item[(iii)] The $Q^\omega_n$ are uniformly bounded in~$L^2(S^2)$, i.e.\
there is a constant~$c_2$ such that
\beq \label{Qnb}
\|Q_n^\omega \| \leq c_2 \qquad \text{for all $n \in \N \cup \{0\}$ and~$\omega \in U$}\:.
\eeq
\end{itemize}

We now choose~$\omega \in \R - 2 i c$ and let~$n \in \N \cup \{0\}$. 
We point out that the range of the operator~$Q^\omega_n$ need not be an
eigenspace of~$\A_\omega$ because there might be Jordan chains.
However, since the length of the Jordan chains is bounded by~$N$, we
can write~$\A_\omega$ on the invariant subspace as
\beq \label{Anil}
\A_\omega \, Q^\omega_n = (\lambda\,\1 + {\mathcal{N}}) \, Q^\omega_n \:,
\eeq
where~${\mathcal{N}}$ is a nilpotent operator with~${\mathcal{N}}^N = 0$.
Let us consider the Teukolsky equation~\eqref{TPhi}
with separated time dependence~$\sim e^{-i \omega t}$
on the invariant subspace.
Using~\eqref{Anil}, the resulting equation is obtained from the radial
equation~\eqref{schroedinger} and~\eqref{Vdef} if the separation constant~$\lambda$
is replaced by the operators~$\lambda\,\1 + {\mathcal{N}}$. This gives the equation
\[ \left( -\frac{d^2}{du^2} + V + \frac{\Delta}{(r^2+a^2)^2}\: {\mathcal{N}} \right) X(u) = 0\:, \]
where~$X(u)$ is now vector-valued, taking values in the invariant subspace.
We want to construct a Green's function~$g_\omega(u,v)$ of this equation, defined by
\beq \label{gdef}
\left( -\frac{d^2}{du^2} + V + \frac{\Delta}{(r^2+a^2)^2}\: {\mathcal{N}} \right) g_\omega(u,v) =  \delta(u-v) \:\1\:.
\eeq
If the nilpotent operator~${\mathcal{N}}$ is absent, this Green's function
is given just as in~\cite[Section~4]{schdecay} by a function which we now denote by~$s_\omega(u,v)$,
\beq \label{sdef}
s_\omega(u,v) = \frac{1}{w(\acute{\phi}, \grave{\phi})} \:\times\:
\left\{ \begin{array}{cl} \acute{\phi}(u)\, \grave{\phi}(v) & {\mbox{if~$v \geq u$}} \\[0.3em]
\grave{\phi}(u)\, \acute{\phi}(v) & {\mbox{if~$v < u$\:.}} \end{array} \right.
\eeq
Namely, a straightforward computation yields
\[ \bigg( -\frac{d^2}{du^2} + V(u) \bigg) \,s_\omega(u,v) = \delta(u-v)\:. \]
We also regard~$s_\omega$ as an operator with corresponding integral kernel~$s_\omega(u,v)$.
Then we can multiply~\eqref{gdef} by~$s_\omega$ to obtain the operator equation
\[ \left( \1 + s_\omega\:\frac{\Delta}{(r^2+a^2)^2}\: {\mathcal{N}} \right) g_\omega =  s_\omega\:\1 \:. \]
Since~${\mathcal{N}}$ is nilpotent, this equation can be solved by a finite Neumann series,
\beq \label{gkerdef}
g_\omega =  \sum_{l=0}^N \Big(-s_\omega\:\frac{\Delta}{(r^2+a^2)^2}\: {\mathcal{N}} \Big)^l \:s_\omega \:.
\eeq
The existence of powers of~$s_\omega$ is proved exactly as in~\cite[Lemma~5.2]{weq}.

After these preparations, we can now decompose the resolvent into angular modes:
\begin{Thm} \label{thmsepres}
For any~$\omega \in \R -2ic$, the resolvent~$R_\omega=(H-\omega)^{-1}$
has the representation
\[ R_\omega = \sum_{n=0}^\infty R_{\omega,n} \:Q_n^\omega \:, \]
where the operators~$R_{\omega,n}$ can be written as integral operators,
\beq \label{Romn}
\big(R_{\omega,n} \Psi)(u, \vartheta) = \int_{-\infty}^\infty \frac{\rho(v, \vartheta)}{r(v)^2+a^2}\: 
\mathfrak{R}_{\omega, n}(u,v)\, \Psi(v, \vartheta)\: dv \:,
\eeq
with integral kernels given by
\beq \label{Rwndef}
\mathfrak{R}_{\omega, n}(u,v) = \frac{r^2+a^2}{\rho} \: \delta(u-v) \begin{pmatrix} 0 & 0 \\ 1 & 0 \end{pmatrix}
+ g_\omega(u,v)\, \begin{pmatrix} \omega - \beta(v) & 1 \\[0.2em] \omega\, \big( \omega - \beta(v) \big) & \omega \end{pmatrix} .
\eeq
\end{Thm}
\Proof It is obvious from~\eqref{Hamform} that
\beq \label{res1}
(H- \omega) \;\frac{r^2+a^2}{\rho} \: \delta(u-v) \begin{pmatrix} 0 & 0 \\ 1 & 0 \end{pmatrix} Q^\omega_n
= \frac{r^2+a^2}{\rho} \: \delta(u-v) \begin{pmatrix} 1 & 0 \\ \beta(v)-\omega & 0 \end{pmatrix} Q^\omega_n\:.
\eeq
We next compute the operator product
\beq \label{Hgprod}
(H_u - \omega) \: g_\omega(u,v)\, \begin{pmatrix} \omega
- \beta(v) & 1 \\[0.2em] \omega\, \big( \omega - \beta(v) \big) & \omega \end{pmatrix} Q_n^\omega \:,
\eeq
where the index~$u$ at the Hamiltonian clarifies that its derivatives act on the variable~$u$.
If~$u \neq v$, we know from~\eqref{gdef} that~$g_\omega(u,v)$ is a solution of the radial Teukolsky equation.
Moreover, using the fact that the second row in the matrix is~$\omega$ times the first row,
one sees that every column of this matrix is of the form~\eqref{twocomp}.
This implies that~\eqref{Hgprod} vanishes. If~$u=v$, the only additional contribution is obtained
when the operator~$\partial_u^2$ contained in~$H$ acts on~$g_\omega$.
In view of~\eqref{Hamform} and~\eqref{A1}, we obtain
\begin{align*}
(H_u - \omega) &\: g_\omega(u,v)\, \begin{pmatrix} \omega
- \beta(v) & 1 \\[0.2em] \omega\, \big( \omega - \beta(v) \big) & \omega \end{pmatrix} Q_n^\omega \\
&= \frac{r^2+a^2}{\rho} \begin{pmatrix} 0 & 0 \\ 1 & 0 \end{pmatrix}
\delta(u-v)\, \begin{pmatrix} \omega
- \beta(v) & 1 \\[0.2em] \omega\, \big( \omega - \beta(v) \big) & \omega \end{pmatrix} Q_n^\omega \\
&= \frac{r^2+a^2}{\rho} \:\delta(u-v)\: \begin{pmatrix} 0 & 0 \\
\omega-\beta(v) & 1 \end{pmatrix} Q_n^\omega \:.
\end{align*}
Adding~\eqref{res1} and using~\eqref{Rwndef}, we conclude that
\[ (H_u - \omega) \,\mathfrak{R}_{\omega, n}(u,v)\:Q^\omega_n = \frac{r^2+a^2}{\rho}\, \delta(u-v)\, Q^\omega_n \:. \]
Summing over~$n$ and using the completeness relation~\eqref{strongcomplete} gives the result.
\QED

We conclude this section with a lemma which shows that the vector~$Q^\omega_n \Psi$
decays rapidly in the corresponding angular eigenvalues.

\begin{Lemma} \label{lemmanilpotent}
For any~$q \in \N$, there is a constant~$C(q)$ such that
for all~$\Psi \in C^\infty(S^2)$ of the form~$\Psi = e^{-i k \varphi}\: Y(\vartheta)$
as well as for all~$n \geq 1$ and all~$\omega \in U$ in the strip~\eqref{stripdef},
\[ \|Q^\omega_n \Psi \|_{L^2(S^2)}
\leq \frac{C(q)}{\displaystyle \inf_{0 \neq \lambda \in \sigma(\A_\omega Q^\omega_n)} |\lambda|^q} \sum_{\ell=0}^{2q} \|Q^\omega_n \A_\omega^\ell \Psi \|_{L^2(S^2)} \:. \]
\end{Lemma}
\Proof As shown in~\cite[Theorem~1.1]{tspectral} (see also~(ii) on page~\pageref{pageii}),
the operator~$Q^\omega_n$ with~$n \geq 1$ has rank at most two.
If it has rank two, we denote the nonzero spectrum of~$\A_\omega Q^\omega_n$ (counting algebraic multiplicities)
by~$\lambda$ and~$\tilde{\lambda}$. Otherwise, if it has rank one, we choose~$\lambda=\tilde{\lambda}$
as the nonzero eigenvalue of~$\A_\omega Q^\omega_n$. Then
\[ Q^n_\omega \,(\A_\omega - \lambda)(\A_\omega - \tilde{\lambda})  = 0 \:. \]
Multiplying out and solving for~$Q^n_\omega$, we obtain
\[ \lambda \tilde{\lambda}\:  Q^n_\omega = Q^n_\omega\, \big( -\A_\omega^2 + (\lambda+\tilde{\lambda}) \A_\omega \big)
 \:. \]
Next, we multiply by the phase factor~$|\lambda \tilde{\lambda}|/\lambda \tilde{\lambda}$ to obtain
\[ |\lambda \tilde{\lambda}| \,Q^n_\omega =
Q^n_\omega\, \frac{\lambda \tilde{\lambda}}{|\lambda \tilde{\lambda}|}
\Big(-\A_\omega^2 + (\lambda+\tilde{\lambda}) \A_\omega\Big) 
 \:. \]
 As a consequence,
\[ \big( 1 + |\lambda \big) \big(1+ \tilde{\lambda}| \big) \,Q^n_\omega =
Q^n_\omega\, \bigg( \frac{\lambda \tilde{\lambda}}{|\lambda \tilde{\lambda}|}
\Big(-\A_\omega^2 + (\lambda+\tilde{\lambda}) \A_\omega\Big) + |\lambda|+ |\tilde{\lambda}| + 1 \bigg)
 \:. \]
Iterating gives the identity
\[ Q^n_\omega = \frac{1}{(1 + |\lambda|)^q (1+|\tilde{\lambda}|)^q}\:
Q^n_\omega\, \bigg( \frac{\lambda \tilde{\lambda}}{|\lambda \tilde{\lambda}|}
\Big(-\A_\omega^2 + (\lambda+\tilde{\lambda}) \A_\omega\Big) + |\lambda|+ |\tilde{\lambda}| + 1 \bigg)^q \:. \]
We now apply this operator to the wave function~$\Psi$ and take the norm.
Multiplying out and using the triangle inequality, we obtain
\begin{align*}
\| &Q^n_\omega \Psi \|
\leq \sum_{a_1+ \cdots +a_6=q}
\begin{pmatrix} q \\ a_1 \cdots a_6 \end{pmatrix}\: \frac{|\lambda|^{a_2+a_4} \: |\tilde{\lambda}|^{a_3+a_5}}
{(1 + |\lambda|)^q (1+|\tilde{\lambda}|)^q}\:
\big\| Q^n_\omega \A_\omega^{2a_1+a_2+a_3} \Psi \big\| \\
&\leq \sum_{a_1+\cdots+a_4=q} \frac{c(q)}{(1 + |\lambda|)^{q-a_2-a_4} \:(1+|\tilde{\lambda}|)^{q-a_3-a_5}}\:
\big\| Q^n_\omega \A_\omega^{2a_1+a_2+a_3} \Psi \big\| \\
&\leq \sum_{a_1+\cdots+a_4=q} \frac{c(q)}{\big(1 + \min(|\lambda|, |\tilde{\lambda}|)\big)^{2q-a_2-a_3-a_4-a_5}}\:
\big\| Q^n_\omega \A_\omega^{2a_1+a_2+a_3} \Psi \big\| \\
&\leq \sum_{a_1+\cdots+a_4=q} \frac{c(q)}{\big(1 + \min(|\lambda|, |\tilde{\lambda}|)\big)^q}\:
\big\| Q^n_\omega \A_\omega^{2a_1+a_2+a_3} \Psi \big\| \\
&\leq \frac{c'(q)}{\big(1 + \min(|\lambda|, |\tilde{\lambda}|)\big)^q}\:
\sum_{\ell=1}^{2q}
\big\| Q^n_\omega \A_\omega^\ell \Psi \big\|
\leq \frac{c'(q)}{\min(|\lambda|, |\tilde{\lambda}|)^q}\:
\sum_{\ell=1}^{2q}
\big\| Q^n_\omega \A_\omega^\ell \Psi \big\|
\end{align*}
with combinatorial factors~$c(q)$ and~$c'(q)$. This gives the result.
\QED

\section{Contour Deformations}
Using the result of Theorem~\ref{thmsepres} in the integral representations of
Corollary~\ref{corprop}, we obtain an integral over an infinite sum of angular modes.
Since it is not clear a-priori whether the integration and summation can be interchanged,
our method is to first analyze the partial sums defined by
\begin{align}
\Psi^N(t) &= -\frac{1}{2 \pi i} \sum_{n=0}^N \int_{\R - 2 i c} e^{-i \omega t}\:
\Big( R_{\omega,n}\:Q_n^\omega \,\Psi_0 + Q_n^\omega \:\frac{\Psi_0}{\omega+3 i c} \Big)\:
d\omega \label{ps1} \\
\Psi^{N,p}(t) &= -\frac{1}{2 \pi i} \sum_{n=0}^N
\int_{\R - 2 i c} \frac{e^{-i \omega t}}{(\omega + 3 i c)^p} \;
\Big( R_{\omega,n}\:Q_n^\omega \,\big(H + 3 i c \big)^p \,\Psi_0  \Big)\: d\omega \label{ps2}
\end{align}
(where again~$p \geq 1$ and~$t < 0$).
After getting suitable estimates, we will be able to prove that the limit~$N \rightarrow \infty$
of the partial sums exists, both with the summation inside and outside the integral
(see Section~\ref{seclarge}).

We now use Whiting's mode stability result~\cite{whiting} to move the contour for the
partial sums up to the real axis:
\begin{Lemma} \label{lemmadeform}
For any~$\Psi_0 \in \D(H)$ and any integer~$p \geq 1$,
the partial sums~\eqref{ps1} and~\eqref{ps2} can be written for any~$t < 0$ as
\begin{align}
\Psi^N(t) &= -\frac{1}{2 \pi i} \sum_{n=0}^N \:\lim_{\varepsilon \searrow 0}
\int_{\R - i \varepsilon} e^{-i \omega t}\:
\Big( R_{\omega,n}\:Q_n^\omega \,\Psi_0 + Q_n^\omega \:\frac{\Psi_0}{\omega+3 i c} \Big)\: d\omega 
\label{psR1} \\
\Psi^{N,p}(t) &= -\frac{1}{2 \pi i} \sum_{n=0}^N\:\lim_{\varepsilon \searrow 0}
\int_{\R - i \varepsilon} \frac{e^{-i \omega t}}{(\omega + 3 i c)^p} \;
\Big( R_{\omega,n}\:Q_n^\omega \,\big(H + 3 i c \big)^p \,\Psi_0  \Big)\: d\omega \:. \label{psR2}
\end{align}
\end{Lemma}
\Proof We first verify that the above integrands decay so fast near~$\omega=\pm \infty$
that the integrals~\eqref{ps1}, \eqref{ps2} and~\eqref{psR1}, \eqref{psR2} converge.
To this end, given~$n$, we need to analyze the asymptotics for large~$\omega$.
In this asymptotic region, the angular eigenvalue~$\lambda$ scales like~$|\lambda| \lesssim |\omega|$
(see Lemma~\ref{lemmaangular1} in the appendix). Therefore, for large~$\omega$
the summand~$-\omega^2$ dominates all the other terms in~\eqref{Vdef}, so that the
potential goes over to a constant potential. In this limiting case, the solutions~$\acute{\phi}$
and~$\grave{\phi}$ go over to plane waves. By direct computation, one verifies
that in this limiting case, the kernels~$s_\omega(u,v)$
and~$g_\omega(u,v)$ in~\eqref{sdef} and~\eqref{gkerdef} are bounded,
uniformly in~$\omega$ and in~$u,v \in \R$.
Using that the matrix entries in~\eqref{Rwndef} involve~$\omega$ at most quadratically,
we can estimate the integral in~\eqref{Romn} with the Schwarz inequality to obtain
\beq \label{Rsimp}
\Big| \big(R_{\omega,n} \,Q^\omega_n \,\Psi \big)(u, \vartheta) \Big| \leq 
C\big(n, \text{supp}\, \Psi \big) \:\big(1 + |\omega|^2 \big)\:
\sup_{v \in \R} \big\| (Q^n_\omega \,\Psi)(v) \big\|_{L^2(S^2)} \:.
\eeq
Next, we can use Lemma~\ref{lemmanilpotent} as well
as the lower bound for~$|\lambda|$ in Lemma~\ref{lemmaangular1}
to generate factors of~$1/|\omega|$,
\beq \label{lamsimp}
\begin{split}
\big\| (Q^\omega_n \,\Psi)(v) \big\|_{L^2(S^2)} &\lesssim
\frac{C(q)}{\displaystyle \inf_{\lambda \in \sigma(\A_\omega Q^\omega_n)} |\lambda|^q} \sum_{\ell=0}^{2q}
\big\| (Q^\omega_n \,\A_\omega^\ell \Psi)(v) \big\|_{L^2(S^2)} \\
&\lesssim \frac{C(q)\, c(n)^q}{(1+|\omega|)^q}\:
\sum_{\ell=0}^{2q}
\big\| (Q^\omega_n \,\A_\omega^\ell \Psi)(v) \big\|_{L^2(S^2)} \:.
\end{split}
\eeq
Choosing~$q$ sufficiently large, we obtain the desired decay for large~$|\omega|$.

Since the angular operators~$Q^\omega_n$ as well as the Jost solutions
are holomorphic in~$\omega$, the integrand in the above contour integrals clearly is
meromorphic in~$\omega$. In order to rule out poles of the integrand, assume
the integrand has a pole at some~$\omega_0 \in \C \setminus \R$.
Then the operator~$R_{\omega, n}$ has a pole at~$\omega_0$. Consequently, its
kernel~$\mathfrak{R}_{\omega, n}$ in~\eqref{Rwndef} has a pole at~$\omega_0$.
This in turn implies that the kernel~$g_\omega$ has a pole at~$\omega_0$.
Using~\eqref{gkerdef}, it follows that the kernel~$s_\omega$ has a pole at~$\omega_0$.
Using~\eqref{sdef}, one sees that the Wronskian~$w(\acute{\phi}, \grave{\phi})$ vanishes at~$\omega_0$.
This gives rise to a mode solution at~$\omega_0$, in contradiction to Whiting's result~\cite{whiting}. We conclude that
the integrands in~\eqref{ps1} and~\eqref{ps2} are holomorphic in the strip~$-2c < \im \omega < 0$.
This makes it possible to deform the contours, giving the result.
\QED

\section{Estimates of the Potential} \label{secestimate}
Before entering the detailed ODE estimates, we give a brief outline of what needs to be done.
Generally speaking, our task is to show that the Jost solutions~$\acute{\phi}$ and~$\grave{\phi}$
are well-approximated by functions obtained by ``glueing together'' WKB, Airy and parabolic cylinder functions.
To this end, we need to construct the approximate solutions and derive rigorous error bounds.
For the construction of the approximate solutions, one needs to identify regions where the different
approximations (WKB, Airy and parabolic cylinder) apply. For the choice of these regions,
one must distinguish different cases. This analysis is carried out in this section.
The following section (Section~\ref{secinvregion}) is devoted to the derivation of the
error estimates.

Recall that our equations involve the
parameters~$k$, $s$, $\omega$ and~$\lambda$.
We always keep~$k$ and~$s$ fixed. The parameters~$\omega$ and~$\lambda$, however,
may vary in a certain parameter range to be specified later on,
and we must make sure that our estimates are uniform in these parameters.
In order to keep track of the dependence on~$\omega$ and~$\lambda$,
we use the same conventions and notation as in~\cite{tspectral}. Namely, we adopt the convention that
\[ \text{all constants are independent of~$\omega$ and~$\lambda$} \]
(but they may depend on~$k$ and~$s$).
Moreover, in order to have a compact and clear notation,
we always denote constants which may be increased during our constructions
by capital letters~$\Const_1, \Const_2, \ldots$.
However, constants with small letters~$\const_1, \const_2, \ldots$
are determined at the beginning and are fixed throughout. We use the symbol
\[ \lesssim \cdots \qquad \text{for} \qquad \leq \const\, \cdots \]
with a constant~$\const$ which is independent of the capital constants~$\Const_l$
(and may thus be fixed right away, without the need to increase it later on).

When increasing the constants~$\Const_l$, we must keep track of the mutual dependences
of these constants. We adopt the convention that the constant~$\Const_l$ may depend
on all previous constants~$\Const_1, \ldots, \Const_{l-1}$, but is independent of the subsequent
constants~$\Const_{l+1}, \ldots$. In particular, we may choose the capital constants such
that~$\Const_1 \ll \Const_2 \ll \cdots$.
This dependence of the constants implies that increasing~$\Const_l$ may also
make it necessary to increase the subsequent constants~$\Const_{l+1}, \Const_{l+2}, \ldots$.
For brevity, when we write ``possibly after increasing~$\Const_l$'' 
we implicitly mean that the subsequent constants~$\Const_{l+1}, \Const_{l+2}, \ldots$
are also suitably increased.

\subsection{Different Cases and Regions}
In Lemma~\ref{lemmadeform} we could deform the integration contours up to the real axis.
With this in mind, it suffices to consider the case that~$\omega$ is real.
Then the potential in the angular Teukolsky operator~$\A_\omega$ in~\eqref{Aop} is real.
Consequently, its eigenvalues are also real.
They have the properties as worked out in~\cite[Sections~5 and~7]{tspectral}
(see also the appendix of the present paper).
In particular, the angular operator has simple eigenvalues, which we order as
\[ \lambda_0 < \lambda_1 < \cdots \:. \]
In view of~\cite[Theorem~1.1]{tspectral} (see also~(ii) on page~\pageref{pageii}),
we know that the spectral points in the image of the operator~$Q^\omega_n$ are in the range
\beq \label{lambdarange}
\sigma(\A_\omega Q^\omega_n) \subset \big\{ \lambda_n, \ldots, \lambda_{2n+N+1} \big\} \:.
\eeq
Moreover, the imaginary part of~$\Omega$ as defined by~\eqref{Omegadef} is a negative constant,
\beq \label{varpidef}
\im \Omega = -\varpi \qquad \text{with} \qquad \varpi := \frac{(r_1-M) \,s}{r_1^2+a^2} > 0 \:.
\eeq

In preparation for proving convergence of our sum of contour integrals
(see Section~\ref{seclarge}), we need to
estimate the behavior of the fundamental solutions of the Sturm-Liouville equation~\eqref{schroedinger}
for large~$\lambda$ and~$|\omega|$. With this mind, we now restrict attention to the parameter range
\beq \label{range}
\boxed{ \quad \omega^2 \geq \Const_6 \qquad \text{and} \qquad
\lambda \geq \Const_7 \:. \quad }
\eeq
Expanding the potential in~\eqref{Vdef}, we obtain
\beq \label{Vsimp}
\begin{split}
V &= - \omega^2 + \frac{\lambda\, \Delta}{(r^2+a^2)^2} -\frac{2  \omega a k}{r^2+a^2}  \bigg(1 -\frac{2 \Delta}{r^2+a^2} \bigg) \\
&\qquad 
-\frac{2 i \omega s}{r^2+a^2} \bigg( M-r + \frac{2 r \Delta}{r^2+a^2} \bigg) 
+ \O\big( \omega^0 \big) + \O \big(\lambda^0 \big) \:.
\end{split}
\eeq
From this formula one sees in particular that the potential is almost constant if~$|\omega|$ is large.
This implies that the WKB conditions are satisfied, as is quantified in the next lemma.
\begin{Lemma} \label{lemma91}
Assume that
\[ \lambda < \Const_5 \,|\omega|^\frac{3}{2} \]
for a given constant~$\Const_5$. Then for any~$\varepsilon>0$, we can arrange by choosing
the constants~$\Const_6$ and~$\Const_7$ sufficiently large that
\[ \frac{|V'(u)|}{|V(u)|^\frac{3}{2}}, \frac{|V''(u)|}{|V(u)|^2} \leq \varepsilon \qquad \text{for all~$u \in \R$}\:, \]
uniformly in~$\omega$ and~$\lambda$.
\end{Lemma}
\Proof From the form of the potential~\eqref{Vsimp} it is obvious that
\beq \label{Velem}
\big| V + \omega^2 \big| \lesssim \lambda+|\omega| \qquad \text{and} \qquad
\big| V' \big|,  \big| V'' \big| \lesssim \lambda+|\omega| \:.
\eeq
As a consequence, $|V| \gtrsim \omega^2$ and thus
\begin{align*}
\frac{|V'|}{|V|^\frac{3}{2}} &\lesssim \frac{\lambda+|\omega|}{|\omega|^3} \leq \frac{\Const_5+|\omega|^{-\frac{1}{2}}}
{|\omega|^\frac{3}{2}} \lesssim \frac{\Const_5}{\Const_6^\frac{3}{4}} \\
\frac{|V''|}{|V|^2} &\lesssim \frac{\lambda+|\omega|}{|\omega|^4} \leq \frac{\Const_5+|\omega|^{-\frac{1}{2}}}
{|\omega|^\frac{5}{2}}
\leq \frac{\Const_5+1}{\Const_6^\frac{5}{4}} \:.
\end{align*}
This concludes the proof.
\QED
In view of this result, in what follows we may assume that
\beq \label{range1}
\boxed{ \quad \lambda \geq \Const_5 \,|\omega|^\frac{3}{2}\:, \quad }
\eeq
because otherwise the potential is nearly constant and can be treated easily with the
WKB approximation. Then, choosing~$\Const_5$ sufficiently large, we can sometimes
work with the simpler approximation
\beq \label{Vapprox}
V = - \omega^2 + \frac{\lambda\, \Delta}{(r^2+a^2)^2} + \O\big( \omega \big) + \O \big(\lambda^0 \big) \:.
\eeq

Discussing the form of the approximate potential immediately gives the following result:
\begin{Lemma} For any~$\omega$ and~$\lambda$ in the range~\eqref{range} and~\eqref{range1},
we have the bounds
\beq \label{imVes}
\big| \im V \big|, \big| \im V' \big|,  \big| \im V'' \big| \lesssim |\omega|\:.
\eeq

Moreover, for sufficiently large~$\Const_5$, the real part of the potential has a unique maximum at a
point~$\umax$ with~$\frac{12}{5}\,m \leq r(\umax) \leq 3m$. The maximal value is bounded by
\[ \re V(\umax) \lesssim \lambda \:. \]
The function~$\re V$ is concave near near~$\umax$.
More quantitatively, there is a constant~$\const$ such that
\beq \label{concave}
\frac{\lambda}{\const} \leq -\re V''(u) \leq \const \lambda \qquad \text{on} \qquad
\Big[ \umax-\frac{1}{2}, \umax+\frac{1}{2} \Big] \:.
\eeq
\end{Lemma}
\Proof We substitute the formula for~$\Delta$, \eqref{UDdef},
into~\eqref{Vapprox} and compute the first and and second $u$-derivatives
with the help of the formula (see also~\eqref{RW})
\[ \frac{\partial}{\partial u} = \frac{\Delta}{r^2+a^2} \:\frac{\partial}{\partial r} \:. \]
Then the result follows by a direct computation.
\QED

The value of the real part of the potential at the point~$\umax$ distinguishes different {\em{cases}}:
\beq \label{cases}
\left\{ \begin{array}{lcl} \text{WKB case} && \text{if~$\re V(\umax) < -\Const_4\, \sqrt{\lambda}$} \\[0.3em]
\text{parabolic cylinder (PC) case} && \text{if~$-\Const_4\, \sqrt{\lambda} \leq \re V(\umax) <
\Const_4\,\sqrt{\lambda}$} \\[0.3em]
\text{Airy case} && \text{if~$\re V(\umax) \geq \Const_4\, \sqrt{\lambda}\:.$}\end{array} \right.
\eeq
In each of these above cases, we estimate the solution by considering different {\em{regions}},
which we now introduce. To this end, we work with the zeros of the function~$\re V$
characterized in the next lemma.
\begin{Lemma} By increasing the constant~$\Const_5$ in~\eqref{range1} we can arrange
that whenever~$\re V(\umax)>0$, there are unique points~$u^L_0, u^R_0 \in \R$ with
\[ \re V(u^L_0) = 0 = \re V(u^R_0) \qquad \text{and} \qquad u^L_0 \leq \umax \leq u^R_0\:. \]
Furthermore, the function~$\re V$ is monotone increasing on~$(-\infty, \umax)$,
and it is monotone decreasing on~$(\umax, \infty)$.
\end{Lemma}
\Proof Since~$\omega^2$ enters the potential~\eqref{Vdef} only as a constant,
for large~$\Const_5$ the derivative~$\re V'$ is dominated by the term~$\lambda \partial_u(\Delta/(r^2+a^2)^2)$.
An asymptotic expansion shows that~$\re V'$ is positive near~$u=-\infty$ and negative near~$u=\infty$.
Moreover, the function~$\Delta/(r^2+a^2)^2$ is monotone increasing
up to a turning point where its second derivative is negative, and from then on is monotone decreasing.
This gives the result.
\QED
If~$\re V(\umax)>0$, we denote the zeros of~$\re V$ by~$u^L_0$ and~$u^R_0$.
If~$\re V(\umax)<0$ (as is always true in the WKB case and may be true in the
PC case), we set~$u^{L\!/\!R}_0 = \umax$. We introduce
\begin{align}
\hspace*{-0.3em} u^L_- &= \left\{ \begin{array}{ll}
u^L_0 & \text{in the WKB case} \\
u^L_0 - \Const_3 \,\Const_1^{-\frac{1}{6}} \:|\omega|^{-\frac{1}{2}} & \text{in the PC case} \\
u^L_0 - \Const_3 \max \Big( |\omega|^{-\frac{2}{3}}, \big(\Const_1 \re V(\umax)
\big)^{-\frac{1}{6}}\, |\omega|^{-\frac{1}{3}} \Big) &\text{in the Airy case} 
\end{array} \right. \label{uLminus} \\
\hspace*{-0.3em} u^R_- &= \left\{ \begin{array}{ll}
u^R_0 & \text{in the WKB case} \\
u^R_0 + \Const_3 \,\Const_1^{-\frac{1}{6}} \:|\omega|^{-\frac{1}{2}} & \text{in the PC case} \\[0.2em]
u^R_0 + \Const_3 \, \lambda^{\frac{1}{6}}\, |\omega|^{-\frac{1}{3}} \\
\qquad\;\;\, \times 
\max \Big( |\omega|^{-\frac{2}{3}}, \big(\Const_1 \re V(\umax)
\big)^{-\frac{1}{6}}\, |\omega|^{-\frac{1}{3}} \Big) &\text{in the Airy case} \:.
\end{array} \right. \label{uRminus}
\end{align}
Moreover, in the Airy case we set
\begin{align}
u^L_+ &= u^L_0 + \Const_3 \max \Big( |\omega|^{-\frac{2}{3}}, \big(\Const_1 \re V(\umax)
\big)^{-\frac{1}{6}}\, |\omega|^{-\frac{1}{3}} \Big) \label{uLplus} \\
u^R_+ &= u^R_0 - \Const_3 \,\lambda^{\frac{1}{6}}\, |\omega|^{-\frac{1}{3}}\,
\max \Big( |\omega|^{-\frac{2}{3}}, \big(\Const_1 \re V(\umax)
\big)^{-\frac{1}{6}}\, |\omega|^{-\frac{1}{3}} \Big) \:.  \label{uRplus}
\end{align}
We thus obtain the following regions:
\beq \label{regions}
\left\{ \begin{array}{lll} 
\text{WKB regions} &(-\infty, u^L_-),\;(u^R_-, \infty) & \text{in all cases} \\[0.3em]
\text{PC region} &(u^L_-, u^R_-) & \text{in the PC case} \\[0.3em]
\text{Airy regions} &(u^L_-, u^L_+),\;(u^R_+, u^R_-) & \text{in the Airy case} \\[0.3em]
\text{WKB region with~$\re V > 0$} &(u^L_+, u^R_+) & \text{in the Airy case}\:. \\[0.3em]
\end{array} \right.
\eeq

\subsection{Estimates in the WKB Regions}
In this section we shall prove the following results:
\begin{Prp} \label{prpWKB} For any~$\varepsilon>0$, we can arrange by choosing
the constants~$\Const_1, \ldots, \Const_4$ sufficiently large that
for all~$\omega$ and~$\lambda$ in the range~\eqref{range},
\[ \frac{|V'|}{|V|^\frac{3}{2}}, \frac{|V''|}{|V|^2} \leq \varepsilon \qquad \text{on~$(-\infty, u^L_-)$}\:. \]
\end{Prp}

\begin{Prp} \label{prpWKBR} For any~$\varepsilon>0$, we can arrange by choosing
the constants~$\Const_1, \ldots, \Const_4$ sufficiently large that
for all~$\omega$ and~$\lambda$ in the range~\eqref{range},
\[ \frac{|V'|}{|V|^\frac{3}{2}}, \frac{|V''|}{|V|^2} \leq \varepsilon \qquad \text{on~$(u^R_-, \infty)$}\:. \]
\end{Prp}

\Proof[Proof of Propositions~\ref{prpWKB} and~\ref{prpWKBR} in the WKB case in~\eqref{cases}]
From~\eqref{Vapprox} and~\eqref{range1}, it is obvious that
\beq \label{Vplam}
|V'|, |V''| \lesssim \lambda \:.
\eeq
Therefore, using~\eqref{cases}, we find that
\[ \frac{|V''|}{|V|^2} \lesssim \frac{1}{\Const_4^2} \:. \]
Moreover, using~\eqref{imVes} and~\eqref{range1}, we get
\[ \frac{|\im V'|}{|V|^\frac{3}{2}} \lesssim \frac{|\omega|}{\Const_4^\frac{3}{2}\, \lambda^\frac{3}{4}}
\lesssim \frac{|\omega|}{\Const_4^\frac{3}{2}\, \Const_5^\frac{3}{4}\: |\omega|^\frac{9}{8}}
= \frac{1}{\Const_4^\frac{3}{2}\, \Const_5^\frac{3}{4}\: |\omega|^\frac{1}{8}}
\lesssim \frac{1}{\Const_4^\frac{3}{2}\, \Const_5^\frac{3}{4}\: \Const_6^\frac{1}{16}}
\:. \]
In order to bound~$|\re V'|/|V|^\frac{3}{2}$, we consider the three different cases
\begin{itemize}
\item[(A)] $|u-\umax| \leq \frac{1}{2}$ and~$|u-\umax| > \Const_0 \lambda^{-\frac{1}{4}}$
\item[(B)] $|u-\umax| \leq \Const_0 \lambda^{-\frac{1}{4}}$
\item[(C)] $|u-\umax| > \frac{1}{2}$,
\end{itemize}
where the constant~$\Const_0$ will be specified below.
In cases~(A) and~(B), we can use~\eqref{cases} and integrate the inequality~\eqref{concave} to obtain
\begin{align}
|\re V'(u)| &\lesssim \lambda \;|u-\umax| \label{VpWKB} \\
\re V(u) &\leq -\Const_4\, \sqrt{\lambda} - \frac{\lambda}{2\const}\, (u-\umax)^2 \:. \label{VWKB}
\end{align}
In case~(A), we drop the first summand in~\eqref{VWKB} and use~\eqref{VpWKB} to obtain
\[ \frac{|V'|}{|V|^\frac{3}{2}} \lesssim \frac{\lambda}{\lambda^\frac{3}{2}\, (u-\umax)^2} 
\lesssim \frac{\lambda}{\lambda^\frac{3}{2}\: \Const_0^2 \lambda^{-\frac{1}{2}}} = \frac{1}{\Const_0^2} \:, \]
which can be made arbitrarily small by choosing~$\Const_0$ sufficiently large.
In case~(B), we drop the second summand in~\eqref{VWKB} and again use~\eqref{VpWKB},
\[ \frac{|V'|}{|V|^\frac{3}{2}} \lesssim \frac{\lambda\, |u-\umax|}{\Const_4^\frac{3}{2} \, \lambda^{\frac{3}{4}}} 
\lesssim \frac{\Const_0}{\Const_4^\frac{3}{2}} \:, \]
which can be made arbitrarily small by increasing~$\Const_4$.

In the remaining case~(C), we know from the monotonicity of~$\re V$ on the intervals~$(-\infty, \umax-\frac{1}{2})$
and~$(\umax+\frac{1}{2}, \infty)$ that
\[ \re V(u) \leq \re V \Big( \umax \pm \frac{1}{2} \Big) \lesssim -\lambda\:, \]
where in the last step we used~\eqref{VWKB}. Hence, using again~\eqref{Vplam},
\[ \frac{|\re V'|}{|V|^\frac{3}{2}} \lesssim \frac{\lambda}{\lambda^\frac{3}{2}} \lesssim \frac{1}{\sqrt{\Const_7}}\:. \]
This concludes the proof.
\QED
It remains to prove Propositions~\ref{prpWKB} and~\ref{prpWKBR} in the Airy and PC cases in~\eqref{cases}.
We distinguish the two cases
\beq \label{abcase}
\textbf{(a)} \quad \omega^2 > \Const_1 \re V(\umax) \qquad \text{and} \qquad
\textbf{(b)} \quad \omega^2 \leq \Const_1 \re V(\umax) \:.
\eeq
We begin with two preparatory lemmas.
\begin{Lemma} \label{lemmaprep} In the Airy case, by increasing~$\Const_7$ we can arrange that
\beq \label{lup}
\lambda \gtrsim \omega^2 \:.
\eeq
Moreover,
\beq \label{llow}
\lambda \lesssim \omega^2 \qquad \text{in case~{\bf{(a)}}}\:.
\eeq
\end{Lemma}
\Proof From~\eqref{cases} and the form of the potential~\eqref{Vsimp}, we obtain
\beq \label{Vesairy}
-\Const_4 \,\sqrt{\lambda} \leq \re V(\umax) \leq -\omega^2 + \const \lambda \:.
\eeq
Applying~\eqref{range} and increasing~$\Const_7$, we obtain~\eqref{lup}.

In case~{\bf{(a)}}, we know that
\[ \omega^2 > \Const_1 \re V(\umax) \geq \Const_1 \,\Big( -\omega^2 + \frac{\lambda}{\const} \Big) , \]
implying that
\[ \lambda \lesssim \Big( 1 + \frac{1}{\Const_1} \Big) \omega^2 \lesssim \omega^2 \:. \]
This gives~\eqref{llow} in case~{\bf{(a)}}.
\QED

\begin{Lemma} \label{lemmaPCa}
In the PC case, by increasing~$\Const_6$ and~$\Const_7$
we can arrange that we are in case~{\bf{(a)}}.
Moreover, the inequalities~\eqref{lup} and~\eqref{llow} again hold.
\end{Lemma}
\Proof The inequality~\eqref{lup} again follows from~\eqref{Vesairy}.
Moreover, we know from~\eqref{cases} that
\[ \Const_4 \sqrt{\lambda} > \re V(\umax) \geq -\omega^2 + \frac{\lambda}{\const}\:. \]
Choosing the constant~$\Const_7$ in~\eqref{range} sufficiently large, we obtain~\eqref{llow}.

Next, by combining~\eqref{cases} with~\eqref{llow} and~\eqref{range}, we obtain
\[ \Const_1 \re V(\umax) < \Const_1 \Const_4\, \sqrt{\lambda} \lesssim 
\Const_1 \Const_4\, |\omega| < \frac{\Const_1 \Const_4}{\sqrt{\Const_6}}\: \omega^2 \:. \]
This concludes the proof.
\QED

The next lemma gives an alternative characterization of the cases in~\eqref{abcase}.
\begin{Lemma} The maximal value of~$\re V$ satisfies the bounds
\beq \label{reVes}
\left\{ \begin{array}{cl} \displaystyle \re V(\umax) \lesssim \frac{\lambda}{\Const_1} & \text{in case~{\bf{(a)}}} \\[1em]
\displaystyle \re V(\umax) \gtrsim \frac{\lambda}{\Const_1} & \text{in case~{\bf{(b)}}}\:. \end{array} \right.
\eeq
Moreover,
\beq \label{u0es}
\left\{ \begin{array}{cl} \displaystyle \big(\umax - u^L_0\big), \,
\big(u^R_0 - \umax\big) \lesssim \frac{1}{\sqrt{\Const_1}} & \text{in case~{\bf{(a)}}} \\[1em]
\displaystyle \big(\umax - u^L_0\big), \, \big(u^R_0 - \umax\big)
\gtrsim \frac{1}{\sqrt{\Const_1}} & \text{in case~{\bf{(b)}}}\:. \end{array} \right.
\eeq
\end{Lemma}
\Proof In case~{\bf{(a)}}, it follows that
\[ \frac{\re V(\umax)}{\lambda} < \frac{\omega^2}{\Const_1\, \lambda} \lesssim \frac{1}{\Const_1}\:, \]
where in the last step we applied Lemma~\ref{lemmaprep}. Likewise, in case~{\bf{(b)}}, from~\eqref{abcase}
and~\eqref{Vapprox} we have the alternative estimates
\[ \re V(\umax) \gtrsim \frac{\omega^2}{\Const_1} \qquad \text{and} \qquad
\re V(\umax) \gtrsim -\omega^2 + \frac{\lambda}{\const} \:. \]
Multiplying the first inequality by~$\Const_1$ and adding them, we obtain
\[ (1+\Const_1) \re V(\umax) \gtrsim \frac{\lambda}{\const} \:, \]
giving~\eqref{reVes}.

In order to derive~\eqref{u0es}, we approximate the function~$\re V$ near its maximum by a
parabola. More precisely, integrating~\eqref{concave}, we obtain
\beq \label{Vines}
- \frac{\const \lambda}{2}\: (u-\umax)^2 \leq \re V(u) - \re V(\umax) \leq - \frac{\lambda}{2\const}\: (u-\umax)^2 \:,
\eeq
valid if~$|u-\umax| \leq \frac{1}{2}$.
In case~{\bf{(a)}}, applying~\eqref{reVes} and increasing~$\Const_1$, one sees
that the function~$\re V$ has zeros~$u^L_0$ and~$u^R_0$
in a neighborhood of~$\umax$, as is made precise in~\eqref{u0es}.
Likewise, in case~{\bf{(b)}} we find that the function~$\re V$ has no zero in a $1/\sqrt{\Const_1}$-neighborhood
of~$\umax$, implying~\eqref{u0es}.
\QED

For the remaining estimates, we consider the regions~$(-\infty, u^L_-)$ and~$(u^R_-, \infty)$ separately.
We begin with the region~$(-\infty, u^L_-)$ as considered in Proposition~\ref{prpWKB}.
We treat the PC case and the Airy case in the two subcases in~\eqref{abcase} after each other.

\Proof[Proof of Proposition~\ref{prpWKB} in the PC case]
We first consider the region~$|u-\umax| > \frac{1}{2}$.
Since we are in case~{\bf{(a)}} in~\eqref{abcase}, the inequality~\eqref{Vines} gives
\beq
\re V\Big(\umax - \frac{1}{2} \Big) \lesssim -\lambda + \frac{\omega^2}{\Const_1}
\overset{\eqref{lup}}{\lesssim} -\lambda \:. \label{ularge1}
\eeq
Using the monotonicity of~$\re V$ on the interval~$(-\infty, \umax-\frac{1}{2})$, we conclude that
\[ \re V \lesssim -\lambda \qquad \text{on~$\Big(-\infty, \umax-\frac{1}{2} \Big)$}\:. \]
Hence, using~\eqref{Vplam},
\beq
\frac{|V'|}{|V|^\frac{3}{2}} \lesssim \frac{1}{\sqrt{\lambda}} 
\overset{\eqref{range}}{\lesssim} \frac{1}{\sqrt{\Const_7}} 
\:,\qquad
\frac{|V''|}{|V|^2} \lesssim \frac{1}{\lambda} \overset{\eqref{range}}{\lesssim} \frac{1}{\Const_7}  \:. \label{ularge2}
\eeq

On the remaining interval~$[\umax-\frac{1}{2}, u^L_-)$,
by integrating~\eqref{concave}, we obtain for any~$u < u^L_-$,
\begin{gather*}
|\re V'(u)| \leq \const \lambda\, |u-\umax| \\
-\re V(u) \geq -\re V(u) + \re V(u^L_0) \geq \frac{\lambda}{2 \const} \:\Big( (u-\umax)^2 - (u^L_0 - \umax)^2 \Big) \:.
\end{gather*}
Setting~$v=\umax-u$ and~$v_0 = \umax - u^L_0$, we obtain
\begin{gather}
\big|\re V'(u) \big| \leq \const \lambda\, v \label{Vpes} \\
-\re V(u) \geq \frac{\lambda}{2\const} \,\big( v^2 - v_0^2 \big) = \frac{\lambda}{2\const} \,(v-v_0)(v+v_0) 
\label{Vlowera} \:.
\end{gather}
Moreover, from~\eqref{uLminus} we know that
\[ v-v_0 = u^L_0 - u \geq u^L_0 - u^L_- = \Const_3 \,\Const_1^{-\frac{1}{6}} \:|\omega|^{-\frac{1}{2}} \:. \]
Combining these inequalities, we obtain
\begin{align*}
\frac{|\re V'|^2}{|V|^3} &\lesssim \frac{\lambda^2 v^2}{\lambda^3\, (v-v_0)^3 (v+v_0)^3}
\leq \frac{v^2}{\lambda\, (v-v_0)^4 (v+v_0)^2} \\
&\leq \frac{1}{\lambda\, (v-v_0)^4}
\leq \frac{\omega^2 \,\Const_1^\frac{2}{3}}{\lambda\, \Const_3^4} \overset{\eqref{lup}}{\lesssim}
\frac{\Const_1^\frac{2}{3}}{\Const_3^4} \:.
\end{align*}
Similarly, using~\eqref{imVes},
\begin{align*}
\frac{|\im V'|^2}{|V|^3} &\lesssim \frac{\omega^2}{\lambda^3\, (v-v_0)^3 (v+v_0)^3}
\leq \frac{\omega^2}{\lambda^3\, (v-v_0)^6} \\
&\leq \frac{\omega^2\, \Const_1\, |\omega|^3}{\lambda^3\, \Const_3^6}
\overset{\eqref{lup}}{\lesssim}
\frac{\Const_1}{\Const_3^6} \: \frac{1}{|\omega|} \lesssim \frac{\Const_1}{\Const_3^6\, \sqrt{\Const_6}} \:.
\end{align*}
This concludes the proof for the term involving the first derivatives.

The second derivatives can be handled similarly by using~\eqref{Vplam},
\begin{align*}
\frac{|V''|}{|V|^2} &\lesssim \frac{\lambda}{\lambda^2\, (v-v_0)^2 (v+v_0)^2}
\leq \frac{1}{\lambda\, (v-v_0)^4}
\leq \frac{\Const_1^\frac{2}{3}\, |\omega|^2}{\lambda\, \Const_3^4}
\overset{\eqref{lup}}{\lesssim}
\frac{\Const_1^\frac{2}{3}}{\Const_3^4} \:.
\end{align*}
This concludes the proof.
\QED

\Proof[Proof of Proposition~\ref{prpWKB} in the Airy case in subcase~{\bf{(a)}}]
If~$|u-\umax| > \frac{1}{2}$, we can again use the estimates in~\eqref{ularge1}--\eqref{ularge2}.
Therefore, it suffices to consider the region~$[\umax-\frac{1}{2}, u^L_-]$.
Then,  by integrating~\eqref{concave}, one finds exactly as in the PC case that~\eqref{Vpes} and~\eqref{Vlowera}
again hold. Moreover, in case~{\bf{(a)}}, equation~\eqref{uLminus} reduces to
\beq
\label{uLa}
u^L_- = u^L_0 - \Const_3 \big(\Const_1 \re V(\umax) \big)^{-\frac{1}{6}}\, |\omega|^{-\frac{1}{3}} \:.
\eeq
Combining these inequalities, we conclude that
\[ \frac{|\re V'(u)|^2}{|V(u)|^3} \lesssim \frac{\lambda^2\, v^2}{\lambda^3\, (v-v_0)^3\, (v+v_0)^3}
\leq \frac{\big(\Const_1 \re V(\umax) \big)^{\frac{1}{2}}\, |\omega|\: v^2}{\lambda\, 
\Const_3^3 \,(v+v_0)^3} \:. \]
Next, we can estimate~$\re V(\umax)$ by
\beq \label{Vup}
0 = \re \big( V(u^L_0) \big) \geq \re V(\umax) - \frac{\const \lambda}{2}\: \, v_0^2 
\qquad \text{and thus} \qquad
\re V(\umax) \lesssim \lambda\, v_0^2 \:.
\eeq
We thus obtain
\begin{align*}
\frac{|\re V'(u)|^2}{|V(u)|^3} \lesssim \frac{\sqrt{\Const_1}}{\Const_3^3}\; \frac{|\omega|}{\sqrt{\lambda}}\:
\frac{v_0\: v^2}{(v+v_0)^3} \lesssim \frac{\sqrt{\Const_1}}{\Const_3^3}\:,
\end{align*}
where in the last step we used~\eqref{lup}.
The imaginary part of~$V'$ can be handled similarly. Namely,
from~\eqref{imVes} and~\eqref{lup}, we know that
\[ \frac{|\im V'(u)|^2}{|V(u)|^3} \lesssim \frac{\lambda}{\lambda^3\, (v-v_0)^3\, (v+v_0)^3} 
\leq \frac{1}{\lambda^2\, (v-v_0)^\frac{9}{2}\, (v+v_0)^\frac{3}{2}} \:. \]
Applying again~\eqref{uLa} and~\eqref{Vup}, we obtain
\begin{align*}
\frac{|\im V'(u)|^2}{|V(u)|^3} &\lesssim
\frac{\big(\Const_1 \re V(\umax) \big)^{\frac{3}{4}} |\omega|^{\frac{3}{2}}}{\lambda^2\, \Const_3^\frac{9}{2}\,
(v+v_0)^\frac{3}{2}} \lesssim
\frac{\big(\Const_1 \lambda\, v_0^2 \big)^{\frac{3}{4}} |\omega|^{\frac{3}{2}}}{\lambda^2\, \Const_3^\frac{9}{2}\,
(v+v_0)^\frac{3}{2}} \\
&\lesssim \frac{\big(\Const_1 \lambda \big)^{\frac{3}{4}} |\omega|^{\frac{3}{2}}}{\lambda^2\, \Const_3^\frac{9}{2}}
\overset{\eqref{lup}}{\lesssim} \frac{\Const_1^{\frac{3}{4}}}{\sqrt{\lambda}\, \Const_3^\frac{9}{2}}
\leq \frac{\Const_1^{\frac{3}{4}}}{\sqrt{\Const_7}\, \Const_3^\frac{9}{2}}\:.
\end{align*}
This concludes the proof of the term involving the first derivatives.

The second derivatives can be handled similarly as follows,
\begin{align*}
\big|V''(u) \big| &\leq \big|\re V''(u) \big| + \big|\im V''(u) \big|
\overset{\eqref{imVes}, \eqref{concave}}{\leq}
\const \big( \lambda + |\omega| \big) \overset{\eqref{range1}}{\lesssim} \lambda \\
\frac{|V''(u)|}{|V(u)|^2} &\lesssim \frac{\lambda}{\lambda^2\, (v-v_0)^2\, (v+v_0)^2}
\leq \frac{1}{\lambda\, (v-v_0)^3\, (v+v_0)} \\
&\!\!\!\overset{\eqref{uLa}}{\leq} \frac{\big(\Const_1 \re V(\umax) \big)^{\frac{1}{2}}\, |\omega|}{\lambda\,
\Const_3^3 \, (v+v_0)}
\overset{\eqref{Vup}}{\lesssim} \frac{\big(\Const_1 \lambda v_0^2 \big)^{\frac{1}{2}}\, |\omega|}
{\lambda\, \Const_3^3 \,(v+v_0)}
\overset{\eqref{lup}}{\lesssim} \frac{\Const_1^\frac{1}{2}}{\Const_3^3} \: \frac{v_0}{v+v_0}
\leq \frac{\Const_1^\frac{1}{2}}{\Const_3^3}\:.
\end{align*}
This concludes the proof.
\QED

\Proof[Proof of Proposition~\ref{prpWKB} in the Airy case in subcase~{\bf{(b)}}]
In this case, equation~\eqref{uLminus} reduces to
\beq \label{uLmdef}
u^L_- = u^L_0 - \Const_2 \,|\omega|^{-\frac{2}{3}} \:.
\eeq
Expanding the potential~\eqref{Vapprox} similar to~\eqref{VLasy}, one sees that
for sufficiently large~$\lambda$, our potential satisfies the inequalities
\begin{align}
\big| V'(u) \big| &\leq \const \,\lambda\: e^{\gamma u} &&\hspace*{-1.5cm} \text{on~$(-\infty, \umax)$} \label{Vpup} \\
\big|V''(u) \big| &\leq \const\, \lambda\, e^{\gamma u} &&\hspace*{-1.5cm} \text{on~$(-\infty, \umax)$} \label{Vpp} \\
\re V'(u) &\geq \frac{\lambda}{\Const_2}\: e^{\gamma u} &&\hspace*{-1.5cm} \text{on~$\Big(-\infty, \umax-
\frac{1}{\Const_1} \Big)$} \label{Vplow}
\end{align}
for a suitable choice of the constants~$\const, \Const_2>1$.
Using~\eqref{VLasy}, \eqref{Vpup} and~\eqref{Vplow}, it follows that
\begin{align*}
\re \big( \Omega^2 \big) &= \int_{-\infty}^{u^L_0} \re V'
\leq \const \lambda  \int_{-\infty}^{u^L_0} e^{\gamma v}\: dv = \frac{\const \lambda}{\gamma}\: e^{\gamma u^L_0} \\
\re \big( \Omega^2 \big) &= \int_{-\infty}^{u^L_0} \re V'
\geq \frac{\lambda}{\Const_2}  \int_{-\infty}^{u^L_0-1} e^{\gamma v}\: dv = \frac{\lambda e^{-\gamma}}{\Const_2 \gamma}\:
e^{\gamma u^L_0}
\end{align*}
and thus
\[ \frac{\lambda e^{-\gamma}}{\Const_2 \gamma}\: e^{\gamma u^L_0}
\leq \re \big( \Omega^2 \big) \leq \frac{\const \lambda}{\gamma}\: e^{\gamma u^L_0} \:. \]
According to~\eqref{Omegadef}, the imaginary part of~$\Omega$ is uniformly bounded.
Combining this fact with the first inequality in~\eqref{range}, we obtain
\beq \label{reo2}
\frac{\lambda e^{-\gamma}}{2 \Const_2 \gamma}\: e^{\gamma u^L_0}
\leq \omega^2 \leq \frac{2 \const \lambda}{\gamma}\: e^{\gamma u^L_0} \:.
\eeq

Combining the above inequalities, for any~$u<u^L_0$ we obtain
\begin{align*}
-\re V(u) &= \int_u^{u^L_0} \re V' \geq \frac{\lambda}{\Const_2} \int_u^{u^L_0} e^{\gamma v}\: dv 
= \frac{\lambda}{\Const_2 \gamma} \:\big( e^{\gamma u^L_0} - e^{\gamma u} \big) \\
&= \frac{\lambda}{\Const_2 \gamma} e^{\gamma u^L_0}\: \big(1 - e^{\gamma (u-u^L_0)} \big)
\geq \frac{1}{2 \const \,\Const_2} \:\big(1 - e^{\gamma (u-u^L_0)} \big)\: \omega^2 \:,
\end{align*}
where in the last step we applied~\eqref{reo2}.
Using~\eqref{uLmdef}, we conclude that for any~$u<u^L_-$ and
for sufficiently large~$|\omega|$,
\beq \label{reVesL}
-\re V(u) \geq \frac{1}{2 \const\, \Const_2} \:\frac{\gamma}{2}\: \Const_3\, |\omega|^{-\frac{2}{3}} \: \omega^2
\gtrsim \frac{\gamma\, \Const_3}{\const\, \Const_2}  \;|\omega|^{\frac{4}{3}}\:.
\eeq

It follows that
\begin{align*}
\frac{|V'|}{|V|^\frac{3}{2}} &\leq \frac{\const\,\lambda\, e^{\gamma u}}{|V|^\frac{3}{2}}
\leq \frac{\const\,\lambda\, e^{\gamma u^L_0}}{|V|^\frac{3}{2}}
\lesssim \frac{\const\, \Const_2 \gamma\:|\omega|^2}{|V|^\frac{3}{2}}
\lesssim \const\, \Const_2 \gamma\, \left( \frac{\const \, \Const_2}{\gamma\, \Const_3} \right)^\frac{3}{2} \\
\frac{|V''|}{|V|^2} &
\leq \frac{\const\,\lambda\, e^{\gamma u^L_0}}{|V|^2}
\lesssim \const\, \Const_2 \gamma\, \left( \frac{\const \, \Const_2}{\gamma\, \Const_3}
\right)^2\, |\omega|^{-\frac{2}{3}} \:.
\end{align*}
Obviously, the right side of these inequalities can be made arbitrarily small by increasing~$\Const_3$.
\QED

It remains to consider the region~$(u^R_-, \infty)$ as considered in Proposition~\ref{prpWKBR}.
We again treat PC case and the Airy case in the two subcases in~\eqref{abcase} after each other.

\Proof[Proof of Proposition~\ref{prpWKBR} in the PC case]
If~$|u-\umax| > \frac{1}{2}$, we can again use the estimates in~\eqref{ularge1}--\eqref{ularge2}.
Therefore, it suffices to consider the region~$[u^R_-, \umax+\frac{1}{2}]$.
In this region, we can proceed exactly as in the proof of Proposition~~\ref{prpWKB}
in the PC case.
\QED

\Proof[Proof of Proposition~\ref{prpWKBR} in the Airy case in subcase~{\bf{(a)}}]
If~$|u-\umax| > \frac{1}{2}$, we can again use the estimates in~\eqref{ularge1}--\eqref{ularge2}.
Therefore, it suffices to consider the region~$[u^R_-, \umax+\frac{1}{2}]$.

In case~{\bf{(a)}}, equation~\eqref{uRminus} reduces to
\[ u^R_- = u^R_0 + \Const_3 \big(\Const_1 \re V(\umax) \big)^{-\frac{1}{6}}\, 
\lambda^{\frac{1}{6}}\, |\omega|^{-\frac{2}{3}} \:. \]
Using Lemma~\ref{lemmaprep}, we know that~$\lambda^{\frac{1}{6}} \eqsim \omega^{\frac{1}{3}}$, so that
\[ u^R_- \eqsim u^R_0 + \Const_3 \big(\Const_1 \re V(\umax) \big)^{-\frac{1}{6}} \, |\omega|^{-\frac{1}{3}} \:. \]
Therefore, the identity~\eqref{uLa} again holds up to a uniform constant.
This makes it possible to proceed just as in the proof of Proposition~\ref{prpWKB} after~\eqref{uLa}.
\QED

\Proof[Proof of Proposition~\ref{prpWKBR} in the Airy case in subcase~{\bf{(b)}}]
In this case, equation~\eqref{uRminus} reduces to
\beq \label{uRmdef}
u^R_- = u^R_0 + \Const_3 \,\lambda^{\frac{1}{6}}\, |\omega|^{-1} \:.
\eeq
Using the form of the potential~\eqref{Vdef} and~\eqref{VRasy}, we obtain the estimates
\begin{align}
\re V + \omega^2 &\eqsim \frac{\tilde{\lambda}}{u^2} &&\hspace*{-1.5cm} \text{on~$(\umax, \infty)$} \label{reV1} \\
\Longrightarrow \;u^R_0 &\eqsim \frac{\tilde{\lambda}^\frac{1}{2}}{|\omega|} \label{u0es1} \\
-\re V'(u) &\lesssim \frac{\tilde{\lambda}}{u^3} &&\hspace*{-1.5cm} \text{on~$(\umax, \infty)$} \label{Vpup2} \\
\big|V''(u) \big| &\lesssim \frac{\tilde{\lambda}}{u^4} &&\hspace*{-1.5cm} \text{on~$(\umax, \infty)$} \label{Vpp2} \\
-\re V'(u) &\gtrsim \frac{\tilde{\lambda}}{\Const_2 u^3} &&\hspace*{-1.5cm} \text{on~$\Big(\umax+
\frac{1}{\Const_1}, \infty \Big)$} \label{Vplow2} \:,
\end{align}
where we introduced the abbreviation
\[ \tilde{\lambda} := \lambda + s^2 + 2 a k \omega \overset{\eqref{range1}}{\eqsim} \lambda \:. \]

Setting $u_0= u_0^R$, for any~$u > u^R_-$ we obtain the estimates
\begin{align}
\re V(u) - \re V&(u_0) \lesssim -\frac{\tilde{\lambda}}{\Const_2} \int_{u_0}^u \frac{1}{u^3} \: du
\lesssim \frac{\tilde{\lambda}}{\Const_2} \left( \frac{1}{u^2} - \frac{1}{u_0^2} \right)
= \frac{\tilde{\lambda}}{\Const_2} \:\frac{u_0^2-u^2}{u^2\, u_0^2} \\
\Longrightarrow \; |\re V(u)| &\geq \frac{\tilde{\lambda}}{\Const_2} \:\frac{u^2-u_0^2}{u^2\, u_0^2} \label{reVes1} \\
\frac{|\re V'|}{|\re V|^\frac{3}{2}} &\lesssim \frac{\tilde{\lambda}}{u^3}\;
\frac{\Const_2^\frac{3}{2}}{\tilde{\lambda}^\frac{3}{2}} \:\frac{u^3\, u_0^3}{(u^2-u_0^2)^\frac{3}{2}}
= \frac{\Const_2^\frac{3}{2}}{\tilde{\lambda}^\frac{1}{2}}\: \frac{u_0^3}{(u^2-u_0^2)^\frac{3}{2}} 
= \frac{\Const_2^\frac{3}{2}}{\tilde{\lambda}^\frac{1}{2}}\: \frac{u_0^3}{(u-u_0)^\frac{3}{2} (u+u_0)^\frac{3}{2}} \notag \\
& \leq \frac{\Const_2^\frac{3}{2}}{\tilde{\lambda}^\frac{1}{2}}\: \frac{u_0^\frac{3}{2}}{(u-u_0)^\frac{3}{2}}
\overset{\eqref{u0es1}, \eqref{uRmdef}}{\lesssim}
\frac{\Const_2^\frac{3}{2}}{\tilde{\lambda}^\frac{1}{2}}\: \frac{\tilde{\lambda}^\frac{3}{4}}{|\omega|^\frac{3}{2}}
\frac{|\omega|^\frac{3}{2}}{\Const_3^\frac{3}{2}\, \lambda^\frac{1}{4}}
\lesssim \frac{\Const_2^\frac{3}{2}}{\Const_3^\frac{3}{2}} \:.
\end{align}

In order to estimate the imaginary part, we first note that if~$s=0$,
the potential is real, so that there is nothing to do.
Therefore, we may assume that~$s \neq 0$.
We again use the form of the potential~\eqref{Vdef}
and~\eqref{VRasy} to obtain
\begin{align}
|\im V(u)| &\lesssim \frac{|\omega|}{u} \:,\qquad
|\im V'(u)| \lesssim \frac{|\omega|}{u^2} \label{imVpes1} \\
\Longrightarrow\; \frac{|\im V'(u)|}{|\re V(u)|^\frac{3}{2}} &
\lesssim \frac{|\omega|}{u^2} \:
\frac{\Const_2^\frac{3}{2}}{\tilde{\lambda}^\frac{3}{2}}
\:\frac{u^3\, u_0^3}{(u^2-u_0^2)^\frac{3}{2}}
= \frac{\Const_2^\frac{3}{2}\,|\omega|}{\tilde{\lambda}^\frac{3}{2}}\: \frac{u \,u_0^3}
{(u-u_0)^\frac{3}{2} (u+u_0)^\frac{3}{2}} \\
&\!\!\!\overset{\eqref{lup}}{\lesssim} \frac{\Const_2^\frac{3}{2}}{\tilde{\lambda}}\: \frac{u\, u_0^\frac{3}{2}}{(u-u_0)^\frac{3}{2}} 
\overset{\eqref{u0es1}, \eqref{uRmdef}}{\lesssim}
\frac{\Const_2^\frac{3}{2}}{\lambda}\: \frac{\lambda^\frac{3}{4}}{|\omega|^\frac{3}{2}}\:
\frac{|\omega|^\frac{3}{2}}{\Const_3^\frac{3}{2}\, \lambda^\frac{1}{4}}\:u
\leq \frac{\Const_2^\frac{3}{2}}{\Const_3^\frac{3}{2}} \: \frac{u}{\sqrt{\lambda}}\:.
\end{align}
This gives the desired estimate provided that~$u \leq \const$ (for a constant~$\const>0$
which is independent of the parameters~$\lambda$ and~$\omega$). 
For large~$u$, on the other hand, we know that
\begin{align}
|\im V(u)| &\eqsim \frac{|\omega|}{u} \qquad \text{for~$u \geq \const$} 
\qquad \qquad \text{(if~$s \neq 0$)}
\label{imVes1} \\
\Longrightarrow\; \frac{|\im V'(u)|}{|V(u)|^\frac{3}{2}} &\leq \frac{|\im V'(u)|}{|\im V(u)|^\frac{3}{2}}
\lesssim \frac{|\omega|}{u^2} \: \frac{u^{\frac{3}{2}}}{|\omega|^\frac{3}{2}}
\leq \frac{1}{|\omega|^\frac{1}{2} \, u_0^\frac{1}{2}} \overset{\eqref{u0es1}}{\eqsim} \frac{1}{\tilde{\lambda}^\frac{1}{4}}\:.
\end{align}

The second derivatives are estimated similarly:
\begin{align}
|\re V''| &\eqsim \frac{\tilde{\lambda}}{u^4} \label{reVppes} \\
\Longrightarrow\; \frac{|\re V''|}{|\re V|^2} \;&\!\!\!\overset{\eqref{reVes1}}{\lesssim}
\frac{\tilde{\lambda}}{u^4} \;\frac{\Const_2^2}{\tilde{\lambda}^2}\: \frac{u^4\, u_0^4}{(u^2-u_0^2)^2}
\leq \frac{\Const_2^2\, u_0^2}{\tilde{\lambda}\, (u-u_0)^2}
\overset{\eqref{u0es1}, \eqref{uRmdef}}{\lesssim}
\frac{\Const_2^2}{\tilde{\lambda}}\: \frac{\tilde{\lambda}}{\omega^2}\: 
\frac{\omega^2}{\Const_3^2\, \lambda^{\frac{1}{3}}}
= \frac{\Const_2^2}{\Const_3^2\, \lambda^{\frac{1}{3}}} \notag \\
|\im V''| &\eqsim \frac{|\omega|}{u^3} \label{imVppes} \\
\Longrightarrow\; \frac{|\im V''|}{|\re V|^2} &\lesssim
\frac{|\omega|}{u^3}\: \;\frac{\Const_2^2}{\tilde{\lambda}^2}\: \frac{u^4\, u_0^4}{(u^2-u_0^2)^2}
\overset{\eqref{lup}}{\lesssim} \frac{\Const_2^2\, u\, u_0^2}{\tilde{\lambda}^\frac{3}{2}\, (u-u_0)^2}
\overset{\eqref{u0es1}, \eqref{uRmdef}}{\lesssim}
\frac{\Const_2^2}{\Const_3^2\, \lambda^{\frac{1}{3}}} \: \frac{u}{\sqrt{\lambda}} \:,
\end{align}
giving the desired estimate if~$u \leq \const$. On the other hand, if~$u \geq \const$,
we again use~\eqref{imVes1} to obtain (again it suffices to consider the case~$s \neq0$
because otherwise~$\im V \equiv 0$)
\[ \frac{|\im V''|}{|\im V|^2} \lesssim
\frac{|\omega|}{u^3}\: \frac{u^2}{\omega^2} \leq \frac{1}{|\omega|\, u_0}
\overset{\eqref{u0es1}}{\eqsim} \frac{1}{\tilde{\lambda}^\frac{1}{2}}\:. \]
This concludes the proof.
\QED

\subsection{Estimates in the WKB Region with~$\re V > 0$} \label{secWKBpos}
In this section we shall prove the following results:
\begin{Prp} \label{prpWKBpos} For any~$\varepsilon>0$, we can arrange by choosing
the constants~$\Const_1, \ldots, \Const_4$ sufficiently large that
for all~$\omega$ and~$\lambda$ in the range~\eqref{range} where we are in the Airy case
(see~\eqref{regions}), the following WKB estimates hold:
\[ \frac{|V'|}{|V|^\frac{3}{2}}, \frac{|V''|}{|V|^2} \leq \varepsilon \qquad \text{on~$(u^L_+, u^R_+)$}\:. \]
\end{Prp}

For the proof, we again consider the cases~{\bf{(a)}} and~{\bf{(b)}} in~\eqref{abcase}
after each other.
\Proof[Proof of Proposition~\ref{prpWKBpos} in case~{\bf{(a)}}]
We proceed similar as in the proofs of Propositions~\ref{prpWKB} and~\ref{prpWKBR} in case~{\bf{(a)}}.
It suffices to consider the region~$(\umax, u^R_+)$, because on the interval~$(u^L_+, \umax)$
the proof is the same with obvious changes.
In case~{\bf{(a)}}, equation~\eqref{uRplus} reduces to
\beq \label{uRpa}
u^R_+ = u^R_0 - \Const_3 \,\lambda^{\frac{1}{6}}\, |\omega|^{-\frac{2}{3}}
\big(\Const_1 \re V(\umax) \big)^{-\frac{1}{6}} \:.
\eeq
Moreover, in view of the inequality~\eqref{u0es}, on the interval~$(\umax, u^R_+)$
the second derivative of~$\re V$ satisfies the inequalities in~\eqref{concave}. Hence,
setting~$v=u-\umax$ and~$v_0 = u^R_0-\umax$, we obtain
for all~$u\in (\umax, u^R_+)$
\begin{gather*}
\frac{\lambda v}{\const} \leq -\re V' \leq \const \lambda v \notag \\
\re V(u) = \re V(u) - \re V(u^R_0) = -\int_u^{u^R_0} \re V'(u) \leq \const \lambda \,\big(v_0^2 - v^2 \big) \notag \\
\Longrightarrow\; \frac{\lambda}{\const} \big(v_0^2 - v^2 \big)
\leq \re V(u) \leq \const \lambda \,\big(v_0^2 - v^2 \big) \:.
\end{gather*}
Combining these estimates with~\eqref{uRpa}, we obtain
\[ \frac{|\re V'(u)|^2}{|V(u)|^3} \lesssim \frac{\lambda^2\, v^2}{\lambda^3\, (v_0-v)^3\, (v_0+v)^3}
\leq \frac{\big(\Const_1 \re V(\umax) \big)^{\frac{1}{2}}\, \omega^2\: v^2}{\lambda^\frac{3}{2}\, 
\Const_3^3 \,(v_0+v)^3} \:. \]
Next, we can estimate~$\re V(\umax)$ by
\[ 0 = \re \big( V(u^R_0) \big) \geq \re V(\umax) - \frac{\const \lambda}{2}\: \, v_0^2 \:, \]
implying that
\beq \label{947a}
\re V(\umax) \lesssim \lambda\, v_0^2 \:.
\eeq
We thus obtain
\begin{align*}
\frac{|\re V'(u)|^2}{|V(u)|^3} \lesssim \frac{\sqrt{\Const_1}}{\Const_3^3}\; \frac{\omega^2}{\lambda}\:
\frac{v_0\: v^2}{(v+v_0)^3} \lesssim \frac{\sqrt{\Const_1}}{\Const_3^3}\:,
\end{align*}
where in the last step we used~\eqref{lup}. The imaginary part of~$V'$ can be handled similar
as in the proof of Proposition~\ref{prpWKB}. Namely, from~\eqref{imVes} and~\eqref{lup}, we know that
\[ \frac{|\im V'(u)|^2}{|V(u)|^3} \lesssim \frac{\lambda}{\lambda^3\, (v_0-v)^3\, (v_0+v)^3} 
\leq \frac{1}{\lambda^2\, (v_0-v)^\frac{9}{2}\, (v_0+v)^\frac{3}{2}} \:. \]
Applying again~\eqref{uRpa} and~\eqref{947a}, we obtain
\begin{align*}
\frac{|\im V'(u)|^2}{|V(u)|^3} &\lesssim
\frac{\big(\Const_1 \re V(\umax) \big)^{\frac{3}{4}} \:|\omega|^3}{\lambda^2\, \Const_3^\frac{9}{2}\,
\lambda^{\frac{3}{4}} \,(v+v_0)^\frac{3}{2}} \lesssim
\frac{\big(\Const_1 \lambda\, v_0^2 \big)^{\frac{3}{4}} |\omega|^3}{\lambda^{\frac{11}{4}}\, \Const_3^\frac{9}{2}\,
(v+v_0)^\frac{3}{2}} \\
&\lesssim \frac{\big(\Const_1 \lambda \big)^{\frac{3}{4}} |\omega|^3}{\lambda^{\frac{11}{4}}\, \Const_3^\frac{9}{2}}
\overset{\eqref{lup}}{\lesssim} \frac{\Const_1^{\frac{3}{4}}}{\sqrt{\lambda}\, \Const_3^\frac{9}{2}}
\leq \frac{\Const_1^{\frac{3}{4}}}{\sqrt{\Const_7}\, \Const_3^\frac{9}{2}}\:.
\end{align*}
This concludes the proof for the first derivatives.

The second derivatives are estimated similarly by
\begin{align*}
\big|V''(u) \big| &\leq \big|\re V''(u) \big| + \big|\im V''(u) \big|
\overset{\eqref{imVes}, \eqref{concave}}{\leq}
\const \,\big( \lambda + 1 + |\omega| \big) \overset{\eqref{range1}}{\lesssim} \lambda \\
\frac{|V''(u)|}{|V(u)|^2} &\lesssim \frac{\lambda}{\lambda^2\, (v_0-v)^2\, (v_0+v)^2}
\leq \frac{1}{\lambda\, (v_0-v)^3\, (v_0+v)} \\
&\leq \frac{\big(\Const_1 \re V(\umax) \big)^{\frac{1}{2}}\, |\omega|^2}{\lambda^\frac{3}{2}\,
\Const_3^3 \, (v_0+v)}
\overset{\eqref{947a}}{\lesssim} \frac{\big(\Const_1 \lambda v_0^2 \big)^{\frac{1}{2}}\, |\omega|^2}
{\lambda^\frac{3}{2}\, \Const_3^3 \,(v+v_0)}
\overset{\eqref{lup}}{\lesssim} \frac{\Const_1^\frac{1}{2}}{\Const_3^3} \: \frac{v_0}{v+v_0}
\leq \frac{\Const_1^\frac{1}{2}}{\Const_3^3} \:.
\end{align*}
This concludes the proof.
\QED

\Proof[Proof of Proposition~\ref{prpWKBpos} in case~{\bf{(b)}}]
In this case, the identities~\eqref{uLplus} and~\eqref{uRplus} simplify to
\beq \label{uLRpsimp}
u^L_+ = u^L_0 + \Const_3 \,|\omega|^{-\frac{2}{3}} \:,\qquad
u^R_+ = u^R_0 - \Const_3 \,\lambda^{\frac{1}{6}}\, |\omega|^{-1} \:.
\eeq

We first consider the interval~$[\umax - \Const_1^{-1}, \umax + \Const_1^{-1}]$
(according to~\eqref{u0es} and~\eqref{uLRpsimp}, this interval is contained in~$(u^L_+, u^R_+)$).
Combining the estimate~\eqref{reVes} with the upper bound in~\eqref{concave},
we know that on~$[\umax - \Const_1^{-1}, \umax + \Const_1^{-1}]$,
\[ \re V(u)
\geq \frac{\lambda}{\Const_1} - \frac{\const}{2}\: \lambda\: (u-\umax)^2
\geq \frac{\lambda}{\Const_1} - \frac{\const\, \lambda}{2 \Const_1^2} \geq \frac{\lambda}{2\Const_1} \:, \]
where in the last step we increased~$\Const_1$.
Moreover, from the right of~\eqref{Velem} we know that~$\big| V' \big|,  \big| V'' \big| \lesssim \lambda+|\omega|$.
Hence
\begin{align*}
\frac{|V'(u)|^2}{|V(u)|^3} &\lesssim
\Const_1^3 \:\frac{(\lambda+|\omega|)^2}{\lambda^3} \overset{\eqref{range1}}{\lesssim} \frac{\Const_1^3}{\lambda}
\overset{\eqref{range}}{\lesssim} \frac{\Const_1^3}{\Const_7} \\
\frac{|V''(u)|}{|V(u)|^2} &\lesssim
\Const_1^2 \:\frac{\lambda+|\omega|}{\lambda^2} \overset{\eqref{range1}}{\lesssim} \frac{\Const_1^2}{\lambda}
\overset{\eqref{range}}{\lesssim} \frac{\Const_1^2}{\Const_7} \:.
\end{align*}
Choosing~$\Const_7$ sufficiently large, we obtain the result.

It remains to consider the regions~$(u^L_+, \umax - \Const_1^{-1})$ and~$(\umax + \Const_1^{-1}, u^R_+)$.
In these regions, we can proceed similar as in the proofs of
Propositions~\ref{prpWKB} and~\ref{prpWKBR} in case~{\bf{(b)}}.
Namely, on the interval~$(u^L_+, \umax - \Const_1^{-1})$ the inequalities~\eqref{Vpup}--\eqref{Vplow} 
and~\eqref{reo2} hold. As a consequence,
\begin{gather}
\re V(u) = \int_{u^L_0}^u \re V' \geq \frac{\lambda}{\Const_2} \int_{u^L_0}^u e^{\gamma v}\: dv 
= \frac{\lambda}{\Const_2 \gamma} \:\big(  e^{\gamma u} - e^{\gamma u^L_0} \big) \label{VlowL} \\
\Longrightarrow\; \frac{|V'(u)|^2}{|V(u)|^3} \overset{\eqref{Vpup}}{\lesssim}
\frac{\Const_2^3}{\lambda}\: \frac{e^{2 \gamma u}}{\big(  e^{\gamma u} - e^{\gamma u^L_0} \big)^3} \:. \notag
\end{gather}
By computing its $u$-derivative, one sees that the last fraction is monotone decreasing in~$u$. Hence
\[ \frac{|V'(u)|^2}{|V(u)|^3} \lesssim
\frac{\Const_2^3}{\lambda}\: \frac{e^{2 \gamma u^L_+}}{\big(  e^{\gamma u^L_+} - e^{\gamma u^L_0} \big)^3}
\leq \frac{\Const_2^3}{\lambda}\: \frac{e^{2 \gamma u^L_+}}{e^{3\gamma u^L_0}
\big(  \gamma\, (u^L_+ - u^L_0) \big)^3} \:, \]
where in the last step we used the mean value inequality.
Applying~\eqref{reo2} and~\eqref{uLRpsimp}, we obtain
\[ \frac{|V'(u)|^2}{|V(u)|^3} \lesssim
\frac{\Const_2^3}{\lambda}\: \frac{e^{2 \gamma (u^L_+-u^L_0)}}{e^{\gamma u^L_0} (u^L_+ - u^L_0)^3}
\lesssim \frac{\Const_2^3}{\lambda}\: \frac{\lambda}{\omega^2 \,\big(\Const_3 \,|\omega|^{-\frac{2}{3}} \big)^3}
= \frac{\Const_2^3}{\Const_3^3} \:, \]
which can be made arbitrarily small by increasing~$\Const_3$ (note that, in view of~\eqref{uLRpsimp},
the factor~$e^{2 \gamma (u^L_+-u^L_0)}$ is uniformly bounded).
The second derivatives can be estimated similarly as follows.
First, using~\eqref{Vpp} and~\eqref{VlowL},
\[ \frac{|V''(u)|}{|V(u)|^2} \lesssim
\frac{\Const_2^2}{\lambda}\: \frac{e^{\gamma u}}{\big(  e^{\gamma u} - e^{\gamma u^L_0} \big)^2} \:. \]
This is again monotone decreasing, implying that
\begin{align*}
\frac{|V''(u)|}{|V(u)|^2} &\lesssim \frac{\Const_2^2}{\lambda}\: \frac{e^{\gamma u^L_+}}{\big(  e^{\gamma u^L_+} - e^{\gamma u^L_0} \big)^2} 
\lesssim \frac{\Const_2^2}{\lambda}\: \frac{e^{\gamma (u^L_+-u^L_0)}} {e^{\gamma u^L_0} (u^L_+ - u^L_0)^2} \\
&\lesssim \frac{\Const_2^2}{\lambda}\: \frac{\lambda} {\omega^2 \,
\big(\Const_3 \,|\omega|^{-\frac{2}{3}} \big)^2}
\lesssim \frac{\Const_2^2}{\Const_3^2}\: \omega^{-\frac{2}{3}}\:,
\end{align*}
where in the last line we again applied~\eqref{reo2} and~\eqref{uLRpsimp}.

In the remaining region~$(\umax + \Const_1^{-1}, u^R_+)$, we can again use the
estimates~\eqref{reV1}--\eqref{Vplow2}. Again omitting the index~$R$,
for any~$u \in (\umax + \Const_1^{-1}, u^R_+)$ we obtain
\begin{align*}
\re V(u) &= \re V(u) - \re V(u_0) = -\int_u^{u_0} \re V' \overset{\eqref{Vplow2}}{\gtrsim}
\frac{\tilde{\lambda}}{\Const_2} \left( \frac{1}{u^2} - \frac{1}{u_0^2} \right) 
= \frac{\tilde{\lambda}}{\Const_2} \frac{u_0^2 - u^2}{u^2\, u_0^2} \\
\frac{|\re V'(u)|}{|\re V(u)|^\frac{3}{2}} &\lesssim
\frac{\tilde{\lambda}}{u^3}\;
\frac{\Const_2^\frac{3}{2}}{\tilde{\lambda}^\frac{3}{2}} \:\frac{u^3\, u_0^3}{(u_0^2-u^2)^\frac{3}{2}}
= \frac{\Const_2^\frac{3}{2}}{\tilde{\lambda}^\frac{1}{2}}\: \frac{u_0^3}{(u_0^2-u^2)^\frac{3}{2}} 
= \frac{\Const_2^\frac{3}{2}}{\tilde{\lambda}^\frac{1}{2}}\: \frac{u_0^3}{(u_0-u)^\frac{3}{2} (u_0+u)^\frac{3}{2}} \notag \\
&\leq \frac{\Const_2^\frac{3}{2}}{\tilde{\lambda}^\frac{1}{2}}\: \frac{u_0^\frac{3}{2}}{(u_0-u)^\frac{3}{2}} 
\overset{\eqref{u0es1}, \eqref{uRmdef}}{\lesssim}
\frac{\Const_2^\frac{3}{2}}{\tilde{\lambda}^\frac{1}{2}}\: \frac{\tilde{\lambda}^\frac{3}{4}}{|\omega|^\frac{3}{2}}
\frac{|\omega|^\frac{3}{2}}{\Const_3^\frac{3}{2}\, \lambda^\frac{1}{4}}
\lesssim \frac{\Const_2^\frac{3}{2}}{\Const_3^\frac{3}{2}} \:.
\end{align*}
For the imaginary part, we again use the estimate~\eqref{imVpes1} to obtain
\begin{align*}
\frac{|\im V'(u)|}{|\re V(u)|^\frac{3}{2}} &\lesssim
\frac{|\omega|}{u^2}\;\frac{\Const_2^\frac{3}{2}}{\tilde{\lambda}^\frac{3}{2}} \frac{u^3\, u_0^3}{(u_0^2 - u^2)^\frac{3}{2}}
= \frac{\Const_2^\frac{3}{2}\, |\omega|}{\tilde{\lambda}^\frac{3}{2}}\;\frac{u\, u_0^3}{(u_0^2 - u^2)^\frac{3}{2}} \\
&\!\!\!\overset{\eqref{lup}}{\lesssim}
\frac{\Const_2^\frac{3}{2}}{\tilde{\lambda}}\: \frac{u\, u_0^3}{(u_0-u)^\frac{3}{2} (u_0+u)^\frac{3}{2}} 
\lesssim \frac{\Const_2^\frac{3}{2}}{\Const_3^\frac{3}{2}}\: \frac{u}{\sqrt{\lambda}} \:.
\end{align*}
This gives the desired estimate if~$u \leq \const$. On the other hand, if~$u > \const$
we again apply~\eqref{imVes1} (again it suffices to consider the case~$s \neq0$
because otherwise~$\im V \equiv 0$). This gives
\[ \frac{|\im V'(u)|}{|\im V(u)|^\frac{3}{2}} \lesssim
\frac{|\omega|}{u^2}\; \frac{u^{\frac{3}{2}}}{|\omega|^\frac{3}{2}}
\lesssim \frac{1}{|\omega|^\frac{1}{2}}\:. \]

The second derivatives are estimated similarly using~\eqref{reVppes} and~\eqref{imVppes},
\begin{align*}
\frac{|\re V''|}{|\re V|^2} &\lesssim
\frac{\tilde{\lambda}}{u^4} \;\frac{\Const_2^2}{\tilde{\lambda}^2}\: \frac{u^4\, u_0^4}{(u_0^2-u^2)^2}
\leq \frac{\Const_2^2\, u_0^2}{\tilde{\lambda}\, (u_0-u)^2}
\overset{\eqref{u0es1}, \eqref{uRmdef}}{\lesssim}
\frac{\Const_2^2}{\tilde{\lambda}}\: \frac{\tilde{\lambda}}{\omega^2}\: 
\frac{\omega^2}{\Const_3^2\, \lambda^{\frac{1}{3}}}
= \frac{\Const_2^2}{\Const_3^2\, \lambda^{\frac{1}{3}}} \\
\frac{|\im V''|}{|\re V|^2} &\lesssim
\frac{|\omega|}{u^3}\: \;\frac{\Const_2^2}{\tilde{\lambda}^2}\: \frac{u^4\, u_0^4}{(u_0^2-u^2)^2}
\leq \frac{\Const_2^2\, u\, u_0^2}{\tilde{\lambda}^\frac{3}{2}\, (u_0-u)^2}
\overset{\eqref{u0es1}, \eqref{uRmdef}}{\lesssim}
\frac{\Const_2^2}{\Const_3^2\, \lambda^{\frac{1}{3}}} \: \frac{u}{\sqrt{\lambda}} \:,
\end{align*}
giving the desired estimate if~$u \leq \const$. On the other hand, if~$u \geq \const$,
we again use~\eqref{imVes1} to obtain (again it suffices to consider the case~$s \neq0$
because otherwise~$\im V \equiv 0$)
\[ \frac{|\im V''|}{|\im V|^2} \lesssim
\frac{|\omega|}{u^3}\: \frac{u^2}{\omega^2}
\lesssim \frac{1}{|\omega|}\:. \]
This concludes the proof.
\QED

\subsection{Estimates in the Airy Regions}
\begin{Lemma} \label{lemmaairy1}
In the Airy case, one can arrange by suitably increasing
the constants $\Const_1, \ldots, \Const_4$ that
\[ \sup_{[u^L_-, u^L_+]} |V|\, \big(u^L_+ - u^L_-)^2 \lesssim \Const_4 \:. \]
\end{Lemma}
\Proof We consider the two cases in~\eqref{abcase} separately.
In case~{\bf{(a)}}, we know from~\eqref{u0es} and~\eqref{uLminus}
that the interval~$[u^L_-, u^L_+]$ is contained in the interval~$[\umax-\frac{1}{2}, \umax+\frac{1}{2}]$.
Therefore, we can integrate the inequality~\eqref{concave} to conclude that for any~$u \in [u^L_-, u^L_+]$,
\begin{align}
\re V(\umax) &\gtrsim \lambda\, \big(\umax-u^L_0 \big)^2 \label{reVlower} \\
|\re V(u)| &\lesssim \big|\re V'(u^L_0)\big| \, \big(u^L_+ - u^L_- \big)
+ \lambda \, \big(u^L_+ - u^L_- \big)^2
\end{align}
and thus
\beq \label{airy1es1}
|\re V(u)| \,\big(u^L_+ - u^L_- \big)^2 \lesssim \big|\re V'(u^L_0)\big| \, \big(u^L_+ - u^L_- \big)^3
+ \lambda \, \big(u^L_+ - u^L_- \big)^4 \:.
\eeq
Moreover, similar to~\eqref{uLa}, the identities~\eqref{uLminus} and~\eqref{uLplus} imply that
\[ u^L_+ - u^L_- = 2 \,\Const_3 \big(\Const_1 \re V(\umax) \big)^{-\frac{1}{6}}\, |\omega|^{-\frac{1}{3}} \:. \]
Using this equation in~\eqref{airy1es1}, we obtain two contributions, which can be estimated as follows,
\begin{align*}
\big|\re V'(u^L_0)\big| \, \big(u^L_+ - u^L_- \big)^3 \;\;&
\!\!\!\overset{\eqref{Vpes}}{\lesssim} \lambda\, v_0\;
\Const_3^3 \big(\Const_1 \re V(\umax) \big)^{-\frac{1}{2}}\, |\omega|^{-1} \\
&\!\!\!\overset{\eqref{reVlower}}{\lesssim} \lambda\, v_0\;
\Const_3^3 \big(\Const_1 \, \lambda\, v_0^2 \big)^{-\frac{1}{2}}\, |\omega|^{-1} \\
&= \Const_3^3 \:\Const_1^{-\frac{1}{2}} \, \lambda^\frac{1}{2}\, |\omega|^{-1} 
\overset{\eqref{llow}}{\lesssim} \Const_3^3 \:\Const_1^{-\frac{1}{2}} \\
\lambda \, \big(u^L_+ - u^L_-\big)^4 &\lesssim \lambda\, \Const_3^4 \:\Const_1^{-\frac{2}{3}}\: \re V(\umax)^{-\frac{2}{3}}\, |\omega|^{-\frac{4}{3}} \\
&\!\!\overset{\eqref{cases}}{\leq}
\Const_3^4 \:\Const_1^{-\frac{2}{3}}\:\Const_4^{-\frac{2}{3}}\: \lambda^{\frac{2}{3}}
\, |\omega|^{-\frac{4}{3}} \overset{\eqref{llow}}{\lesssim} \Const_3^4 \:\Const_1^{-\frac{2}{3}}\:\Const_4^{-\frac{2}{3}} \:,
\end{align*}
where we used the abbreviation~$v_0 := \umax - u^L_0$.
The imaginary part of the potential is estimated similarly with the help of~\eqref{imVes},
\begin{align*}
\sup_{[u^L_-, u^L_+]} |\im V|\, \big(u^L_+ - u^L_- \big)^2 &\lesssim |\omega|\, \big(u^L_+ - u^L_-\big)^2
\lesssim |\omega| \:\Const_3^2 \:\Const_1^{-\frac{1}{3}}\: \re V(\umax)^{-\frac{1}{3}}\, |\omega|^{-\frac{2}{3}} \\
&\!\!\overset{\eqref{cases}}{\leq}
\Const_3^2 \:\Const_1^{-\frac{1}{3}}\:\Const_4^{-\frac{1}{3}}\: \lambda^{-\frac{1}{6}}\, |\omega|^{\frac{1}{3}} 
\overset{\eqref{lup}}{\lesssim} \Const_3^2 \:\Const_1^{-\frac{1}{3}}\:\Const_4^{-\frac{1}{3}} \:.
\end{align*}
This can be made arbitrarily small by increasing~$\Const_4$. This completes the proof in case~{\bf{(a)}}.

In case~{\bf{(b)}}, similar to~\eqref{uLmdef}, the identities~\eqref{uLminus} and~\eqref{uLplus} imply that
\beq \label{uLpmb}
u^L_+ - u^L_- = 2 \,\Const_2\, |\omega|^{-\frac{2}{3}} \:.
\eeq
We integrate the inequality~\eqref{Vpup} to obtain for any~$u \in [u^L_-, u^L_+]$
\begin{gather*}
|\re V(u)| \lesssim \lambda\, e^{\gamma u^L_0}\, \big(u^L_+ - u^L_- \big)
\overset{\eqref{reo2}}{\lesssim} \Const_2\: \omega^2 \,\big(u^L_+ - u^L_- \big) \\
\Longrightarrow \quad |\re V(u)| \:\big(u^L_+ - u^L_- \big)^2 \lesssim \Const_2\,\omega^2 \,
\big(u^L_+ - u^L_- \big)^3 \overset{\eqref{uLpmb}}{\lesssim}  \Const_2^4 \\
\intertext{Moreover,}
|\im V(u)| \:\big(u^L_+ - u^L_- \big)^2 \overset{\eqref{imVes}}{\lesssim} |\omega| \: \big(u^L_+ - u^L_- \big)^2
\overset{\eqref{uLpmb}}{\lesssim} \Const_2^2\: |\omega|^{-\frac{1}{3}} \:.
\end{gather*}
This concludes the proof.
\QED

\begin{Lemma} \label{lemmaairy2}
In the Airy case, one can arrange by suitably increasing
the constants $\Const_1, \ldots, \Const_4$ that
\[ \sup_{[u^R_+, u^R_-]} |V|\, \big(u^R_- - u^R_+)^2 \lesssim \Const_3^3 \:. \]
\end{Lemma}
\Proof We consider the two cases in~\eqref{abcase} separately.
In case~{\bf{(a)}}, we can proceed exactly as in the proof of Lemma~\ref{lemmaairy1}.
In the remaining case~{\bf{(b)}}, similar to~\eqref{uRmdef}, the identities~\eqref{uRminus} and~\eqref{uRplus}
imply that
\beq \label{uRpmb}
u^R_- - u^R_+ = 2 \,\Const_3 \,\lambda^{\frac{1}{6}}\, |\omega|^{-1}\:.
\eeq
Integrating~\eqref{Vpup2}, we obtain for any~$u \in [u^R_+, u^R_-]$ that
\begin{gather*}
|\re V(u)| \lesssim \frac{\tilde{\lambda}}{\big(u^R_0\big)^3} \: \big( u^R_- - u^R_+ \big)
\overset{\eqref{u0es1}}{\lesssim}
\frac{|\omega|^3}{\tilde{\lambda}^\frac{1}{2}} \: \big( u^R_- - u^R_+ \big) \\
\Longrightarrow \quad |\re V(u)|\, \big(u^R_- - u^R_+)^2 \lesssim
\frac{|\omega|^3}{\tilde{\lambda}^\frac{1}{2}} \: \big( u^R_- - u^R_+ \big)^3
\overset{\eqref{uRpmb}}{\lesssim} \Const_3^3\:.
\end{gather*}
Similarly, using~\eqref{imVpes1},
\begin{align*}
|\im V(u)|\, \big(u^R_- - u^R_+)^2 &\lesssim \frac{|\omega|}{u_0}\: \big(u^R_- - u^R_+)^2
\overset{\eqref{u0es1}}{\lesssim} \frac{|\omega|^2}{\tilde{\lambda}^\frac{1}{2}}\: \big(u^R_- - u^R_+)^2 \\
&\!\!\!\overset{\eqref{uRpmb}}{\lesssim} \frac{|\omega|^2}{\tilde{\lambda}^\frac{1}{2}}\:
\Const_3^2\, \lambda^\frac{1}{3}\: |\omega|^{-2}
\lesssim \Const_3^2\, \lambda^{-\frac{1}{6}}\:.
\end{align*}
This concludes the proof.
\QED

\subsection{Estimates in the Parabolic Cylinder Region}
\begin{Lemma} \label{lemmaPC}
In the PC case, one can arrange by suitably increasing
the constants $\Const_1, \ldots, \Const_4$ that
\[ \sup_{[u^L_-, u^R_-]} |V|\, \big(u^R_- - u^L_-)^2 \lesssim \Const_4^2 \:. \]
\end{Lemma}
\Proof We first estimate~$u^R_- - u^L_-$. In the case~$\re V(\umax)>0$, we know that~$\re V(u^L_0)$
vanishes. Hence, using~\eqref{cases} and again integrating~\eqref{concave}, we obtain
\[ \lambda\, (\umax - u^L_0)^2 \lesssim \re V(\umax) \leq \Const_4\, \sqrt{\lambda} \]
and thus
\beq \label{espar1}
(\umax - u^L_0) \lesssim \frac{\sqrt{\Const_4}}{\lambda^\frac{1}{4}}\:.
\eeq
Now we can use~\eqref{uLminus} to obtain
\beq
(\umax - u^L_-) \lesssim \frac{\sqrt{\Const_4}}{\lambda^\frac{1}{4}}
+\Const_3 \,\Const_1^{-\frac{1}{6}} \:|\omega|^{-\frac{1}{2}}
\overset{\eqref{llow}}{\lesssim} \frac{\sqrt{\Const_4}}{\lambda^\frac{1}{4}} \:.  \label{espar2}
\eeq
In the case~$\re V(\umax) \leq 0$, on the other hand, we know that~$u^L_0=\umax$,
so that~\eqref{espar1} and consequently also~\eqref{espar2} again hold.

Combining the last inequality with~\eqref{concave} and~\eqref{cases}, we obtain
for any~$u \in [u^L_-, u^R_-]$ the estimate
\begin{align*}
|V(u)|\, \big(u^R_- - u^L_-)^2 &\lesssim |V(\umax)|\, \big(u^R_- - u^L_-)^2
+ \sup_{[u^L_-, u^R_-]} |V''|\: \big(u^R_- - u^L_-)^4 \\
&\lesssim \Const_4\,\sqrt{\lambda}\: \big(u^R_- - u^L_-)^2
+ \lambda\, \big(u^R_- - u^L_-)^4 \lesssim \Const_4^2 \:.
\end{align*}
This concludes the proof.
\QED

\subsection{Estimates of the Zeros of~$\im V$}
For the $T$-estimates introduced in~\cite[Section~3.2]{tinvariant}, the sign of~$\im V$
is of particular importance. More precisely, if~$y$ is in the upper half plane and~$\im V>0$
(and similarly if~$y$ is in the lower half plane and~$\im V<0$), then we can use these
estimates setting~$g \equiv 0$ (the ``good'' sign).
If, however, the imaginary part of~$V$ has the opposite ``bad'' sign,
then the estimates apply only if~$|\im V|$ is small in a quite restrictive sense.
In the next lemma, we identify the regions where~$\im V$ has a ``good'' sign
and estimate~$\im V$ in the regions where the sign is ``bad.''

\begin{Lemma} \label{lemmasignimV}
Assume that we are in the Airy or PC case. Then by choosing the
constants~$\Const_5$, ~$\Const_6$ and~$\Const_7$ in~\eqref{range} and~\eqref{range1}
sufficiently large, one can arrange that
\beq \left\{  \;\; \begin{split}
\omega \, \im V(u) &\geq 0 \qquad \text{on~$\big(-\infty, \umax - \Const_1^{-\frac{1}{2}} \big]$} \\
\omega \, \im V(u) &\leq 0 \qquad \text{on~$\big(\umax + \Const_1^{-\frac{1}{2}} , \infty \big)$} \:.
\end{split} \right. \label{omim1}
\eeq
Moreover, on the remaining intervals, the function~$\im V$ satisfies the inequalities
\beq \left\{  \;\; \begin{split}
\omega \, \im V(u) &\gtrsim -|\omega| \qquad \text{on~$\big(\umax - \Const_1^{-\frac{1}{2}} , \umax\big]$} \\
\omega \, \im V(u) &\lesssim |\omega| \qquad \;\;\;\text{on~$\big[\umax, \umax + \Const_1^{-\frac{1}{2}}\big)$} \:.
\end{split} \right. \label{omim2}
\eeq
\end{Lemma}
\Proof Using the form of the potential~\eqref{Vdef}, one obtains the expansions
\begin{align}
\omega\, \im V &= -\frac{2 \omega^2 s}{(r^2+a^2)^2} \:\Big( r^2 \,(r-3M) + a^2\,(r+M) \Big) +
\O\big( \omega \big) \label{omimV} \\
\frac{d}{dr} \re V &= \lambda \:\frac{d}{dr} \frac{\Delta}{(r^2+a^2)^2}
+ \O\big( \lambda^0 \big) + \O(\omega) \nonumber \\
&= -\frac{2 \lambda}{(r^2+a^2)^3} \:\Big( r^2 \,(r-3M) + a^2\,(r+M) \Big) + \O\big( \lambda^0 \big) + \O(\omega) \:. \label{drReV}
\end{align}
Comparing these formulas, one sees that the leading contributions to both functions have the
opposite sign as the factor~$r^2 (r-3M) + a^2\,(r+M)$.
In order to control the error term, we denote the zero of the function~$r^2 (r-3M) + a^2\,(r+M)$
by~$u_\text{f}$. Then the zero~$u_{\text{im}}$ of~$\im V$ and the
zero~$\umax$ of the function~$\partial_r \re V$ are given by
\[ u_\text{im} = u_\text{f} + \O\Big( \frac{1}{\omega} \Big) \qquad \text{and} \qquad
\umax = u_\text{f} + \O \Big( \frac{\omega}{\lambda} \Big) \:. \]
This proves~\eqref{omim1}. In order to derive~\eqref{omim2}, we need to take
into account two contributions: First, the error term in~\eqref{omimV}, which is uniformly bounded
and thus unproblematic. Second, we need to take into account that the 
deviations of the zeros~$u_\text{im}$ and~$\umax$ from~$u_\text{f}$
may have the effect that the leading contribution
to~$\omega \im V$ in~\eqref{omimV} has the wrong sign. This contribution scales like
\[  |\im V'|\: \big| u_\text{im} - \umax \big| \lesssim |\omega| \left( \O\Big( \frac{1}{\omega} \Big) + \O \Big( \frac{\omega}{\lambda} \Big) \right) = \O(1) \:, \]
where in the last step we used~\eqref{range} and~\eqref{lup}. This concludes the proof.
\QED

\section{Invariant Region Estimates} \label{secinvregion}
\subsection{Estimates of~$\acute{\phi}$} \label{secacute}
We again restrict attention to the parameter range~\eqref{range}.
We consider the solutions~$\acute{\phi}_-$ constructed in Theorem~\ref{thm31}.
For ease in notation, we shall omit the index~$-$.
We denote the corresponding solution of the Riccati equation by
\[ \acute{y}(u) := \frac{\acute{\phi}'(u)}{\acute{\phi}(u)} \:. \]
According to Proposition~\ref{prpWKB}, the WKB approximation holds on
the interval~$(-\infty, u^L_-)$, meaning that
\beq \label{phiyWKB}
\acute{\phi}
\:\approx\: \frac{1}{\sqrt[4]{-V}} \: \exp\left( \pm i \int^u \sqrt{-V} \right) \qquad \text{and} \qquad
\acute{y} \:\approx\: \pm i \,\sqrt{-V} - \frac{V'}{4V}
\eeq
with an arbitrarily small error, where the sign is chosen such that
\[ \lim_{u \rightarrow -\infty} \pm \sqrt{-V} =\Omega \:. \]
Moreover, the integration constant is chosen in agreement with the normalization
convention in~\eqref{acutephiasy}.
The goal of this section is to estimate the solution~$\acute{\phi}$ all the way to~$u=\umax$.

We begin with the parabolic cylinder case:
\begin{Lemma} \label{lemmaPCacute} In the parabolic cylinder case, there is a constant~$\Const_9$
such that on the interval~$[u^L_-, \umax]$,
the solutions~$\acute{y}$ and~$\acute{\phi}$ are bounded in terms of its values at~$u^L_-$ by
\begin{gather}
|\acute{y}(u)| \leq \Const_9\,\big| \acute{y}(u^L_-) \big| \label{yes1PC} \\
\im \acute{y}(u) \geq \frac{\im \acute{y}(u^L_-)}{\Const_9} \label{yes2PC} \\
\frac{\big| \acute{\phi}(u^L_-) \big|}{\Const_9}
\leq \big| \acute{\phi}(u) \big| \leq \Const_9\, \big| \acute{\phi}(u^L_-) \big| \:. \label{pesPC}
\end{gather}
\end{Lemma}
\Proof Our strategy is to estimate~$y$ using the $T$-method
as introduced in~\cite[Section~3.2]{tinvariant}. More precisely, we shall
apply~\cite[Theorem~3.3]{tinvariant} for a suitable function~$g$.
Moreover, setting
\[ \nu = \sup_{[u^L_-, \umax]} |V| \:, \]
we choose
\beq \label{abchoice}
\alpha = \sqrt{2 \nu} \qquad \text{and} \qquad \tilde{\beta}=0 \:.
\eeq
Then~$\tilde{V}$ and~$U$ are given by
\beq \label{tVUform}
\tilde{V} = \alpha^2 = 2 \nu \qquad \text{and} \qquad
U = \re V - \alpha^2 \leq -\nu \:.
\eeq
Moreover, the error terms~$E_1, \ldots, E_4$ are bounded by
\beq \label{Eerr}
\begin{split}
|E_1| &\lesssim \frac{1}{\sqrt \nu}\: |\re V - \re \tilde{V}| + \frac{\re V'}{\nu} 
\lesssim \sqrt{\nu} + \frac{\re V'}{\nu} \\
E_2 &= 0 \:,\qquad
|E_3|+|E_4| \lesssim \frac{|\im V|}{\sqrt{\nu}} \:\big(1+g \big) \:.
\end{split} \eeq
The integral over the error term~$E_1$ can estimated by
\[ \int_{u^L_-}^{\umax} |E_1| \lesssim
\sqrt{\nu} \,(\umax - u^L_-) \bigg(1 + \sup_{[u^L_-, \umax]} \frac{|\re V'|}{\nu^\frac{3}{2}} \bigg) . \]
The factor~$\sqrt{\nu} \,(\umax - u^L_-)$ was estimated in Lemma~\ref{lemmaPC}.
The factor~$\re V'|/\nu^\frac{3}{2}$, on the other hand, is bounded at the left end point~$u^L_-$
because of the WKB estimate in Proposition~\ref{prpWKB}.
This bound can be extended to the interval~$(u^L_-, \umax)$ by using that~$\re V'$ is
monotone decreasing according to~\eqref{concave}, i.e.
\beq
0 \leq \re V'(u) \leq \re V'(u^L_-) \qquad \text{for all~$u \in [u^L_-, \umax]$}\:. \label{WKBextend}
\eeq

From Lemma~\ref{lemmasignimV} we know that~$\im V$ and~$\im \acute{y}$
have the same signs, except for the error estimated in~\eqref{omim2}. We choose
\[ g(u) = \left\{ \begin{array}{cl} 0 & \text{if~$\omega\, \im V \geq 0$} \\
\sqrt{\lambda}& \text{if~$\omega\, \im V < 0$}\:. \end{array} \right. \]
Then the function~$g\, \im V$ vanishes unless~$\omega\, \im V < 0$, in which case~\eqref{omim2}
gives the estimate
\beq \label{omim3}
\big| g\, \im V \big| \lesssim |g| \lesssim \sqrt{\lambda}\:.
\eeq
Moreover, using the inequality
\[ \nu \gtrsim \lambda\, \big(\umax - u^L_- \big)^2 \]
(obtained again by integrating~\eqref{concave}), we obtain
\begin{align*}
\int_{u^L_-}^{\umax}& |E_3|+|E_4| \lesssim
\int_{u^L_-}^{\umax} \frac{|\im V|}{\sqrt{\lambda}\, \big(\umax - u^L_- \big)} \:\big(1+g \big) \\
&\lesssim \int_{u^L_-}^{\umax} \frac{|\im V|}{\sqrt{\lambda}\, \big(\umax - u^L_- \big)} 
+ \int_{u^L_-}^{\umax} \frac{|g\,\im V|}{\sqrt{\lambda}\, \big(\umax - u^L_- \big)}  \\
&\lesssim \sqrt{\lambda} 
\int_{u^L_-}^{u^R_-} \frac{1}{\sqrt{\lambda}\: \big(\umax - u^L_- \big)} \lesssim 1 \:,
\end{align*}
where in the last line we used~\eqref{imVes}, \eqref{lup} and~\eqref{omim3}.

Combining the above estimates, we conclude that
\[ \int_{u^L_-}^{\umax} |E_1| + |E_2| + |E_3|+|E_4| \lesssim \Const_8 \:. \]
As a consequence, the function~$T$ is bounded by~$e^{\Const_8}$.
It follows that the inequality
\[ g \geq T-1 \qquad \text{if~$\im V < 0$} \]
is satisfied for large~$\lambda$.

Having verified the hypotheses of~\cite[Theorem~3.3]{tinvariant}, we can apply this theorem
to obtain~\eqref{yes1PC} and~\eqref{yes2PC}.
The inequality~\eqref{pesPC} is derived as follows. At~$u^L_-$, this inequality clearly holds
in view of the WKB approximation~\eqref{phiyWKB}. Expressing~$\acute{\phi}(u)$ as
\[ \phi(u) = \phi\big( u^L_- \big)\: \exp \bigg( \int_{u^L_-}^u y \bigg) , \]
it remains to show that the integral in the exponent is uniformly bounded. To this end, we use~\eqref{yes1PC}
to obtain the estimate
\[ \int_{u^L_-}^\umax |y| \leq \Const_9 \,\big|\acute{y}(u^L_-) \big| \,\big(\umax - u^L_- \big)
\lesssim \Const_9 \,\sqrt{|V(u^L_-)|} \,\big(\umax - u^L_- \big)  \lesssim \Const_9 \Const_4 \:, \]
where we employed the WKB approximation at~$u^L_-$ and applied Lemma~\ref{lemmaPC}.
This concludes the proof.
\QED

In the remaining Airy case, we need to consider the Airy region and the
WKB region with~$\re V>0$. We begin with the Airy region.
\begin{Lemma} \label{lemmaairyacute} In the Airy case, there is a constant~$\Const_9$
such that on the interval~$[u^L_-, u^L_+]$,
the solutions~$\acute{y}$ and~$\acute{\phi}$ are bounded in terms of its values at~$u^L_-$ by
\begin{gather*}
|\acute{y}(u)| \leq \Const_9\,\big| \acute{y}(u^L_-) \big| \\
\im \acute{y}(u) \geq \frac{\im \acute{y}(u^L_-)}{\Const_9} \\
\frac{\big| \acute{\phi}(u^L_-) \big|}{\Const_9}
\leq \big| \acute{\phi}(u) \big| \leq \Const_9\, \big| \acute{\phi}(u^L_-) \big| \:.
\end{gather*}
\end{Lemma}
\Proof Our strategy is to estimate~$y$ using the $T$-method
as introduced in~\cite[Section~3.2]{tinvariant}. More precisely, we shall
apply~\cite[Theorem~3.3]{tinvariant} for a suitable function~$g$.
Moreover, setting
\[ \nu = \sup_{[u^L_-, u^L_+]} |V| \:, \]
we again choose~$\alpha$ and~$\tilde{\beta}$ according to~\eqref{abchoice}.
Then~$\tilde{V}$ and~$U$ are again given by~\eqref{tVUform}.
Moreover, the error terms~$E_1, \ldots, E_4$ can again be estimated
as in~\eqref{Eerr}.
The integral over the error term~$E_1$ can estimated by
\beq \label{E1es}
\int_{u^L_-}^{u^L_+} |E_1| \lesssim
\sqrt{\nu} \,(u^L_+ - u^L_-) \bigg(1 + \sup_{[u^L_-, u^L_+]} \frac{|\re V'|}{\nu^\frac{3}{2}} \bigg) .
\eeq
The factor~$\sqrt{\nu} \,(u^L_+ - u^L_-)$ was estimated in Lemma~\ref{lemmaairy1}.
The factor~$\re V'|/\nu^\frac{3}{2}$, on the other hand, is bounded at the left end point~$u^L_-$
because of the WKB estimate in Proposition~\ref{prpWKB}.
This bound can be extended to the interval~$(u^L_-, u^L_+)$ again by using the estimate~\eqref{WKBextend}.

In order to control the error terms~$E_3$ and~$E_4$, we again distinguish the two cases in~\eqref{abcase}.
In case~{\bf{(b)}}, we know from Lemma~\ref{lemmasignimV} that~$\im V$ and~$\im \acute{y}$
have the same signs, making it possible to choose~$g \equiv 0$. Hence
\[ \int_{u^L_-}^{u^L_+} |E_3|+|E_4| \lesssim \int_{u^L_-}^{u^L_+} 
\frac{|\im V|}{\sqrt{\nu}} \overset{\eqref{imVes}}{\lesssim} 
\frac{|\omega|}{\sqrt{\nu}}\: \big(u^L_+ - u^L_- \big)
= \frac{|\omega|}{\nu}\: \sqrt{\nu} \:\big(u^L_+ - u^L_- \big) \:. \]
The factor~$\sqrt{\nu} \,(u^L_+ - u^L_-)$ was estimated in Lemma~\ref{lemmaairy1}.
Moreover, the factor~$\nu$ can be estimated with the help of~\eqref{reVesL} by
\[ \nu \geq \frac{\gamma\, \Const_3}{\const\, \Const_2} \;|\omega|^{\frac{4}{3}} \:. \]
It follows that
\[ \int_{u^L_-}^{u^L_+} |E_3|+|E_4| 
\lesssim \frac{\const\, \Const_2}{\gamma\, \Const_3} \frac{|\omega|}{|\omega|^{\frac{4}{3}}}
\lesssim \frac{\Const_2}{\Const_3} \: |\omega|^{-\frac{1}{3}}\:, \]
which can be made arbitrarily small by increasing~$\Const_3$.

The remaining case~{\bf{(a)}} is more subtle. We begin by proving the inequality
\beq \label{reVuLm}
-\re V \big( u^L_- \big) \gtrsim \Const_3^\frac{3}{2}\, \Const_1^{-\frac{1}{4}}\: \sqrt{\lambda} \:.
\eeq
To this end, we integrate~\eqref{concave} to obtain
\[ \re V(\umax) \eqsim \lambda\, v_0^2 \:. \]
As a consequence, using~\eqref{uLminus}, \eqref{uLplus}, \eqref{lup} and~\eqref{llow}, we obtain
\begin{align*}
v_0-v_- &= \Const_3 \big(\Const_1 \re V(\umax) \big)^{-\frac{1}{6}}\, |\omega|^{-\frac{1}{3}} \\
&\eqsim \Const_3\, \Const_1^{-\frac{1}{6}}\: \lambda^{-\frac{1}{6}}\: |v_0|^{-\frac{1}{3}}\: |\omega|^{-\frac{1}{3}}
\eqsim \Const_3\, \Const_1^{-\frac{1}{6}}\: |v_0|^{-\frac{1}{3}}\: \lambda^{-\frac{1}{3}}
\end{align*}
and thus
\[ (v_0-v_-)\, |v_0|^\frac{1}{3} \eqsim \Const_3\, \Const_1^{-\frac{1}{6}}\: \lambda^{-\frac{1}{3}} \:. \]
Hence
\begin{align*}
-\re V(v_-) &\eqsim \lambda\, |v_0-v_-|\,|v_0+v_-| \geq 
\lambda\, |v_0-v_-|^\frac{3}{2}\: |v_0+v_-|^\frac{1}{2} \\
&\eqsim \lambda\, \Big( (v_0-v_-)\: |v_0|^\frac{1}{3} \Big)^\frac{3}{2}
\eqsim \lambda\, \Const_3^\frac{3}{2}\, \Const_1^{-\frac{1}{4}}\: \lambda^{-\frac{1}{2}}
= \Const_3^\frac{3}{2}\, \Const_1^{-\frac{1}{4}}\: \sqrt{\lambda} \:,
\end{align*}
giving~\eqref{reVuLm}.

Next, we know from Lemma~\ref{lemmasignimV} that~$\im V$ and~$\im \acute{y}$
have the same signs, except for the error estimated in~\eqref{omim2}. We choose
\[ g(u) = \left\{ \begin{array}{cl} 0 & \text{if~$\omega\, \im V \geq 0$} \\
\sqrt{\lambda}& \text{if~$\omega\, \im V < 0$}\:. \end{array} \right. \]
Then, using~\eqref{omim2} and~\eqref{Eerr}, we obtain
\beq \label{E34es}
\int_{u^L_-}^{u^L_+} |E_3|+|E_4| \lesssim \sqrt{\nu} \,(u^L_+ - u^L_-) 
\left( \sup_{[u^L_-, u^L_+]} \frac{|\im V|}{\nu} + \frac{\sqrt{\lambda}}{\nu} \right) \:.
\eeq
The prefactor~$\sqrt{\nu} \,(u^L_+ - u^L_-)$ was estimated in Lemma~\ref{lemmaairy1}.
Moreover, using~\eqref{reVuLm}, we find that the factor~$\sqrt{\lambda}/\nu$ in~\eqref{E34es}
is bounded by a constant.
In order to estimate the term~$|\im V|/\nu$, we compare the equations~\eqref{omimV}
and~\eqref{drReV} to obtain
\[ \frac{|\im V|}{\nu} \lesssim \frac{|\omega|}{\nu \lambda}\:\Big( |\re V'| + \O(\omega) \Big) \:. \]
The error term is uniformly bounded in view of~\eqref{lup} and~\eqref{reVuLm}. The other summand can be estimated by
\[ \frac{|\omega|}{\nu \lambda}\: |\re V'| \lesssim
\frac{|\omega|}{\lambda}\: \sqrt{\nu}\; \frac{|\re V'|}{\nu^\frac{3}{2}} \overset{\eqref{lup}}{\lesssim}
\frac{\sqrt{\nu}}{\sqrt{\lambda}}\; \frac{|\re V'|}{\nu^\frac{3}{2}} \:. \]
Here the factor~$\nu/\lambda$ can be estimated with the help of~\eqref{imVes} and~\eqref{Vdef} by
\begin{align*}
\frac{\nu}{\lambda} &\leq \frac{1}{\lambda}\: \sup_{[u^L_-, u^L_+]} |\im V| + 
\frac{u^L_+ - u^L_-}{\lambda}\: \sup_{[u^L_-, u^L_+]} |\re V'| \\
&\lesssim \frac{|\omega|}{\lambda} + \frac{\lambda + |\omega|}{\lambda}\: \big(u^L_+ - u^L_- \big) \:,
\end{align*}
which is uniformly bounded in view of~\eqref{lup} and~\eqref{uLminus}.
In order to estimate the remaining factor~$|\re V'|/{\nu^\frac{3}{2}}$, we first note that this factor
is uniformly bounded at~$u^L_-$ because of the WKB approximation (see Lemma~\ref{prpWKB}).
In order to extend this inequality to~$u \in [u^L_-, u^L_+]$, we make use of the monotonicity
(see~\eqref{u0es}, \eqref{uLminus} and~\eqref{concave})
\[ 0 \leq \re' V(u) \leq \re V(u^L_-) \:. \]
We conclude that~\eqref{E34es} is uniformly bounded.

Combining the above estimates, we obtain in case~{\bf{(a)}},
\[ \int_{u^L_-}^{u^L_+} |E_1| + |E_2| + |E_3|+|E_4| \lesssim \Const_8 \:. \]
As a consequence, the function~$T$ is bounded by~$e^{\Const_8}$.
It follows that the inequality
\[ g \geq T-1 \qquad \text{if~$\im V < 0$} \]
is satisfied for large~$\lambda$.

Having verified the hypothesis of~\cite[Theorem~3.3]{tinvariant}, we can proceed
exactly as in the proof of Lemma~\ref{lemmaPCacute}.
This concludes the proof.
\QED

\begin{Lemma} \label{lemmaWKBmatch}
There are constants~$\Const_{10}$ and~$c=c(\lambda, \omega)$ such that in the WKB region
with~$\re V>0$ the following inequality holds on the interval~$[u^L_+, \umax]$,
\[ \frac{|\acute{\phi}(u)|}{\Const_{10}}
\leq \frac{c(\lambda, \omega)}{(\re V(u))^\frac{1}{4}} \: e^{\int_{u^L_+}^u \re \sqrt{V}} \leq \Const_{10}\,
|\acute{\phi}(u)| \:. \]
\end{Lemma}
\Proof In Proposition~\ref{prpWKBpos} it was shown that the WKB conditions
are satisfied on the interval~$[u^L_+, \umax]$.
Thus the solution is well-approximated by the WKB solution
\beq \label{WKBgen}
\acute{\phi} \approx \frac{1}{(\re V)^\frac{1}{4}} \left( C_1\: e^{\int_{u^L_+}^u \sqrt{V}}
+ C_2\: e^{-\int_{u^L_+}^u \sqrt{V}} \right) ,
\eeq
with error terms which are under control in view of the estimates in~\cite{tinvariant}.
Choosing the sign convention for the square roots such that~$\re \sqrt{V}>0$,
the first fundamental solutions in~\eqref{WKBgen} is exponentially increasing,
whereas the second is exponentially decaying.

It remains to show that the quotient~$C_1/C_2$ is bounded away from zero.
To this end, we compute the derivative of~\eqref{WKBgen} at~$u^L_+$ to obtain
\[ \acute{y}(u^L_+) \approx \sqrt{V}\; \frac{C_1-C_2}{C_1+C_2} - \frac{V'}{4 V}\:, \]
again with an arbitrarily small error.
On the other hand, the estimate of Proposition~\ref{prpWKB} implies that the
WKB approximation~\eqref{phiyWKB} holds at~$u^L_-$ with an arbitrarily small error.
Moreover, the estimates of Lemma~\ref{lemmaairyacute} give uniform control of
the solution on the interval~$[u^L_-, u^L_+]$. The resulting estimate shows that~$\acute{y}(u^L_+)$
is not well-approximated by~$-\sqrt{V}$. This gives the desired lower bound for~$|C_1/C_2|$,
concluding the proof.
%
%
\QED

\subsection{Estimates of~$\grave{\phi}$} \label{secgrave}
We again restrict attention to the parameter range~\eqref{range}.
We consider the solutions~$\grave{\phi}_-$ constructed in Theorem~\ref{thm34}.
For ease in notation, we shall omit the index~$-$.
We denote the corresponding solution of the Riccati equation by
\[ \grave{y}(u) := \frac{\grave{\phi}'(u)}{\grave{\phi}(u)} \:. \]
According to Proposition~\ref{prpWKBR}, the WKB approximation holds on
the interval~$(u^R_-, \infty)$, meaning that
\beq \label{phiyWKBR}
\grave{\phi}
\:\approx\: \frac{1}{\sqrt[4]{-V}} \: \exp\left( \pm i \int^u \sqrt{-V} \right) \qquad \text{and} \qquad
\grave{y} \:\approx\: \pm i \,\sqrt{-V} - \frac{V'}{4V}
\eeq
with an arbitrarily small error, where the sign is chosen such that
\[ \lim_{u \rightarrow \infty} \pm \sqrt{-V} = -\omega \:. \]
Moreover, the integration constant is chosen in agreement with the normalization
convention in~\eqref{phiasy}.
The goal of this section is to estimate the solution~$\grave{\phi}$ backwards in~$u$
all the way to~$u=\umax$.

We again begin with the parabolic cylinder case:
\begin{Lemma} \label{lemmaPCgrave} In the parabolic cylinder case, there is a constant~$\Const_9$
such that on the interval~$[\umax, u^R_-]$,
the solutions~$\grave{y}$ and~$\grave{\phi}$ are bounded in terms of its values at~$u^R_-$ by
\begin{gather*}
|\grave{y}(u)| \leq \Const_9\,\big| \grave{y}(u^R_-) \big| \\
\im \grave{y}(u) \geq \frac{\im \grave{y}(u^R_-)}{\Const_9} \\
\frac{\big| \grave{\phi}(u^R_-) \big|}{\Const_9}
\leq \big| \grave{\phi}(u) \big| \leq \Const_9\, \big| \grave{\phi}(u^R_-) \big| \:.
\end{gather*}
\end{Lemma}
\Proof Follows exactly as in the proof of Lemma~\ref{lemmaPCacute}.
\QED

In the remaining Airy case, we again need to consider the Airy region and the
WKB region with~$\re V>0$:
\begin{Lemma} \label{lemmaairygrave} In the Airy case, there is a constant~$\Const_9$
such that on the interval~$[u^R_+, u^R_-]$,
the solutions~$\grave{y}$ and~$\grave{\phi}$ are bounded in terms of its values at~$u^R_-$ by
\begin{gather*}
|\grave{y}(u)| \leq \Const_9\,\big| \grave{y}(u^R_-) \big| \\
\im \grave{y}(u) \geq \frac{\im \grave{y}(u^R_-)}{\Const_9} \\
\frac{\big| \grave{\phi}(u^R_-) \big|}{\Const_9}
\leq \big| \grave{\phi}(u) \big| \leq \Const_9\, \big| \grave{\phi}(u^R_-) \big| \:.
\end{gather*}
\end{Lemma}
\Proof Our strategy is to estimate~$y$ using the $T$-method
in~\cite[Theorem~3.3]{tinvariant}. We set
\[ \nu = \sup_{[u^R_+, u^R_-]} |V| \]
and choose~$\alpha$ and~$\tilde{\beta}$ again according to~\eqref{abchoice}.
The function~$g$ will be specified below.
Then~$\tilde{V}$ and~$U$ are again given by~\eqref{tVUform}.
Moreover, the error terms~$E_1, \ldots, E_4$ are again estimated by~\eqref{Eerr}.
The integral over the error term~$E_1$ can be estimated exactly
as explained after~\eqref{E1es}.

In order to control the error terms~$E_3$ and~$E_4$, we again distinguish the two cases in~\eqref{abcase}.
In case~{\bf{(a)}}, we can proceed exactly as in the proof of Lemma~\ref{lemmaairyacute}.
In the remaining case~{\bf{(b)}}, we know from Lemma~\ref{lemmasignimV} that~$\im V$ and~$\im \grave{y}$
have opposite signs, making it possible to choose~$g \equiv 0$.
Next, from~\eqref{reVes1}, we know that
\[ \nu \gtrsim \frac{\lambda}{\Const_1} \: \frac{\big(u^R_- - u^R_0 \big) \big( u^R_- + u^R_0 \big)}{
\big(u^R_- \big)^2 \big(u^R_0 \big)^2}
\overset{\eqref{u0es1}, \eqref{uRmdef}}{\gtrsim}
\frac{\lambda}{\Const_1} \: \frac{u^R_- - u^R_0}{\big(u^R_0 \big)^3}\:. \]
It follows that
\begin{align*}
\int_{u^R_+}^{u^R_-} &\frac{|\im V|}{\sqrt{\nu}} \overset{\eqref{imVpes1}}{\lesssim}
|\omega|\: \frac{\sqrt{\Const_1}}{\sqrt{\lambda}}\: \sqrt{u^R_0}
\:\sqrt{u^R_- - u^R_+} \\
&\!\!\!\!\!\!\!\!\!\!\overset{\eqref{u0es1}, \eqref{uRmdef}, \eqref{uLRpsimp}}{\lesssim}
|\omega|\: \frac{\sqrt{\Const_1}}{\sqrt{\lambda}}\: \frac{\lambda^\frac{1}{4}}{|\omega|^\frac{1}{2}}\;
\sqrt{\Const_3} \;\lambda^{\frac{1}{12}}\: |\omega|^{-\frac{1}{2}}
= \sqrt{\Const_1 \Const_3}\: \lambda^{-\frac{1}{6}}\:,
\end{align*}
which can be made arbitrarily small by increasing~$\Const_7$.
This concludes the proof.
\QED

\begin{Lemma} \label{lemmaWKBmatchgrave}
There are constants~$\Const_{10}$ and~$c=c(\lambda, \omega)$ such that in the WKB region
with~$\re V>0$ the following inequality holds on the interval~$[\umax, u^R_+]$,
\[ \frac{|\grave{\phi}(u)|}{\Const_{10}} \lesssim \frac{c(\lambda, \omega)}{(\re V(u))^\frac{1}{4}} \: e^{-\int_{u^R_+}^u \re \sqrt{V}}  \leq \Const_{10}\, |\grave{\phi}(u)| \:. \]
\end{Lemma}
\Proof We proceed similar as in Lemma~\ref{lemmaWKBmatch}.
According to Proposition~\ref{prpWKBpos} we know that on the interval~$[\umax, u^R_+]$
the WKB approximation
\[ \grave{\phi} \approx \frac{1}{(\re V)^\frac{1}{4}} \left( C_1\: e^{\int_{u^R_+}^u \sqrt{V}}
+ C_2\: e^{-\int_{u^R_+}^u \sqrt{V}} \right) \]
holds with an arbitrarily small error. By combining the results of Proposition~\ref{prpWKBR}
and Lemma~\ref{lemmaairygrave}, we conclude that~$C_2/C_1$ is bounded away from zero.
\QED

\subsection{Estimates for Bounded~$\omega$ and Large~$\lambda$} \label{seccompact}
In Section~\ref{secestimate} we restricted attention to the case that~$|\omega|$
is large (see~\eqref{range}). We now consider the complementary region that~$\omega$ is in a bounded
set, but again for large~$\lambda$. We exclude the case~$\omega=0$, which will be
considered separately in Section~\ref{secomegazero}. We thus consider the parameter range
\beq \label{rangecompact}
0 \neq \omega^2 < \Const_6 \qquad \text{and} \qquad \lambda \geq \Const_7\:.
\eeq
Choosing~$\Const_7$ sufficiently large, the potential looks qualitatively as in the
Airy case in Section~\ref{secestimate}. The real part of the potential is negative both at~$u=\pm \infty$
with the asymptotics~\eqref{VRasy} and~\eqref{VLasy}.
Since the summand involving~$\lambda$ in~\eqref{schroedinger} is non-negative,
for large~$\lambda$ the real part of the potential will be non-negative on an
interval~$(u^L_0, u^R_0)$ whose size tends to infinity as~$\lambda \rightarrow \infty$.
We now work out the resulting estimates in detail. The only major change compared to the
estimates in Section~\ref{secestimate} and Sections~\ref{secacute} and~\ref{secgrave}
is that for large negative~$u$ we must approximate~$\acute{\phi}$ by Bessel functions and
must derive suitable error estimates.

Clearly, the real part of the potential again has a unique maximum~$\umax$.
We begin with the estimates in the region~$(\umax, \infty)$. Expanding the potential~\eqref{Vdef}
gives (see also~\eqref{VRasy} and~\eqref{reV1})
\begin{align}
\re V &= -\omega^2 + \frac{\tilde{\lambda}}{u^2} + \O\big( \lambda u^{-3} \big) \label{reVasp} \\
\im V &= -\frac{2 s \omega}{u} +  \O\big( u^{-2} \big) \:. \label{imVasp}
\end{align}
The connection to the situation of large~$\omega$ as considered in Section~\ref{secestimate} can be understood
directly from the following scaling argument. Suppose that~$\phi(u)$ is a solution of
the Sturm-Liouville equation~\eqref{schroedinger}. Then, introducing the new
variable~$\tilde{u} = \omega u$ and the function~$\tilde{\phi}(\tilde{u}) = \phi(u)$, we obtain
\[ \frac{d^2}{d\tilde{u}^2} \tilde{\phi} \big( \tilde{u} \big) = \frac{1}{\omega^2}\: \phi''(u) = 
\frac{1}{\omega^2}\: V(u) \:\phi(u) = 
\tilde{V}(\tilde{u}) \: \tilde{\phi}\big(\tilde{u} \big)\:, \]
where the new potential~$\tilde{V}(\tilde{u})$ has the asymptotics
\begin{align*}
\re \tilde{V}(\tilde{u}) &= \frac{1}{\omega^2} \: \re V\Big(\frac{\tilde{u}}{\omega} \Big)
= -1 + \frac{\tilde{\lambda}}{\tilde{u}^2} + \O\Big( \frac{\lambda \omega}{\tilde{u}^3} \Big) \\
\im \tilde{V}(\tilde{u}) &= -\frac{2 s}{\tilde{u}} +  \O\big( \tilde{u}^{-2} \big) \:.
\end{align*}
This means that the potential looks just as before, but with~$\omega$ replaced by one.
Hence the above scaling argument makes it possible to change~$\omega$ arbitrarily.
This explains why the methods in Section~\ref{secestimate} and Section~\ref{secgrave} again
apply. The situation is even a bit easier because we are in case~{\bf{(b)}} in~\eqref{abcase}.
For clarity, we summarize these estimates: According to~\eqref{reVasp},
the function~$\re V$ has a unique zero for large~$u$,
\[ \re V(u^R_0) = 0 \:, \qquad u^R_0 = \frac{\tilde{\lambda}}{\omega^2} + \O(1)\:. \]
As in~\eqref{uRmdef} and~\eqref{uLRpsimp} we set
\[ u^R_\pm = u^R_0 \mp \Const_3 \,\lambda^{\frac{1}{6}}\, |\omega|^{-1} \:. \]
Then the results of Proposition~\ref{prpWKBR} and Lemma~\ref{lemmaairy2}
remain true for all~$\omega$ and~$\lambda$ in the range~\eqref{rangecompact}.
As a consequence, the fundamental solution~$\grave{\phi}$ satisfies
on the interval~$(u^R_-, \infty)$ the WKB approximation~\eqref{phiyWKBR},
again with an arbitrarily small error. Moreover, the behavior
on the interval~$(u^R_+, u^R_-)$ can be estimated as in Lemma~\ref{lemmaairygrave}.
Finally, on the interval~$(\umax, u^R_+)$ one can estimate the solution exactly
as in Lemma~\ref{lemmaWKBmatchgrave}.

We come to the estimates in the region~$(-\infty, \umax)$.
Near~$u=-\infty$, the potential has the asymptotic form (see also~\eqref{VLasy})
\beq \label{Vapproxminf}
V(u) = -\Omega^2 + c_1 \big( \lambda + \nu) \: e^{\gamma u} + 
c_2 \,\lambda\, e^{2 \gamma u} + \O\big( e^{2 \gamma u} \big) \:,
\eeq
where~$c_1$ is the positive constant
\[ c_1 = \frac{r_1-r_0}{a^2+r_1^2} \:, \]
$c_2$ is a real constant,
and~$\nu=\nu(\omega)$ is a linear polynomial in~$\omega$ with complex coefficients.
The real part of the potential again has a unique zero for large negative~$u$,
\[ \re V(u^L_0) = 0 \:, \qquad u^L_0 = \frac{1}{\gamma} \:\log \bigg( \frac{\Omega^2}{c_1 \,(\lambda+\nu)} \bigg)
\Big( 1 + \O \big(\lambda^{-1} \big) \Big) \:. \]
Obviously, $u^L_0$ tends to $-\infty$ as~$\lambda \rightarrow \infty$.
In order to see the basic difference to the estimates in Section~\ref{secestimate},
let us consider the situation that~$u \ll u^L_0$ and~$\Omega=0$.
Then~$V \approx c_1 \big( \lambda + \nu) \: e^{\gamma u}$.
Evaluating the expressions in the WKB conditions, we obtain
\[ \frac{|V'|}{|V|^\frac{3}{2}} \approx \frac{c_2}{(\lambda + \nu)^\frac{1}{2}}\: e^{-\frac{\gamma u}{2}} \:, \qquad
\frac{|V''|}{|V|^2} \approx \frac{c_3}{\lambda + \nu}\: e^{-\gamma u} \:. \]
Since the exponential factors increase exponentially as~$u \rightarrow -\infty$, the
WKB conditions fail if~$u \lesssim -\log \lambda$.
In particular, in the limiting case~$\Omega=0$, the WKB approximation does not apply near~$u=-\infty$.
This is why the methods in Section~\ref{secestimate} and Section~\ref{secacute} no longer apply.
Instead, we use the $\kappa$-method introduced in~\cite[Section~3.3]{tinvariant} to obtain the
following result:
\begin{Prp} \label{prpayes}
There are constants~$\const_0, \const_1>0$ such that for all~$\omega$ and~$\lambda$
in the range~\eqref{rangecompact}, the solution~$\acute{y}(u)$ in Theorem~\ref{thm31}
satisfies the estimate
\beq \label{ayes}
\re \acute{y}(u) \geq \frac{\sqrt{\lambda}}{\const_1}\:e^{\frac{\gamma u}{2}} - \const_1
\eeq
for all~$u$ in the interval
\beq \label{c0def}
u \in (-\infty, \umin] \qquad \text{with} \qquad \umin =  -\frac{\log \lambda}{2 \gamma} - \const_0 \:.
\eeq
\end{Prp}
Before coming to the proof of this proposition, we explain its significance and
work out an application. To this end, we evaluate the inequality~\eqref{ayes} at~$u=\umin$. Then
\beq \label{WKBcompact}
\lambda\, e^{\gamma u} = \lambda\, e^{-\gamma \const_0} \, e^{-\frac{\log \lambda}{2}} =
\frac{\sqrt{\lambda}}{e^{\gamma c_0}} \:,
\eeq
which can be made arbitrarily large by increasing~$\lambda$
(note that the factor~$1/2$ in~\eqref{c0def} is essential).
As a consequence, at~$\umin$ the summand~$\lambda\, e^{\gamma u}$ dominates
the potential~\eqref{Vapproxminf}. This also implies that we are in the WKB regime.
The inequality~\eqref{ayes} shows that in this regime, the solution~$\acute{y}$
has a large real part, meaning that~$\acute{\phi}$ is well-approximated by the
exponentially {\em{increasing}} fundamental solution.
This gives rise to the following result:

\begin{Lemma}  \label{lemmaWKBplus} 
There are constants~$\Const_{10}$ and~$c=c(\lambda, \omega)$ such that on the interval $(\umin, \umax)$
the following inequalities hold
\[ \frac{|\acute{\phi}(u)|}{\Const_{10}} \leq
\frac{c(\lambda, \omega)}{(\re V(u))^\frac{1}{4}} \: e^{\int_{-\infty}^u \re \sqrt{V}} 
\leq \Const_{10}\, |\acute{\phi}(u)|
\:, \]
for all~$\omega$ and~$\lambda$ in the range~\eqref{rangecompact}.
\end{Lemma}
\Proof We proceed similar as in the proof of Lemma~\ref{lemmaWKBmatch}.
As shown after~\eqref{WKBcompact}, the WKB conditions are satisfied at~$\umin$.
Moreover, as in Proposition~\ref{prpWKBpos} one verifies that the WKB conditions are
also satisfied on the interval~$(\umin, \umax)$. Thus on this interval, the WKB approximation
\[ \acute{\phi} \approx \frac{1}{(\re V)^\frac{1}{4}} \left( C_1\: e^{\int_{u^R_+}^u \sqrt{V}}
+ C_2\: e^{-\int_{u^R_+}^u \sqrt{V}} \right) \]
holds with an error which can be made arbitrarily small by increasing~$\lambda$.
The inequality~\eqref{ayes} implies that the quotient~$C_1/C_2$
is bounded away from zero.
\QED

The remainder of this section is devoted to the proof of Proposition~\ref{prpayes}.
According to~\eqref{varpidef}, we know that~$\im \Omega<0$.
Moreover, it suffices to consider the case
\[ \re \Omega \geq 0 \:, \]
because the case~$\re \Omega <0$ can be treated in exactly the same way
by considering the complex conjugate equation.
Then the solution~$\acute{\phi}$ with the asymptotics~\eqref{acutephiasy}
starts at~$u=-\infty$ in the upper half plane. But, depending on the sign of the imaginary
part of the parameter~$\nu$ in~\eqref{Vapproxminf}, the imaginary part of~$y$ could change signs.
Thus there might be~$\uflip \in (-\infty, \umin)$ such that
\beq \label{flip}
\im y \big|_{(-\infty, \uflip)} \geq 0 \qquad \text{and} \qquad \im y \big|_{(\uflip, \umin)} \leq 0 \:.
\eeq
In order to treat all possible cases at once, in the case that~$\im y$ does not change signs,
we again work with~\eqref{flip} but choose~$\uflip = \umin$.

We choose the approximate potential as
\beq \label{Vtdef}
\tilde{V}(u) = -\tilde{\Omega}^2 +  c_1 \big( \lambda + \tilde{\nu} \big) \: e^{\gamma u} 
\qquad \text{with} \qquad  \tilde{\nu} := i \,|\im \nu| + i
\eeq
and~$\tilde{\Omega} \in \C$ to be determined below. The
corresponding Sturm-Liouville equation $\tilde{\phi}''(u) = \tilde{V}(u)\, \tilde{\phi}(u)$
can be solved explicitly in terms of Bessel functions.
Similar to~\eqref{acutephiasy}, we want to arrange the asymptotic behavior
\[ \tilde{\phi}(u) \sim e^{i \tilde{\Omega} u} \qquad \text{as~$u \rightarrow -\infty$}\:. \]
This gives the unique solution (see~\cite[\S10.25]{DLMF})
\beq \label{tpBessel}
\tilde{\phi}(u) = \gamma^{\frac{2 i \tilde{\Omega}}{\gamma}}
\: \Gamma\Big( 1 + \frac{2 i \tilde{\Omega}}{\gamma} \Big) \:I_\frac{2 i \tilde{\Omega}}{\gamma}\bigg( 
\frac{2}{\gamma}\: \sqrt{c_1 ( \lambda + \tilde{\nu})\: e^{\gamma u}} \bigg) \:.
\eeq
This solution is well-defined and regular. It is well-behaved in the limit~$\tilde{\Omega} \rightarrow 0$.
The corresponding solution of the Riccati equation
\[ \tilde{y} := \frac{\tilde{\phi}'(u)}{\tilde{\phi}(u)} \]
is also well-defined and smooth and has the asymptotics
\beq \label{ytasy}
\lim_{u \rightarrow -\infty} \tilde{y}(u) = i \tilde{\Omega}\:.
\eeq

Next, we choose
\beq \label{tOmega}
\re \tilde{\Omega} = (1+\delta)\: \re \Omega \geq 0 \qquad \text{and} \qquad
\im \tilde{\Omega} = \frac{\im \Omega}{1+\delta} < 0
\eeq
for a parameter~$\delta>0$ which later on we will choose sufficiently small.
Using these inequalities in~\eqref{ytasy}, one sees that
the solution~$\tilde{y}$ starts at~$u=-\infty$ in the upper half
plane. Moreover, as~$\im \tilde{V}>0$ (see~\eqref{Vtdef} and again~\eqref{tOmega}), we conclude
that~$\tilde{y}$ stays in the upper half plane,
\[ \im \tilde{y} \geq 0 \qquad \text{for all~$u \in \R$}\:. \]
In order to get more detailed information on~$\tilde{y}$, it is useful to again consider the WKB
approximation and expand it for large~$\lambda$,
\begin{align}
\tilde{\phi}_\WKB \;&\!:= \frac{1}{\sqrt[4]{\tilde{V}}} \;e^{\int^u \sqrt{\tilde{V}}} \label{phiWKB} \\
\tilde{y}_\WKB \:&\!:= \frac{\tilde{\phi}'_\WKB}{\tilde{\phi}_\WKB}
= \sqrt{\tilde{V}} -\frac{\tilde{V}'}{4 \tilde{V}} \notag \\
&= \sqrt{-\tilde{\Omega}^2 +  c_1 \big( \lambda + \tilde{\nu})\, e^{\gamma u}}
-\frac{1}{4}\: \frac{c_1 \big( \lambda + \tilde{\nu})\, \gamma\,e^{\gamma u}}
{-\tilde{\Omega}^2 + c_1 \big( \lambda + \tilde{\nu})\, e^{\gamma u}} \notag \\
&= \sqrt{c_1 \lambda} \: e^{\frac{\gamma u}{2}} - \frac{\gamma}{4}
+ \frac{i \mathscr{K}}{\sqrt{\lambda}}\: e^{\frac{\gamma u}{2}}
+ \O\big( \lambda^{-\frac{1}{2}} \big) \label{yWKB} \\
\re \big( \tilde{y}_\WKB \big) \:&\im \big( \tilde{y}_\WKB \big) 
= \frac{1}{2} \, \im \big( \tilde{y}_\WKB^2 \big) \approx \frac{1}{2} \, \im \big( \tilde{V} \big) \:,
\end{align}
where the parameter~$\mathscr{K}$ depends on~$\tilde{\Omega}$.
Clearly, this WKB approximation only applies if~$\lambda e^{\gamma u} \gg 1$.
Even in this regime, it is not at all obvious that~$\tilde{y}_\WKB$ approximates~$\tilde{y}$.
Namely, the function~$\tilde{\phi}$ is in general a linear combination of the WKB solution in~\eqref{phiWKB}
and the other, exponentially decaying WKB solution.
As a consequence, the function~$\tilde{y}$ could have a different form.
It turns out that, using the explicit form of~\eqref{tpBessel}, the function~$\tilde{y}_\WKB$
does describe the qualitative behavior of the solution correctly, as is made precise in the following lemma.

\begin{Lemma} There are constants~$\const_2$, $\const_3$, $\const_4$ such
that for all~$\omega$ and~$\lambda$ in the range \eqref{rangecompact}, the functions
\beq \label{alphadef}
\alpha := \re \tilde{y} \qquad \text{and} \qquad \tilde{\beta} := \im \tilde{y}
\eeq
satisfy on the interval~$(-\infty, \umin]$ (with~$\umin$ according to~\eqref{c0def}) the inequalities
\begin{align}
\frac{\sqrt{\lambda}}{\const_2} \: e^{\frac{\gamma u}{2}} &\leq \alpha \label{bess1} \\
\frac{1}{\const_3\: \alpha} &\leq \tilde{\beta} \leq \const_3 \label{bess2} \\
\bigg| \frac{\tilde{\beta}'}{\tilde{\beta}} \bigg| &\leq \const_4\: e^{\frac{\gamma u}{2}} \:. \label{bess3}
\end{align}
\end{Lemma}
\Proof Writing the potential~$\tilde{V}$, \eqref{Vtdef} as
\[ \tilde{V}(u) = -\tilde{\Omega}^2 +  c_1 \bigg( 1 + \frac{\tilde{\nu}}{\lambda} \Big) \: e^{\gamma u + \log \lambda} 
=  -\tilde{\Omega}^2 +  \tilde{b} \: e^{\gamma v} \]
with
\beq \label{vbdef}
v := u + \frac{\log \lambda}{\gamma} \qquad \text{and} \qquad \tilde{b} := 
c_1 \bigg( 1 + \frac{\tilde{\nu}}{\lambda} \Big) \:,
\eeq
one can arrange that the problem depends on two parameters~$\tilde{\Omega}$ and~$\tilde{b}$
which both lie in a bounded set. Indeed, by increasing the constant~$\Const_7$
in~\eqref{rangecompact} one can even arrange that~$\tilde{b}$ is arbitrarily close to~$c_1$.
Therefore, it suffices to analyze the perturbation of a one-parameter problem.

In the variable~$v$, the fundamental solution~$\tilde{\phi}$, \eqref{tpBessel}, becomes
\[ \tilde{\phi}(v) \sim I_\frac{2 i \tilde{\Omega}}{\gamma}\bigg( 
\frac{2}{\gamma}\: \sqrt{\tilde{b}\: e^{\gamma v}} \bigg) \:. \]
In the limit~$v \rightarrow -\infty$, the argument of the Bessel function tends to zero.
Using the power expansion of the Bessel functions (see~\cite[eqn~(10.25.2)]{DLMF}), one obtains
\begin{align*}
\tilde{\phi}(v) &\sim e^{i \tilde{\Omega} v} \:\Big( a_0 + a_1 \, e^{\gamma v} \Big) + \O\big( e^{2 \gamma v} \big) \\
\tilde{\phi}'(v) &\sim i \tilde{\Omega} \, \tilde{\phi}(v) +
e^{i \tilde{\Omega} v} \: \gamma a_1 \, e^{\gamma v} + \O\big( e^{2 \gamma v} \big) \\
\tilde{y}(v) &= i \tilde{\Omega} + \frac{\gamma a_1}{a_0}\: e^{\gamma v} + \O\big( e^{2 \gamma v} \big) \:,
\end{align*}
where the complex coefficients~$a_0$ and~$a_1$ depend on~$\tilde{\Omega}$ and~$\tilde{b}$.
These coefficients are given in terms of the gamma function, and one verifies explicitly that they are non-zero.
This shows that the relations~\eqref{bess1}--\eqref{bess3} hold for sufficiently small and negative~$v$.

For large~$v$, the Bessel function goes over to the exponentially increasing WKB approximation
(see~\cite[eqn~(10.40.1)]{DLMF}),
\begin{align}
\tilde{\phi}(v) &\approx \frac{c}{\sqrt[4]{\tilde{V}(v)}}\: \exp \left( \int^v \sqrt{\tilde{V}} \right) \\
\tilde{y}(v) &\approx \sqrt{\tilde{V}} - \frac{\tilde{V}'}{4 \tilde{V}} 
= \sqrt{\tilde{b} \: e^{\gamma v}-\tilde{\Omega}^2}
-\frac{\gamma \tilde{b} \: e^{\gamma v}}{4\, \big(\tilde{b} \: e^{\gamma v}-\tilde{\Omega}^2 \big)} \notag \\
&\approx \sqrt{\tilde{b}}\: e^{\frac{\gamma v}{2}} -\frac{\tilde{\Omega}^2}{2 \sqrt{\tilde{b}}}\: e^{-\frac{\gamma v}{2}}
-\frac{\gamma}{4} - \frac{\gamma \tilde{\Omega}^2}{4 \tilde{b}}\: e^{-\gamma v} \:, \label{imtake}
\end{align}
where~$\approx$ means that we neglect higher orders in~$e^{-\gamma v}$.
Taking the real part of the last equation, we immediately obtain
\beq \label{alphaupper}
\alpha \approx e^{\frac{\gamma v}{2}} \:,
\eeq
giving in particular the lower bound~\eqref{bess1}.
Taking the imaginary part of~\eqref{imtake}, we can make use of the fact that, according to~\eqref{vbdef}, the
parameter~$\tilde{b}$ is real up to an error of order~$1/\lambda$, so that
\beq \label{terms}
\tilde{\beta} \approx e^{\frac{\gamma v}{2}}\: \O\Big( \frac{1}{\lambda} \Big)
- \im \left( \frac{\tilde{\Omega}^2}{2 \sqrt{\tilde{b}}}\right) e^{-\frac{\gamma v}{2}}
- \im \left( \frac{\tilde{\gamma \Omega}^2}{4 b} \right) \: e^{-\gamma v} \:.
\eeq
Using~\eqref{WKBcompact}, we obtain at~$\vmin:=\umin+(\log \lambda)/\gamma$ that
\[ e^{\frac{\gamma v_{\min}}{2}}\: \O\Big( \frac{1}{\lambda} \Big) = \O\big( \lambda^{-\frac{3}{4}} \big)\:, \qquad
e^{-\frac{\gamma v_{\min}}{2}} = \O\big( \lambda^{-\frac{1}{4}} \big)\:,\qquad
e^{-\gamma v_{\min}} = \O\big( \lambda^{-\frac{1}{2}} \big) \:, \]
showing that the second summand in~\eqref{terms} dominates.
Combining this estimate with the upper bound~\eqref{alphaupper},
we conclude that also~\eqref{bess2} and~\eqref{bess3} hold at~$v=\vmin$.

Since~$\im \tilde{V} > 0$, we know furthermore that~$\tilde{\beta}$ remains strictly positive.
Using the validity of the inequalities~\eqref{bess1}--\eqref{bess3} asymptotically as~$v \rightarrow -\infty$
and at~$\vmin$, we conclude that for every~$\tilde{\Omega}$ and~$\tilde{\beta}$, there are
constants~$\const_2$, $\const_3$ and~$\const_4$ such that~\eqref{bess1}--\eqref{bess3}
hold for all~$v \in (-\infty, v_{\min}]$.
Since the constants can be chosen continuously in the parameters~$\tilde{\Omega}$
and~$\tilde{\beta}$, it follows that the constants can be chosen uniformly for the parameters
in any compact set. This concludes the proof.
\QED

Comparing~\eqref{Vapproxminf} and~\eqref{Vtdef}, we obtain
\begin{align}
\re (V-\tilde{V}) &= -\re \big( \Omega^2 - \tilde{\Omega}^2 \big)
+ \O \big( e^{\gamma u} \big) + \O\big( \lambda e^{2 \gamma u} \big) \notag \\
&\!\!\!\! \overset{\eqref{tOmega}}{=} \big( 2 \delta + \delta^2 \big) \left( \re^2 \Omega
+ \frac{\im^2 \Omega}{(1+\delta)^2} \right)
+ \O \big( e^{\gamma u} \big) + \O\big( \lambda e^{2 \gamma u} \big) >0 \label{reVVt} \\
\im (V-\tilde{V}) &= -\im \big( \Omega^2 - \tilde{\Omega}^2 \big)
+ c_1 \im \big( \nu - \tilde{\nu} ) \: e^{\gamma u} + \O\big( e^{2 \gamma u} \big) \notag \\
&\!\!\!\!\overset{\eqref{tOmega}}{=}
c_1 \im \big( \nu - \tilde{\nu} ) \: e^{\gamma u} + \O\big( e^{2 \gamma u} \big) \notag \\
&\!\!\!\!\overset{\eqref{Vtdef}}{=}
c_1 \big( \im \nu - |\im \nu| - 1 \big) \: e^{\gamma u} + \O\big( e^{2 \gamma u} \big) < 0 \:, \label{imVVt}
\end{align}
where the inequalities hold for all~$u<\umin$, provided that~$\umin$ is sufficiently small
(as can be arranged by increasing the constant~$\Const_7$ in~\eqref{rangecompact}
as a function of~$\delta$).

We first consider the region~$(-\infty, \uflip)$ where~$\im y \geq 0$.
We apply the $\kappa$-method introduced in~\cite[Section~3.3]{tinvariant}.
Let us choose the function~$\kappa$. Recall that this function is defined by
\beq \label{kset}
\kappa(u) = \frac{g(u)}{\sigma(u)} + \frac{1}{\sigma} \int_{-\infty}^u \sigma \im(V-\tilde{V}) \:,
\eeq
where~$\sigma$ is defined by
\beq \label{sigmadef0}
\sigma(u) := \exp \left( \int^u 2 \alpha \right) ,
\eeq
and~$g$ can be any monotone increasing function.
Since~$\im(V-\tilde{V})<0$, we may choose~$g(u)$ as
\[ g(u) = -\int_{-\infty}^u \sigma \im(V-\tilde{V}) \]
to obtain
\[ \kappa \equiv 0 \:. \]
Then~\cite[Lemma~3.4]{tinvariant} simplifies to
\[ \kappa - R = \frac{- \re (V-\tilde{V})}{2 \tilde{\beta}} \:. \]
As a consequence, the formula for the determinator~\cite[eqn~(3.27)]{tinvariant} can be rewritten as follows,
\begin{align}
{\mathfrak{D}} &= 2 \alpha\, \re (V-\tilde{V}) + \frac{1}{2}\, \re (V-\tilde{V})'
+ \tilde{\beta} \im (V-\tilde{V}) + (\kappa - R) \im V \label{deter} \\
&= \left( 2 \alpha - \frac{\im V}{2 \tilde{\beta}} \right) \re (V-\tilde{V}) 
+ \O \big( e^{\gamma u} \big) + \O\big( \lambda e^{2 \gamma u} \big) \:, \label{deter2}
\end{align}
where in the last line we used~\eqref{bess2} and~\eqref{imVVt}.
Next,
\begin{align*}
2 \alpha - \frac{\im V}{2 \tilde{\beta}} &= \alpha +
\frac{2 \alpha \tilde{\beta}- \im \tilde{V}}{2 \tilde{\beta}}
- \frac{\im (V-\tilde{V})}{2 \tilde{\beta}}
\overset{\eqref{imVVt}}{\geq} \alpha +
\frac{2 \alpha \tilde{\beta}- \im \tilde{V}}{2 \tilde{\beta}} \\
&= \alpha +
\frac{\im \big(\tilde{y}^2 - \im \tilde{V} \big)}{2 \tilde{\beta}}
= \alpha - \frac{\im \tilde{y}'}{2 \tilde{\beta}}
= \alpha - \frac{\tilde{\beta}'}{2 \tilde{\beta}} 
\geq \frac{\sqrt{\lambda}}{2 \const_2} \: e^{\frac{\gamma u}{2}} \:,
\end{align*}
where in the last step we applied~\eqref{bess1} and~\eqref{bess3}
and increased the constant~$\Const_7$ in~\eqref{rangecompact}.
Using~\eqref{reVVt}, we conclude that
\begin{align*}
{\mathfrak{D}} &= \frac{\sqrt{\lambda}}{\const}\:  e^{\frac{\gamma u}{2}}\:
\Big( 1 + \O \big( e^{\gamma u} \big) + \O\big( \lambda\, e^{2 \gamma u} \big) \Big) 
+ \O \big( e^{\gamma u} \big) + \O\big( \lambda e^{2 \gamma u} \big)
\end{align*}
with a positive constant~$\const=\const(\delta)$.
By choosing~$\const_0$ in~\eqref{c0def} sufficiently large, we can arrange that all the error
terms are small on the interval~$(-\infty, \uflip)$.
Thus the determinator is positive. We conclude that the invariant region estimate
in~\cite[Proposition~3.5]{tinvariant} applies, giving the estimate~\eqref{ayes}
on the interval~$(-\infty, \uflip)$.

It remains to consider the interval~$(\uflip, \umin)$. Since we want to apply again~\cite[Proposition~3.5]{tinvariant},
it is most convenient to take the complex conjugate of the equation. This corresponds to the replacements
\[ \im V \rightarrow -\im V \:,\quad \im \tilde{V} \rightarrow -\im \tilde{V} \:,\quad
\im y \rightarrow -\im y \:,\quad \im \tilde{y} \rightarrow -\im \tilde{y}\:, \ldots \:. \]
Then the solution~$\tilde{y}$ is again in the upper half plane. The invariant circle is reflected at the
real axis (corresponding to the transformation~$\beta \rightarrow -\beta$).
The only difference compared to the above analysis is that the factor~$\im (V - \tilde{V})$ in~\eqref{kset}
is now positive, so that the integral in~\eqref{kset} is increasing. Therefore, we now choose~$g \equiv 0$,
implying that~$\kappa \geq 0$. Using the formula for~$\kappa-R$ in~\cite[Lemma~3.4]{tinvariant},
the last summand in~\eqref{deter} can be estimated by
\begin{align*}
(\kappa - R)\, \im V &= \frac{\kappa^2 - \re (V-\tilde{V})}{2 \,(\tilde{\beta} + \kappa)}\: \im V
= \frac{\kappa^2}{2 \,(\tilde{\beta} + \kappa)}\: \im V  
- \frac{\re (V-\tilde{V})}{2 \,(\tilde{\beta} + \kappa)}\: \im V \\
&\geq - \frac{\re (V-\tilde{V})}{2 \,(\tilde{\beta} + \kappa)}\: \im V
\geq - \frac{\re (V-\tilde{V})}{2 \tilde{\beta}}\: \im V \:,
\end{align*}
where in the last line we used the fact that~$\im V \geq 0$ (otherwise the solution~$y$ would
not have crossed the real axis), and that~$\re (V-\tilde{V})$ is positive according to~\eqref{reVVt}.
Thus we have estimated the determinator by the expression in~\eqref{deter2}, making it possible
to proceed just as on the interval~$(-\infty, \uflip)$ above. 
Note that, estimating~\eqref{kset} in the case~$g \equiv 0$
using~\eqref{imVVt}, keeping in mind that~$\sigma$ is monotone increasing in view
of~\eqref{sigmadef0} and~\eqref{bess1}, one sees that
\beq \label{kappabound}
0 \leq \kappa \leq \const \qquad \text{on~$(-\infty, \umin)$}
\eeq
(where~$\const$ is again a constant which is uniform in~$\omega$ and~$\lambda$ in the
range~\eqref{rangecompact}; note that this inequality is trivial on the interval~$(-\infty, \uflip)$
where~$\kappa \equiv 0$).

In the above arguments we concluded that~\cite[Proposition~3.5]{tinvariant}
applies. It follows that the solution~$\acute{y}$ lies inside the circle with center~$m=\alpha + i \beta$
and radius~$R$, with~$\alpha$ as defined in~\eqref{alphadef} and
\beq \label{ninvdisk}
R+\beta = \tilde{\beta}+\kappa \:,\qquad R-\beta = \frac{U}{R+\beta}\:.
\eeq
Here the function~$U$ is given by (see~\cite[eqns~(3.3) and~(3.17)]{tinvariant},
\beq \label{Urel}
U := \re V - \alpha^2 - \alpha' = \re (V-\tilde{V}) - \tilde{\beta}^2 \:.
\eeq
Let us analyze what this estimate means for the radius. Combining~\eqref{ninvdisk}
and~\eqref{Urel}, we obtain
\[ 2 R = \big( \tilde{\beta}+\kappa \big) + \frac{U}{\tilde{\beta}+\kappa}
= \big( \tilde{\beta}+\kappa \big) + \frac{\re (V-\tilde{V})}{\tilde{\beta}+\kappa} 
- \frac{\tilde{\beta}^2}{\tilde{\beta}+\kappa} 
\leq \big( \tilde{\beta}+\kappa \big) + \frac{\re (V-\tilde{V})}{\tilde{\beta}+\kappa} \:, \]
where we used that the summand~$\tilde{\beta}+\kappa$ is non-negative
according to~\eqref{bess2} and~\eqref{kappabound}.
Next, we know from~\eqref{reVVt} that the term~$\re (V-\tilde{V})$ is uniformly bounded
and can be made arbitrarily small by decreasing~$\delta$.
Also using that~$\tilde{\beta}$ and~$\kappa$ are both positive (see again~\eqref{bess2} and~\eqref{kappabound}),
we conclude that
\[ 2R \leq \tilde{\beta}+\kappa + \frac{\const \,\delta}{\tilde{\beta}} \:. \]
Applying the estimates~\eqref{bess2} and~\eqref{kappabound}, we conclude that
\beq \label{Rbound}
R \leq \const \Big( 1 + \frac{\delta}{\tilde{\beta}} \Big) \leq \const \big( 1 + \const_3\,\delta\, \alpha \big) \:.
\eeq
The fact that~$\acute{y}$ lies inside the invariant circle gives the inequality~$\re \acute{y} \geq \alpha - R$.
Combining this inequality with~\eqref{Rbound}, we obtain
\[ \re \acute{y} \geq \big(1 - \const \,\const_3 \,\delta \big) \alpha - \const \:. \]
We choose~$\delta$ so small that~$\const \const_3 \delta < 1/2$.
Using~\eqref{bess1} gives the inequality~\eqref{ayes}.
This concludes the proof of Proposition~\ref{prpayes}.

\subsection{The Limit~$\omega \rightarrow 0$} \label{secomegazero}
In the construction of the Jost solution~$\grave{\phi}:=\grave{\phi}_-$ in Theorem~\ref{thm34} as well
as in all the previous estimates of~$\grave{\phi}$ we always assumed that~$\omega \neq 0$.
We now analyze the behavior of these Jost solutions in the limit~$\omega \rightarrow 0$,
coming from the lower half plane~$\im \omega<0$.
Before beginning, we point out that if~$\lambda$ is sufficiently large, the asymptotics
for small~$\omega$ is obtained immediately by taking the limit~$\omega \rightarrow 0$
in the estimates of Sections~\ref{secgrave} and~\ref{seccompact}. This can be understood directly
by analyzing the WKB conditions: For~$\omega=0$, the asymptotics of the potential in~\eqref{reVasp}
and~\eqref{imVasp} simplifies to
\[ V(u) = \frac{\tilde{\lambda}}{u^2} + \O \big( u^{-3} \big) \:. \]
Hence
\[ \frac{|V'|}{|V|^\frac{3}{2}} = \frac{2}{\sqrt{|\tilde{\lambda}|}} \:\Big( 1 + \O\big( u^{-1} \big) \Big)
\qquad \text{and} \qquad
\frac{|V''|}{|V|^2} = \frac{6}{|\tilde{\lambda}|} \:\Big( 1 + \O\big( u^{-1} \big) \Big) \:, \]
showing that the WKB conditions are satisfied for large~$\lambda$ and~$u$.
Combining this result with the estimates in Section~\ref{secWKBpos}, one finds that
for~$\omega=0$ and large~$\lambda$, the solution~$\grave{\phi}$ is well-approximated
by the WKB solution. Consequently, the behavior for small~$\omega$ and large~$\lambda$
can be described simply by perturbing this WKB solution.

If~$\lambda$ is not large, we can use methods and results in~\cite{schdecay}.
For self-consistency, we now restate these results in our setting and outline the proofs.
\begin{Lemma} \label{lemmaomzero} Setting
\beq \label{sigmadef}
\sigma = \frac{1}{2} \left(\sqrt{1+ 4 \lambda + 4 s^2  +8 a k \omega} - 1 \right) ,
\eeq
the following limit exists,
\beq \label{philim}
\lim_{\omega \rightarrow 0, \;\im \omega \leq 0,\; \omega \neq 0} \omega^{s+\sigma} \grave{\phi} = \grave{\phi}_0\:.
\eeq
The limit function~$\grave{\phi}_0$ is a solution of the Sturm-Liouville equation~\eqref{schroedinger}
for~$\omega=0$ and has the asymptotics
\beq \label{phi0asy}
\lim_{u \rightarrow \infty} \left( u^\sigma\; \grave{\phi}_0 \right) =
\frac{(-4)^{-\frac{\sigma}{4}} \,\Gamma(2 \sigma+2)}{(2i)^s \,\Gamma(\sigma + 1-s)}\;.
\eeq
\end{Lemma}
\Proof We proceed as in the proof of~\cite[Lemma~8.1]{schdecay}.
Again working in the $r$-coordinate and writing the radial equation as
\beq \label{schV}
-\frac{d^2}{dr^2} \psi(r) + {\mathcal{V}}(r)\, \psi(r) = 0\:,
\eeq
the potential~${\mathcal{V}}$ has the following asymptotics near infinity:
\beq \label{calV}
{\mathcal{V}}(r) = - \omega^2 - 2\, \frac{i s \omega + M \omega^2}{r} +
\frac{\lambda + s^2 + 2 a k \omega - 2 i M s \omega - 12 M^2 \, \omega^2}{r^2}\:+\: {\mathcal{O}}(r^{-3})
\eeq
(this differs from the potential in~\cite[eqn~(8.6)]{schdecay} only by the summand~$(2 a k \omega)/r^2$).
Dropping the error term, the equation~\eqref{schV} can be solved explicitly in terms of Whittaker functions.
Satisfying the correct asymptotics at infinity~\eqref{phiasy}, one obtains the unique solution
\[ \tilde{\phi}(r) = \frac{r}{\sqrt{\Delta}}\; (2 i \omega)^{-s + 2 i M \omega} \:W_{\kappa, \mu}\big(
2 i \omega r \big) \:, \]
where the parameters~$\kappa$ and~$\mu$ are given by
\[ \kappa = s - 2 i \omega M \qquad \text{and} \qquad
\mu = \frac{1}{2} \, \sqrt{1 + 4 \lambda + 4 s^2  +8 a k \omega - 8 i M s \omega - 48\, M^2 \omega^2}\:. \]
For small~$\omega$, this solution has the asymptotics
\[ \tilde{\phi}(r) = \frac{r}{\sqrt{\Delta}}\; \omega^{-s-\sigma} r^{-\sigma}\;
\frac{(-4)^{-\frac{\sigma}{4}} \:\Gamma(2 \sigma+2)}{(2i)^s \Gamma(\sigma + 1-s)} \]
with~$\sigma$ as in~\eqref{sigmadef}.
This function obviously satisfies~\eqref{philim} and~\eqref{phi0asy}.

The error term in~\eqref{calV} can be treated exactly as in the proof of~\cite[Lemma~8.1]{schdecay}
by a Jost iteration, taking the solution~$\tilde{\phi}$ as the starting point.
\QED

\subsection{Estimates of the Large Angular Modes} \label{seclarge}
Combining the estimates of Sections~\ref{secacute}--\ref{secomegazero}, we obtain the following a-priori estimate:

\begin{Prp} \label{prpkernes} For any~$u_\infty >0$,
there is a constant~$\Const_{11}>0$ and~$N \in \N$ such that for all~$n>N$, the
kernels of the Green's functions~$s_\omega$ and of the operator~$g_\omega$, \eqref{sdef} and~\eqref{gkerdef},
satisfy for all~$\omega \in \R$ and all~$\lambda>\Const_7$ the bound
\beq \label{varpiform}
\big| e^{-\varpi u}\: s_\omega(u,u') \big|, \big| e^{-\varpi u}\: g_\omega(u,u') \big| \leq \Const_{11} \:,
\eeq
uniformly for all~$u<u_\infty$ and $-u_\infty < u' < u_\infty$.
\end{Prp}
Before coming to the proof, we point out that the exponential factor~$e^{-\varpi u}$
compensates for the exponential decay as~$u \rightarrow -\infty$
of the fundamental solution~$\acute{\phi}(u)$ contained in~$g(u,u')$
(see~\eqref{gkerdef}, \eqref{sdef}, \eqref{acutephiasy} and~\eqref{varpidef}).
In order to verify that this exponential factor really controls the asymptotics uniformly in~$\lambda$ and~$\omega$,
we need to estimate the absolute value of the exponential in the WKB solution~\eqref{phiyWKB}.
This is done in the next lemma.
\begin{Lemma} \label{lemmaintWKB}
There is a constant~$\Const_4$ such that
the following estimate holds in the WKB region~$(-\infty, u^L_-)$ for all~$\lambda$ and~$\omega$
in the range~\eqref{range},
\[ \int_{-\infty}^{u^L_-} \Big( -\varpi \mp \im \sqrt{-V} \Big) < \Const_4 \:. \]
\end{Lemma}
\Proof We begin with the PC and Airy cases.
Then, in view of Lemma~\ref{lemma91}, we know that~\eqref{range1} holds.
We again consider the cases~{\bf{(a)}} and~{\bf{(b)}} in~\eqref{abcase} after each other.
We begin with case~{\bf{(b)}}, where in view of Lemma~\ref{lemmaPCa} we must be in the Airy case.
From~\eqref{imVes} and~\eqref{reVesL}, we know that
on the interval~$(-\infty, u^L_-)$ the inequalities
\[ |\im V| \lesssim |\omega| \qquad \text{and} \qquad -\re V \gtrsim \frac{\Const_3}{\Const_2}\: |\omega|^\frac{4}{3} \]
hold. As a consequence,
the real part of~$V$ dominates its imaginary part, giving rise to the expansion
\[ \im \sqrt{-V} = -\frac{\im V}{2 \sqrt{-\re V}} \:\Big(1 + \O\big( |\omega|^{-\frac{1}{3}} \big) \Big) \:. \]
Using the asymptotic form of the potential~\eqref{Vapproxminf}, we obtain the expansions
\begin{align*}
\sqrt{-V} &= \Omega - \frac{c_1}{2 \Omega} \big( \lambda + \nu) \: e^{\gamma u}\:
\Big( 1 + \O\big( e^{\gamma u} \big) \Big) \\
\re \sqrt{-V} &= \re \Omega - \frac{c_1}{2 |\Omega|^2} \: \re \Omega \: \lambda\: e^{\gamma u}\:
\Big( 1 + \O\big( \omega\,\lambda^{-1} \big) + \O\big( e^{\gamma u} \big) \Big) \\
\im V &= -2 \re \Omega \: \im \Omega + c_1 \: \im \nu \: e^{\gamma u} + \O\big( e^{2 \gamma u} \big)
\end{align*}
and thus
\[ \im \sqrt{-V} = \im \Omega + \Big( - \frac{c_1\: e^{\gamma u}}{2 \re \Omega} \: \im \nu
+ \frac{c_1\: e^{\gamma u}}{2 |\Omega|^2}\: \lambda \Big)
\Big( 1 + \O\big( \omega\, \lambda^{-1} \big)+ \O\big( |\omega|^{-\frac{1}{3}} \big) + \O\big( e^{\gamma u} \big) \Big) \:. \]
Using the notation~\eqref{varpidef}, we obtain the estimate
\beq \label{bigintes}
\Big| -\varpi + \im \sqrt{-V} \Big| \lesssim \Big( \frac{|\im \nu|}{|\omega|} +
\frac{\lambda}{\omega^2} \Big) \,e^{\gamma u}
\lesssim \Big( 1 + \frac{\lambda}{\omega^2} \Big) \,e^{\gamma u}\:,
\eeq
where in the last step we used that the function~$\nu$ in~\eqref{Vapproxminf} is a linear polynomial in~$\omega$.
Integrating this inequality, we obtain
\[ \int_{-\infty}^{u^L_-} \Big( -\varpi \mp \im \sqrt{-V} \Big)
\lesssim \Big( 1 + \frac{\lambda}{\omega^2} \Big) \frac{1}{\gamma}\: e^{\gamma u^L_-} \:. \]
Applying the first inequality in~\eqref{reo2} gives the result.

In case~{\bf{(a)}}, the last estimates apply without changes in the region~$(-\infty, u_{\max}
-\Const^{-\frac{1}{2}})$ away from the maximum of~$\re V$,
and the term~$\lambda/\omega^2$ in~\eqref{bigintes} is uniformly bounded in view of~\eqref{llow}.
On the interval~$(u_{\max}-\Const^{-\frac{1}{2}}, u^L_-)$,
on the other hand, we know from~\eqref{concave} that~$|\re V''| \simeq \lambda$.
Therefore, the WKB inequality for the second derivative in Proposition~\ref{prpWKB} implies
that~$|V| \gtrsim \sqrt{\lambda/\varepsilon} \eqsim |\omega|/\sqrt{\varepsilon}$
(where in the last step we used Lemma~\ref{lemmaprep}). Combining this inequality with~\eqref{imVes},
we see that for sufficiently small~$\varepsilon$,
the real part of the potential again dominates its imaginary part, implying that
\beq \label{varpies}
\Big| -\varpi + \im \sqrt{-V} \Big| \lesssim 1 + \frac{|\im V|}{\sqrt{|\re V|}} \qquad \text{on~$\big(u_{\max}-\Const^{-\frac{1}{2}}, u^L_- \big)$}\:.
\eeq
Comparing the inequalities~\eqref{omimV} and~\eqref{drReV}
in Lemma~\ref{lemmasignimV} and using that~$\re V'(\umax)$ is zero,
we find that~$|\im V(\umax)| \lesssim 1$. Integrating~\eqref{imVes}, we infer the bound
\[ |\im V(u)| \lesssim 1 + |\omega|\, \big|u - \umax \big| \:. \]
Moreover, integrating~\eqref{concave}, we know that
\[ |\re V| \gtrsim \lambda\, \big(u - \umax \big)^2 \:. \]
Hence
\begin{align}
\int_{u_{\max}-\Const^{-\frac{1}{2}}}^{u^L_-}  \frac{|\im V|}{\sqrt{|\re V|}}
&\lesssim  \int_{u_{\max}-\Const^{-\frac{1}{2}}}^{u^L_-}
\bigg( \frac{1}{\sqrt{\lambda}\: |u-\umax|} + \frac{|\omega|}{\sqrt{\lambda}} \bigg) \notag \\
&\lesssim \frac{|\log(\umax-u^L_-)|}{\sqrt{\lambda}} + \frac{|\omega|}{\sqrt{\lambda}} \:, \label{intimes}
\end{align}
which is uniformly bounded in view of~\eqref{uLminus}, \eqref{llow} and~\eqref{range}.
This concludes the proof in case~{\bf{(a)}}.

In the remaining WKB case, we use the following monotonicity argument:
We increase~$\lambda$ until $\re V(\umax)=-\Const_4 \,\sqrt{\lambda}$.
Then we are in the PC case (see~\eqref{cases}), where the above method applies. When decreasing~$\lambda$,
the absolute value of the real part of the potential increases, whereas its imaginary part remains unchanged.
Therefore, the inequality~\eqref{varpies} remains valid, and the integral~\eqref{intimes} decreases.
This concludes the proof.
\QED

\Proof[Proof of Proposition~\ref{prpkernes}] Let us go through the different cases, beginning with the parameter range
that both~$|\omega|$ and~$\lambda$ are large~\eqref{range}:
First, according to Proposition~\ref{prpWKB}, on the interval~$(-\infty, u^L_-)$
the fundamental solution~$\acute{\phi}$ is approximated by the WKB solution in~\eqref{phiyWKB},
up to an arbitrarily small error. From Lemma~\ref{lemmaintWKB} we conclude that the asymptotics
of~$|\acute{\phi}(u)|$ is controlled by the exponential~$e^{\varpi u}$, uniformly in~$\lambda$ and~$\omega$.

We next proceed by analyzing the different cases in~\eqref{cases}. In the WKB case,
the WKB approximation applies on the whole interval~$(-\infty, \umax)$. Likewise,
on the interval~$(\umax, \infty)$ also the fundamental solution~$\grave{\phi}$ is well-approximated
by the WKB solution. Moreover, the fundamental solutions~$\acute{y}$ and~$\grave{y}$
lie in different half planes (see~\eqref{acutephiasy} and~\eqref{phiasy}). This implies that
\beq \label{sges}
\big| s_\omega(u,u') \big|, \big| g_\omega(u,u') \big| \lesssim \frac{e^{\varpi u}}{|\omega|}\:.
\eeq
Next, in the parabolic cylinder case, the estimates of Lemmas~\ref{lemmaPCacute}
and~\ref{lemmaPCgrave} show that~\eqref{sges} again holds.
Finally, in the Airy case, the estimates of Lemmas~\ref{lemmaairyacute}, \ref{lemmaWKBmatch},
\ref{lemmaairygrave} and~\ref{lemmaWKBmatchgrave} imply that~$\acute{\phi}$ is
increasing exponentially in the WKB region with~$\re V>0$, whereas~$\grave{\phi}$ is
exponentially decaying in this region. Hence~$\big| s(u,u') \big|$ and~$\big| g(u,u') \big|$ decay for large~$\lambda$,
uniformly in~$\omega$. This concludes the proof in the parameter range~\eqref{range}.

If~$\omega \neq 0$ is in a bounded set and~$\lambda$ is large~\eqref{rangecompact},
the estimates in Section~\ref{seccompact} show that~$\acute{\phi}$ and~$\grave{\phi}$
behave again just as described in the Airy case. Moreover, as by rescaling one can arrange
a compact parameter range (as explained after~\eqref{vbdef}), it is obvious that
the exponential~$e^{-\varpi u}$ in~\eqref{varpiform} again controls the behavior as~$u \rightarrow -\infty$
uniformly in all parameters. Finally, in Section~\ref{secomegazero}
it is shown that the fundamental solution~$\grave{\phi}$ as well as the Wronskian are continuous at~$\omega=0$
after the rescaling~\eqref{philim}.
Moreover, since the function~$\grave{\phi}_0$ is decreasing at infinity~\eqref{phi0asy}, the
Wronskian is non-zero in the limit. This concludes the proof.
\QED

The estimate of Proposition~\ref{prpkernes}
gives us uniform control of the large angular modes:
\begin{Prp} \label{prphighmode}
For sufficiently large~$p$ and all~$\omega \in \R$, the following estimate holds for all~$u < u_\infty$,
\[  \frac{1}{|\omega + 3 i c|^p}\;
\Big\| \Big( R_{\omega,n}\:Q_n^\omega \,\big(H + 3 i c \big)^p \,\Psi_0\big) \Big)(u)\Big\|_{L^2(S^2)} 
\leq \frac{c(u_\infty,\Psi_0)}{(n+1)^2\, (1+|\omega| )^2} \:. \]
\end{Prp}
\Proof Using Proposition~\ref{prpkernes}, similar to~\eqref{Rsimp} we obtain the estimate
\[ \Big\| \Big( R_{\omega,n}\:Q_n^\omega \,\big(H + 3 i c \big)^p \,\Psi_0 \Big)(u) \Big\|_{L^2(S^2)}
\leq C(u_\infty, \Psi_0)\: (1+\omega^2) \:. \]
Using the method in~\eqref{lamsimp}, one can generate factors of~$1/\lambda$,
\begin{align*}
\frac{1}{|\omega + 3 i c|^p}\;&
\Big\| \Big( R_{\omega,n}\:Q_n^\omega \,\big(H + 3 i c \big)^p \,\Psi_0\big) \Big)(u) \Big\|_{L^2(S^2)} \\
&\leq C\big( u_\infty, \Psi_0, \A_\omega \Psi_0, \ldots, \A_\omega^{2q} \Psi_0 \big)\: \frac{1+\omega^2}{(1+|\omega|)^p\, \lambda_n^q} \:,
\end{align*}
where in the last step we used~\eqref{lambdarange}.
Since the operator~$\A_\omega$ involves~$\omega$ at most quadratically (see~\eqref{Aop}),
we obtain the estimate
\[ \frac{1}{|\omega + 3 i c|^p}\;
\Big\| \Big( R_{\omega,n}\:Q_n^\omega \,\big(H + 3 i c \big)^p \,\Psi_0\big) \Big)(u) \Big\|_{L^2(S^2)}
\leq C(u_\infty, \Psi_0)\: \frac{1+\omega^2}{(1+|\omega|)^{p-4q}\, \lambda_n^q} \:. \]
Choosing~$p$ sufficiently large and estimating the eigenvalues~$\lambda_n$ from below
with the help of Proposition~\ref{prpangular}, we obtain the result.
\QED

\begin{Corollary} \label{cormodes}
For sufficiently large~$p$, the solution of the Cauchy problem
for the Teukolsky equation
with initial data~$\Psi|_{t=0} = \Psi_0 \in \D(H)$
can be written for any~$t < 0$ as
\[ \Psi(t) = -\frac{1}{2 \pi i} \sum_{n=0}^\infty\:\lim_{\varepsilon \searrow 0}
\int_{\R - i \varepsilon} \frac{e^{-i \omega t}}{(\omega + 3 i c)^p} \;
\Big( R_{\omega,n}\:Q_n^\omega \,\big(H + 3 i c \big)^p \,\Psi_0  \Big)\: d\omega \:. \]
Here the series converges absolutely in the sense that for any~$\varepsilon>0$, there is~$N$
such that for all~$t<0$ and all~$u<u_\infty$,
\[ \sum_{n=N}^\infty \bigg\|
\lim_{\varepsilon \searrow 0}  \bigg(
\int_{\R - i \varepsilon} \frac{e^{-i \omega t}}{(\omega + 3 i c)^p} \;
\Big( R_{\omega,n}\:Q_n^\omega \,\big(H + 3 i c \big)^p \,\Psi_0  \Big)\: d\omega \bigg)(u) \bigg\|_{L^2(S^2)} 
< \varepsilon \:. \]
\end{Corollary}
\Proof Our starting point is the integral representation~\eqref{propagatorp2} in Corollary~\ref{corprop}.
Separating the resolvent (Theorem~\ref{thmsepres}), we obtain an integral over an infinite sum
of angular modes. Exactly as explained in Lemma~\ref{lemmadeform} for the first~$N$ angular modes,
for each angular mode we may deform the integral and move it up to the real axis,
without changing the values of the integrals.
Applying the estimates of the large angular modes of Proposition~\ref{prphighmode}, we obtain the result.
\QED

\section{Ruling out Radiant Modes}
In the integral representation of Corollary~\ref{cormodes}, we know that all integrands
are holomorphic for~$\omega$ in the lower half plane, making it possible to move the
contour arbitrarily close to the real axis. However, our analysis so far does not rule
out the possibility that the integrands might have poles on the real axis. We refer to such
poles as {\em{radiant modes}}. In this section we rule out radiant modes.

\subsection{Ruling out Radiant Modes at~$\omega=0$}
For~$\omega=0$, the potential~\eqref{Vdef} simplifies to
\begin{align}
V(u) &= \frac{\lambda\, \Delta}{(r^2+a^2)^2} +  \frac{ \partial_u^2 \sqrt{r^2+a^2}}{\sqrt{r^2+a^2}} -\Big( \frac{ak - i (r-M) \,s}{r^2+a^2} \Big)^2 \notag \\
&= \frac{\lambda\, \Delta}{(r^2+a^2)^2} +  \frac{ \partial_u^2 \sqrt{r^2+a^2}}{\sqrt{r^2+a^2}}
+\frac{(r-M)^2 \,s^2 - a^2 k^2}{(r^2+a^2)^2} +2 i s \; \frac{ak\: (r-M)}{(r^2+a^2)^2}\:. \label{Vradform}
\end{align}
In particular, one sees that the imaginary part of~$V$ has a fixed sign,
\beq \label{imVpos}
k \,\im V(u) = 2 s \; \frac{ak^2\: (r-M)}{(r^2+a^2)^2} \geq 0 \:.
\eeq

\begin{Lemma} \label{lemmaom0}
For every angular mode,
the kernels of the Green's functions~$s_\omega$ and of the operator~$g_\omega$, \eqref{sdef} and~\eqref{gkerdef},
are uniformly bounded in a neighborhood of~$\omega=0$ (here again~$\im \omega \leq 0$).
\end{Lemma}
\Proof In view of the continuity results of Theorem~\ref{thm31} and Lemma~\ref{lemmaomzero},
it remains to show that choosing~$\omega=0$,
the Wronskian~$w(\acute{\phi}, \grave{\phi}_0)$ (with~$\grave{\phi}_0$ as in~\eqref{philim})
is non-zero. Assume conversely that this Wronskian were zero.
Then the solutions~$\acute{\phi}$, $\grave{\phi}_0$ are multiples of each other.
Thus there is a non-trivial solution~$\phi$ of the Sturm-Liouville equation~\eqref{schroedinger}
which decays both as~$u \rightarrow \pm \infty$. More precisely, this solution
decays exponentially as~$u \rightarrow -\infty$ (see~\eqref{acutephiasy}, keeping in mind
that~$\Omega$ in~\eqref{Omegadef} has a negative imaginary part),
whereas it decays polynomially as~$u \rightarrow \infty$ (see~\eqref{phi0asy}).
In particular, the solution is in~$L^2(\R)$.

In the case~$k=0$, the potential~\eqref{Vradform} is obviously real and positive.
As a consequence, the solution~$\phi$ is convex (for details see~\cite[Section~5]{angular}),
contradicting the fact that it decays
as~$u \rightarrow \pm \infty$. In the remaining case~$k \neq 0$, we
make use of an observation made previously in~\cite[Section~9]{tspectral}.
Multiplying the differential equation for~$\phi$ by~$\overline{\phi}$ and integrating, we obtain
\[ 0 = \int_0^\pi \overline{\phi} \left( -\frac{d^2}{du^2} + V \right) \phi
\overset{(\star)}{=}  \int_0^\pi \overline{\left( -\frac{d^2}{du^2} + \overline{V} \right)  \phi} \:\phi
= \int_0^\pi (V-\overline{V}) \:\overline{\phi} \phi \:, \]
where in~($\star$) we integrated by parts and used the
decay properties of~$\phi$ to conclude that the boundary terms vanish. We thus obtain the relation
\[ \int_0^\pi \im V\, |\phi|^2 = 0 \:. \]
Using~\eqref{imVpos}, we conclude that~$\phi$ must vanish identically, a contradiction.
\QED

\subsection{A Causality Argument}
In the following proposition we show that the separated resolvent has no poles on the real
axis. The method makes use of finite speed of propagation and is an improvement of the
method first developed for the scalar wave equation in~\cite[Section~7]{wdecay}.
We remark that an alternative method for ruling out radiant modes is given in~\cite{andersson+whiting}.

\begin{Prp} \label{prpnopole}
For any~$n \in \N \cup \{0\}$, the separated resolvent~$R_{\omega, n}$, \eqref{Romn},
is holomorphic in the lower half plane~$\{\im \omega < 0\}$. Moreover, it is continuous
up to the real axis, i.e. the limit
\[ R^-_{\omega,n} \Psi := \lim_{\varepsilon \searrow 0} \big(R_{\omega-i \varepsilon,n} \Psi) \qquad
\text{exists for all~$\omega \in \R$}\:. \]
\end{Prp}
\Proof Let~$\omega_0 \in \R$. We want to show that~$R_{\omega, n}$ is continuous
at~$\omega_0$. In the case~$\omega_0=0$, the result follows immediately
from Lemma~\ref{lemmaom0}.
In the remaining case~$\omega_0 \neq 0$, for test functions~$\eta_1, \eta_2 \in C^\infty_0(\R)$ and a real
parameter~$L$ we set
\begin{align}
\Phi_L(u,\vartheta, \varphi) &= \eta_1(u+2 L)\: e^{-i k \varphi}\: 
e^{-2i \overline{\Omega_0} L}\: \Theta_{\omega_0, n}(\vartheta) \label{PhiLdef} \\
\Phi_\text{test}(u,\vartheta, \varphi) &= \frac{r^2+a^2}{\rho}\: \eta_2(u)\: e^{-i k \varphi}\: \Theta_{\omega_0, n}(\vartheta) \:, \label{Phitestdef}
\end{align}
where~$\Theta_{\omega_0, n}$ is an eigenfunction of the angular operator~$\A_{\omega_0}$
in the image of~$Q^\omega_n$. Moreover, we set
\[ \Psi_L = \begin{pmatrix} \Phi_L \\ 0 \end{pmatrix} \qquad \text{and} \qquad
\Psi_\text{test} = \begin{pmatrix} 0 \\ \Phi_\text{test} \end{pmatrix} \:. \]
Finally, we let~$\Psi_{L,t}$ be the solution of the Cauchy problem with initial data~$\Psi_{L,0} = \Psi_L$.
Then, due to finite propagation speed, it follows that for sufficiently large~$L$,
the functions~$\Psi^t_L$ and~$\Psi_\text{test}$ have disjoint supports if~$t \in [-L,0]$. Hence
for any power~$r \in \N$,
\beq \label{causality}
0 = \frac{1}{L} \int_{-L}^0 e^{i \omega_0 t} \:\big\la (H-\omega_0)^r \,\Psi^t_L, \Psi_\text{test}
\big\ra_{L^2(\R \times S^2)}\: dt \:.
\eeq
Corollary~\ref{cormodes} yields the integral representation 
\begin{align*}
&(H-\omega_0)^r \Psi^t_L \\
&= -\frac{1}{2 \pi i} \sum_{n'=0}^\infty \;\lim_{\varepsilon \searrow 0}
\int_{\R - i \varepsilon} e^{-i \omega t}\: \frac{(\omega-\omega_0)^r}{(\omega + 3 i c)^p} \;
\Big( R_{\omega,n'} \:Q_{n'}^\omega \,\big(H + 3 i c \big)^p \,\Psi_L  \Big)\: d\omega \:,
\end{align*}
where the infinite sum over~$n'$ converges absolutely if for any given~$r \in \N$
we choose~$p$ sufficiently large.
Using this representation in~\eqref{causality} and introducing the short notation
\[ \Xi_{L, \omega} = \frac{(\omega-\omega_0)^r}{(\omega + 3 i c)^p} \:\big(H + 3 i c \big)^p\, \Psi_L \:, \]
we obtain
\begin{align}
0 &= \sum_{n'=0}^\infty \;\lim_{\varepsilon \searrow 0}
\int_{\R - i \varepsilon} \big\la R_{\omega,n'} \:Q_{n'}^\omega \,\Xi_{L, \omega}, \Psi_\text{test} \big\ra_{L^2(\R \times S^2)}\;
\frac{i}{L} \int_{-L}^0 e^{-i (\omega-\omega_0) t} \: dt \notag \\
&= \sum_{n'=0}^\infty \;\lim_{\varepsilon \searrow 0}
\int_{\R - i \varepsilon}\:  \:\frac{e^{i (\omega-\omega_0) L}-1}{(\omega-\omega_0) L}\;
\big\la R_{\omega,n'} \,Q_{n'}^\omega \,\Xi_{L, \omega}, \Psi_\text{test} \big\ra_{L^2(\R \times S^2)} \:.
\label{sumrep}
\end{align}

Let~$\delta>0$. According to Corollary~\ref{cormodes}, we know that for sufficiently large~$N$,
\[ \sum_{n'=N}^\infty \bigg| \lim_{\varepsilon \searrow 0}
\int_{\R - i \varepsilon} \!\!\!\frac{e^{i (\omega-\omega_0) L}-1}{(\omega-\omega_0) L}\;
\big\la R_{\omega,n'} \,Q_{n'}^\omega \,\Xi_{L, \omega}, \Psi_\text{test} \big\ra_{L^2(\R \times S^2)} \bigg|
< \delta \:, \]
uniformly for large~$L$. Moreover, using Proposition~\ref{prphighmode}, we may
choose~$\omega_{\max} > 2 \,|\omega_0|$ such that
\[ \sum_{n=0}^N \bigg| \Big( \int_{-\infty}^{-\omega_{\max}} \!\!\!+ \int_{\omega_{\max}}^\infty \Big)
\frac{e^{i (\omega-\omega_0) L}-1}{(\omega-\omega_0) L}\;
\big\la R_{\omega,n'} \,Q_{n'}^\omega \,\Xi_{L, \omega}, \Psi_\text{test} \big\ra_{L^2(\R \times S^2)} \bigg|
< \delta \:, \]
again uniformly in~$L$. Using these estimates in~\eqref{sumrep}, we conclude that
\beq \label{intes}
\bigg| \sum_{n'=0}^N \lim_{\varepsilon \searrow 0}
\int_{-\omega_{\max}-i \varepsilon}^{\omega_{\max}-i \varepsilon}
\frac{e^{i (\omega-\omega_0) L}-1}{(\omega-\omega_0) L}\;
\big\la R_{\omega,n'} \,Q_{n'}^\omega \,\Xi_{L, \omega}, \Psi_\text{test} \big\ra_{L^2(\R \times S^2)} \bigg|
< 2 \delta \:,
\eeq
uniformly for large~$L$.

\begin{figure}
\psscalebox{1.0 1.0} 
{
\begin{pspicture}(0,-1.4103028)(6.52,1.4103028)
\rput[bl](6.76,-0.6693652){\normalsize{$\re \omega$}}
\psline[linecolor=black, linewidth=0.04, arrowsize=0.02cm 8.0,arrowlength=1.4,arrowinset=0.4]{->}(0.0,-0.8593652)(6.8,-0.8593652)
\psline[linecolor=black, linewidth=0.04, arrowsize=0.02cm 8.0,arrowlength=1.4,arrowinset=0.4]{->}(3.195,-1.3393652)(3.195,1.4456347)
\rput[bl](3.42,1.1306348){\normalsize{$\im \omega$}}
\rput[bl](0.385,-1.3243653){\normalsize{$-\omega_{\max}$}}
\psbezier[linecolor=black, linewidth=0.06](0.8,-0.8593652)(2.3978846,-0.68924093)(2.64,0.73563474)(3.64,0.73563474)(4.64,0.73563474)(5.257776,0.1277335)(5.59,-0.8593652)
\psline[linecolor=black, linewidth=0.04](0.79,-0.7543652)(0.795,-0.95436525)
\psline[linecolor=black, linewidth=0.04](3.985,-0.7543652)(3.99,-0.95436525)
\psline[linecolor=black, linewidth=0.04](5.59,-0.76436526)(5.595,-0.96436524)
\rput[bl](5.465,-1.3343652){\normalsize{$\omega_{\max}$}}
\rput[bl](3.825,-1.3343652){\normalsize{$\omega_0$}}
\rput[bl](5.005,0.37063476){\normalsize{$\Gamma$}}
\end{pspicture}
}
\caption{The contour~$\Gamma$.}
\label{figupper}
\end{figure}
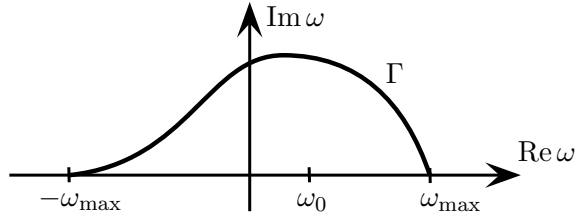

In order to estimate the remaining integrals, we iteratively apply the identity
\begin{align}
&\frac{1}{(\omega+3 i c)^q}\: R_{\omega,n'} \,Q_{n'}^\omega\, (H+3ic)^q \notag \\
&=  \frac{1}{(\omega+3 i c)^q}\: R_{\omega,n'} \,Q_{n'}^\omega\, \Big((H - \omega) + (\omega + 3 ic) \Big)
(H+3ic)^{q-1} \notag \\
&= \frac{1}{(\omega+3 i c)^q}\: Q_{n'}^\omega\, + \frac{1}{(\omega+3 i c)^{q-1}}\: R_{\omega,n'} \,Q_{n'}^\omega\,
(H+3ic)^{q-1} \:. \label{Hid}
\end{align}
Using this relation in~\eqref{intes}, the first summand in~\eqref{Hid} gives rise to integrals of the form
\[ \int_{-\omega_{\max}-i \varepsilon}^{\omega_{\max}-i \varepsilon}
\frac{e^{i (\omega-\omega_0) L}-1}{(\omega-\omega_0) L}\;
\frac{(\omega-\omega_0)^r}{(\omega+3 i c)^q}\:
\big\la Q_{n'}^\omega \,\Psi_L, \Psi_\text{test} \big\ra_{L^2(\R \times S^2)}
\:. \]
As the integrand is holomorphic in a neighborhood of~$\omega_0$, we may deform
the contour into the upper half plane
(keeping the end points fixed) such that~$|\omega-\omega_0|>\omega_{\max}/2$ along the
contour. Taking the limit~$\varepsilon \searrow 0$, we obtain the integral along a contour~$\Gamma$
which joins the points~$-\omega_{\max}$ and~$\omega_{\max}$ and lies in the upper half plane
(see Figure~\ref{figupper}).
Then the bounds~$|e^{i (\omega-\omega_0) L}| \leq 1$ and~$|\omega-\omega_0|>\omega_{\max}/2$
show that the integral tends to zero in the limit~$L \rightarrow \infty$.
Therefore, for large~$L$ only the second summand in~\eqref{Hid} must be taken into account.
We conclude that for large~$L$,
\beq \label{sumsimp}
\bigg| \sum_{n'=0}^N \lim_{\varepsilon \searrow 0}
\int_{-\omega_{\max}-i \varepsilon}^{\omega_{\max}-i \varepsilon}\!
\frac{e^{i (\omega-\omega_0) L}-1}{(\omega-\omega_0) L}\; (\omega-\omega_0)^r \,
\big\la R_{\omega,n'} \,Q_{n'}^\omega \,\Psi_L, \Psi_\text{test} \big\ra_{L^2(\R \times S^2)} \bigg|
< 3 \delta
\eeq
(with~$r$ as in~\eqref{causality}).

We proceed indirectly. Let us assume that the separated resolvents~$R_{\omega,n'}$ have poles
at~$\omega_0$. Since the poles of meromorphic functions are isolated,
there is a small neighborhood of~$\omega_0$ 
on the real axis where the resolvent has no other poles.
Moreover, we may choose~$n$ such that the pole of~$R_{\omega,n}$ has a pole of
order~$q \geq 1$, and that for all~$n' \neq n$, the separated resolvents~$R_{\omega,n'}$
have a pole of order at most~$q$.
By choosing~$r=q-1$, we can arrange that the integrand in~\eqref{sumsimp}
for~$n'=n$ has a pole of order one, whereas all the integrands for~$n' \neq n$ have
a pole of order at most one.

For all modes~$n'$ with~$n' \neq n$, we can make use of the fact that~$\Psi_L$
is an eigenfunction of the angular operator~$\A_{\omega_0}$ in the image of~$Q^\omega_n$
(see~\eqref{PhiLdef}).
As a consequence, $Q_{n'}^{\omega_0} \Psi_L=0$, and thus
\[ \big| \big\la R_{\omega,n'} \,Q_{n'}^\omega \,\Psi_L, \Psi_\text{test} \big\ra_{L^2(\R \times S^2)} \big|
= \big| \big\la R_{\omega,n'} \,\big( Q_{n'}^\omega - Q_{n'}^{\omega_0} \big)\,\Psi_L, \Psi_\text{test}
\big\ra_{L^2(\R \times S^2)} \big| \:. \]
Next, since~$\omega_0$ is real, the angular operator~$\A_{\omega_0}$ is self-adjoint
and has simple eigenvalues (for details see~\cite[Section~7]{tspectral}).
Therefore, the operator~$Q_{n'}^\omega - Q_{n'}^{\omega_0}$ is given linearly in~$(\omega-\omega_0)$
by a standard first order perturbation calculation without degeneracies
(see~\cite{kato}). We thus obtain the estimate
\[  \big| \big\la R_{\omega,n'} \,Q_{n'}^\omega \,\Psi_L, \Psi_\text{test} \big\ra_{L^2(\R \times S^2)} \big|
\leq c\big(\eta_1, \eta_2, \omega_0 \big) \:\big|\omega - \omega_0 \big| \:. \]
Using this estimate in~\eqref{sumsimp}, the factor~$|\omega-\omega_0|$
has the effect that the integrand is bounded near~$\omega=\omega_0$.
Due to the factor~$1/L$ in~\eqref{sumsimp}, the corresponding summand in~\eqref{sumsimp}
tends to zero as~$L \rightarrow \infty$. We conclude that for sufficiently large~$L$,
\beq \label{nsimp}
\bigg| \int_{-\omega_{\max}-i \varepsilon}^{\omega_{\max}-i \varepsilon}\!
\frac{e^{i (\omega-\omega_0) L}-1}{(\omega-\omega_0) L}\; (\omega-\omega_0)^{q-1} \,
\big\la R_{\omega,n} \,Q_{n}^\omega \,\Psi_L, \Psi_\text{test} \big\ra_{L^2(\R \times S^2)} \bigg|
< 4 \delta \:.
\eeq

Thus it remains to analyze the angular mode~$n$:
The integrand can be simplified with the relations 
\begin{align*}
\big\la & R_{\omega,n} \,Q_{n}^\omega \,\Psi_L, \Psi_\text{test} \big\ra_{L^2(\R \times S^2)} \\
&= \la Q_{n}^\omega \,\Theta_{\omega_0, n}, \Theta_{\omega_0, n} \ra_{L^2(S^2)}
\int_{-\infty}^\infty du \int_{-\infty}^\infty dv\; \overline{\Phi_L(u)}\:
\big(\mathfrak{R}_{\omega, n}(u,v)\big)^1_2\: \Phi_\text{test}(v) \\
&= \la Q_{n}^\omega \,\Theta_{\omega_0, n}, \Theta_{\omega_0, n} \ra_{L^2(S^2)}
\int_{-\infty}^\infty du \int_{-\infty}^\infty dv\;\overline{\Phi_L(u)}\: 
g_\omega(u,v)\: \Phi_\text{test}(v) \:,
\end{align*}
where in the last step we used the explicit form of the kernel~$\mathfrak{R}_{\omega, n}$
in~\eqref{Rwndef}. Since~$\omega_0$ is real, the angular operator~$\A_{\omega_0}$ is self-adjoint
and has no degeneracies. A standard perturbation argument implies that if~$|\omega-\omega_0|$
is sufficiently small, the operator~$\A_\omega$ is diagonalizable.
Therefore, the nilpotent matrix~${\mathcal{N}}$ in~\eqref{gkerdef} vanishes, so
that~$g_\omega = s_\omega$ with~$s_\omega$ given by~\eqref{sdef}. We conclude that
\begin{align*}
\big\la & R_{\omega,n} \,Q_{n}^\omega \,\Psi_L, \Psi_\text{test} \big\ra_{L^2(\R \times S^2)} \\
&= \frac{\la Q_{n}^\omega \,\Theta_{\omega_0, n}, \Theta_{\omega_0, n} \ra_{L^2(S^2)}}{w(\acute{\phi}, \grave{\phi})}
\bigg( \int_{-\infty}^\infty \overline{\Phi_L(u)} \,\acute{\phi}(u)\: du \bigg)
\bigg( \int_{-\infty}^\infty \Phi_\text{test}(v) \,\grave{\phi}(v)\: dv \bigg)  \:.
\end{align*}
Using this relation in~\eqref{nsimp}, the integral can be computed with residues.
Since the Wronskian is assumed to have a zero of order~$q$, we obtain
\begin{align*}
&\lim_{\varepsilon \searrow 0} \int_{-\omega_{\max}-i \varepsilon}^{\omega_{\max}-i \varepsilon}\!
\frac{e^{i (\omega-\omega_0) L}-1}{(\omega-\omega_0) L}\; (\omega-\omega_0)^{q-1} \,
\big\la R_{\omega,n} \,Q_{n}^\omega \,\Psi_L, \Psi_\text{test} \big\ra_{L^2(\R \times S^2)} \\
&= \bigg( \lim_{L \rightarrow \infty} \int_{-\infty}^\infty \overline{\Phi_L(u)} \,\acute{\phi}(u)\: du \bigg)
\bigg( \int_{-\infty}^\infty \Phi_\text{test}(v) \,\grave{\phi}(v)\: dv \bigg) \;
\la \Theta_{\omega_0, n}, \Theta_{\omega_0, n} \ra_{L^2(S^2)} \\
&\qquad \times (-i \pi) \:
\text{Res}_{\omega_0} \bigg(
\frac{e^{i (\omega-\omega_0) L}-1}{(\omega-\omega_0) L}\;
\frac{(\omega-\omega_0)^{q-1}}{w(\acute{\phi}, \grave{\phi})}
\bigg) + \O\big( L^{-1} \big) \:.
\end{align*}
Clearly, the residue is non-zero. Moreover, the limit~$L \rightarrow \infty$ of the first
integral exists in view of the asymptotics of the fundamental solution~\eqref{acutephiasy},
as is obvious after a change of variables,
\[ \int_{-\infty}^\infty \eta_1(u+2 L)\: \overline{e^{-2i \overline{\Omega_0} L}} \: e^{i \Omega_0 u}\: du
= \int_{-\infty}^\infty \eta_1(u+2 L)\: e^{i \Omega_0 (u+2L)}\: du 
= \int_{-\infty}^\infty \eta_1(\tau)\: e^{i \Omega_0 \tau}\: d\tau \:. \]
By choosing the test functions~$\eta_1$ and~$\eta_2$ in~\eqref{PhiLdef} and~\eqref{Phitestdef}
appropriately, we can clearly arrange that this limit as well as the second integral are non-zero.
We conclude that the integral in~\eqref{nsimp} has a non-zero limit as~$\varepsilon \searrow 0$.
Since~$\delta$ can be chosen arbitrarily small, we obtain a contradiction.
This concludes the proof.
\QED

\section{Integral Representation and Proof of Decay}
We now consider the Cauchy problem for the Teukolsky equation~\eqref{teukolsky}
with smooth and compactly supported
initial data~$\Psi_0 = (\Phi|_{t=0}, \partial_t \Phi|_{t=0}) \in C^\infty_0(\R \times S^2, \C^2)$
(we always work in the Regge-Wheeler variable~$u \in \R$ (see~\eqref{RW})
and the function~$\Phi := \sqrt{r^2+a^2} \,\phi$ (see~\eqref{PhiR}).
We decompose the initial data into a Fourier series of azimuthal modes (cf.~\eqref{ksep}),
\[ \Psi_0(u, \vartheta, \varphi) = \sum_{k \in \Z} e^{-i k \varphi}\: \Psi_0^{(k)}(u, \vartheta) \:. \]
By linearity, the solution of the Cauchy problem for~$\Psi_0$ is obtained by solving
the Cauchy problem for each azimuthal mode~$k$ and taking the sum of
all the resulting solutions. In the next theorem an integral representation for the solution of each azimuthal mode
is given, and it is shown that the solution decays pointwise.
\begin{Thm} \label{thmrep}
For any~$k \in \Z/2$, there is a parameter~$p>0$ such that for any~$t<0$, the solution of the Cauchy problem 
for the Teukolsky equation with initial data
\[ \Psi|_{t=0} = e^{-i k \varphi}\: \Psi_0^{(k)}(r, \vartheta) \qquad \text{with} \qquad
\Psi^{(k)}_0 \in C^\infty_0(\R \times S^2, \C^2) \]
has the integral representation
\beq \begin{split}
\Psi&(t,u,\vartheta, \varphi) \\
&= -\frac{1}{2 \pi i} \:e^{-i k \varphi}\: \sum_{n=0}^\infty \int_{-\infty}^\infty \frac{e^{-i \omega t}}{(\omega + 3 i c)^p} \;
\Big( R^-_{\omega,n} \:Q_n^\omega
\big(H + 3 i c \big)^p \,\Psi_0^{(k)}  \Big)(u, \vartheta)\: d\omega \:.
\end{split} \label{propfinal}
\eeq
Moreover, the integrals in~\eqref{propfinal} all exist in the Lebesgue sense.
Furthermore, for every~$\varepsilon>0$ and~$u_\infty \in \R$, there is~$N$ such that
for all~$u<u_\infty$,
\beq \label{biges}
\sum_{n=N}^\infty \int_{-\infty}^\infty \bigg\| \frac{1}{(\omega + 3 i c)^p} \;
\Big( R^-_{\omega,n} \:Q_n^\omega
\,\big( H + 3 i c )^p \,\Psi_0^{(k)} \Big)(u) \bigg\|_{L^2(S^2)} \: d\omega < \varepsilon \:.
\eeq
\end{Thm}
\Proof Starting from the result of Corollary~\ref{cormodes}, we apply Proposition~\ref{prpnopole}
to move the contour up to the real axis.
\QED

\begin{Corollary} \label{cordecay}
For every~$k \in \Z/2$, the solution of the Cauchy problem 
for the Teukolsky equation with initial data~$\Psi|_{t=0} = \Psi_0^{(k)} \in \D(H)$
decays pointwise, i.e.
\[ \lim_{t \rightarrow -\infty} \Psi(t,u,\vartheta, \varphi) = 0 \quad
\text{in~$L^\infty_\text{\rm{loc}}(\R \times S^2)$} \:. \]
\end{Corollary}
\Proof Given~$\varepsilon>0$, we choose~$N$ such that~\eqref{biges} holds.
For each of the angular modes~$n=0,\ldots, N-1$, the Riemann-Lebesgue lemma
gives pointwise decay as~$t \rightarrow -\infty$, locally uniformly in the spatial variables.
We conclude that~$\Psi(t)$ decays in~$L^2_\text{loc}(\R \times S^2, \C^2)$.
Differentiating the equation with respect to~$t$, we conclude that
all time derivatives~$\partial_t^q \Psi(t)$ decay in~$L^2_\text{loc}(\R \times S^2, \C^2)$.
Using the Teukolsky equation~\eqref{partH} and applying the Sobolev embedding theorem,
we obtain pointwise decay in~$L^\infty_\text{loc}$.
\QED

For clarity, we point out that, applying the Teukolsky-Starobinsky identities (see
for example~\cite{chandra}), one also gets decay of all other components of the
spin~$s$ wave. Applying the above corollary to the lowest component of the spin wave
with reversed time direction, one also gets decay of the Teukolsky solution~$\Psi$ in the limit~$t \rightarrow +\infty$.
In the case~$s=2$ of gravitational waves, the corresponding metric perturbations can be
constructed as explained in~\cite{lousto+whiting}.

\section{Concluding Remarks}
We close with a few remarks.
We first point out that the integral representation of Theorem~\ref{thmrep}
opens the door to a detailed analysis of the dynamics of the Teukolsky waves.
In particular, one can study decay rates (similar as worked out for massive Dirac waves in~\cite{wdecay})
and derive uniform energy estimates outside the ergosphere (similar as for scalar waves in~\cite{sobolev}).
Moreover, using the methods in~\cite{penrose}, one could analyze superradiance phenomena
for wave packets in the time-dependent setting.

We finally comment on the limitations of our methods.
First, we do not aim for minimal regularity assumptions on the initial data, and we do
not analyze decay in weighted Sobolev spaces. Also, we do not study to which extent our estimates
are uniform in the support of the initial data.
Moreover, we do not consider whether our estimates are uniform in the azimuthal separation constant~$k$,
and we do not analyze the convergence and decay properties of the infinite series of azimuthal modes.
Indeed, the analysis of the infinite sum of azimuthal modes is closely related to the analysis of
optimal regularity. Namely, for smooth initial data, the coefficients of the Fourier
series~\eqref{fouriermode} clearly decay rapidly in~$k$, so that the convergence of the $k$-series
is not an issue. The question of whether this rapid decay in~$k$
also holds for later times is intimately linked to the question of whether the regularity of the solution
(as quantified by suitable weighted Sobolev norms) is preserved under the time evolution. As just mentioned,
such regularity questions are not addressed in this paper.
In order to attack these important open problems, it seems a promising strategy to us
to combine our methods and results with techniques of microlocal analysis as
used in~\cite{hintz+vasy} to study the high frequency behavior in the related Kerr-De Sitter geometry.

Clearly, the next challenge is to prove {\em{nonlinear stability}} of the Kerr geometry.
This will make it necessary to refine our results on the linear problem, in particular
by deriving weighted Sobolev estimates and by analyzing the $k$-dependence of our estimates.

\appendix
\section{Some Estimates of the Angular Eigenvalues}
As in~\cite[Section~2]{tspectral} we rewrite the angular equation in~\eqref{coupled} as the eigenvalue equation
\beq \label{Schrodinger}
H \phi = \lambda \phi \;,
\eeq
where~$H$ has the form of a one-dimensional Hamiltonian
\[ H = -\frac{d^2}{d u^2} + W \]
with the complex potential
\begin{align}
W &= -\frac{1}{4}\: \frac{\cos^2 u}{\sin^2 u} - \frac{1}{2} +\frac{1}{\sin^2 u}(\Omega \sin^2 u + k - s \cos u)^{2} \notag \\
&= \Omega^2\: \sin^2 u + \left(k^2 + s^2 - \frac{1}{4} \right) \frac{1}{\sin^2 u}
\:+\: 2 \Omega k - s^2 - \frac{1}{4} \notag \\
&\quad - 2 s \Omega \cos u - 2 s k\, \frac{\cos u}{\sin^2 u} \label{Wpot}
\end{align}
and~$u=\vartheta$, $\Omega := -a \omega$. 

We begin with estimates for real~$\Omega$. Then, as explained in detail in~\cite[Sections~5 and~7]{tspectral},
the Hamiltonian has non-degenerate eigenvalues~$\lambda_0 < \lambda_1 < \cdots$.

\begin{Lemma} \label{lemmaA1}
There is a constant~$c>0$ such that
\[ \lambda_n \geq \frac{1}{c}\: \big(1 + |\Omega| \big) \qquad \text{for all~$\Omega \in \R$}\:. \]
\end{Lemma}
\Proof It clearly suffices to prove the inequality for the lowest eigenvalues~$\lambda_0$.
In the formulation as an eigenvalue equation for the partial differential operator on the sphere
(see~\cite[eqn~(1.1)]{tspectral}), the angular operator is a sum of two positive operators.
Hence its spectrum is clearly non-negative, so that~$\lambda_0 > 0$.
Assume that the statement of the lemma is false. Then there is a sequence~$(\Omega^\ell)_{\ell \in \N}$
with~$|\Omega^\ell| \rightarrow \infty$, so that the corresponding eigenvalues~$\lambda_0^\ell$
satisfy the relation
\[ \frac{\lambda_0^\ell}{|\Omega^\ell|} \rightarrow 0 \:. \]
Let us derive a contradiction.
Due to the summand~$\Omega^2 \sin^2 u$, the potential is positive except possibly at
a neighborhood of~$u=0$ or~$\pi$. Near the pole at~$u=0$, the potential has the asymptotic form
(see~\cite[eqn~(11.18)]{tspectral}),
\begin{align*}
V(u) &= \frac{\Lambda}{u^2}  + \Omega^2 \,u^2 - 2 s \Omega - \mu + \O \big( |\Omega| u^2 \big)
+ \O \big( |\Omega|^2\, u^4 \big) \\
\Lambda &=  (k-s)^2 - \frac{1}{4} \\
\mu &= \lambda_0 - 2 \Omega k + s^2 + \frac{1}{4} \:.
\end{align*}
Introducing the new variable~$\tilde{u} = \sqrt{|\Omega|}\, u$, the eigenvalue equation~\eqref{Schrodinger}
becomes
\[ \left( -\frac{d^2}{d\tilde{u}^2} + \tilde{V}(\tilde{u}) \right) \phi = 0 \]
with the new potential
\begin{align*}
\tilde{V} &= \frac{\Lambda}{\tilde{u}^2} - \frac{2s\Omega + \mu}{|\Omega|} + \tilde{u}^2 
+ \O \bigg( \frac{\tilde{u}^2}{|\Omega|} \bigg) + \O \bigg( \frac{\tilde{u}^4}{|\Omega|} \bigg) \\
&= - \frac{1}{4 \tilde{u}^2} +  \frac{(k-s)^2}{\tilde{u}^2} \pm2 (k-s) + \tilde{u}^2 -\frac{\lambda_0}{|\Omega|}
+ \O \bigg( \frac{\tilde{u}^2}{|\Omega|} \bigg) + \O \bigg( \frac{\tilde{u}^4}{|\Omega|} \bigg) 
+ \O \bigg( \frac{1}{|\Omega|} \bigg) \:,
\end{align*}
where the plus and minus signs correspond to the cases~$\Omega>0$ and~$\Omega<0$,
respectively. Hence the potential can be regarded as a perturbation of the potential
\[ \tilde{V}_\text{asy} := - \frac{1}{4 \tilde{u}^2} +  \frac{(k-s)^2}{\tilde{u}^2} \pm2 (k-s) + \tilde{u}^2 \:. \]
For this potential, the fundamental solution with the same asymptotics as~$\acute{\phi}$ is
given explicitly in terms of generalized Laguerre polynomials (see~\cite[\S18.8.1]{DLMF}),
\[ \phi_\text{asy}(\tilde{u}) = c \: e^{\frac{\tilde{u}^2}{2}}\: \tilde{u}^{\frac{1}{2} \pm (k-s)} \:
L_{-\frac{1}{2}}^{\pm (k-s)}\big(-\tilde{u}^2\big) \:. \]
Considering the asymptotics for large~$\tilde{u}$, one sees that this function is strictly monotone
increasing for large~$\tilde{u}$. A perturbation argument shows that the same is true
for the eigenfunction~$\phi$ if~$\ell$ is sufficiently large
(this perturbation argument could be carried out in a straightforward way for example by performing
a Jost iteration, taking~$\phi_\text{asy}$ as the unperturbed solution; for details
see~\cite[Section~3]{wdecay} or~\cite{alfaro+regge}). This implies that for any sufficiently large~$c$ and
sufficiently large~$|\Omega|$,
\[ \phi'\big|_{\tilde{u}=c} > 0 \:. \]

Repeating the above argument at the pole at~$u=-\pi$, we conclude the
the eigenfunction~$\phi$ has the properties that for any sufficiently large~$c$ and
sufficiently large~$|\Omega|$,
\[ \phi'\big(c\, |\Omega|^{-\frac{1}{2}} \big) > 0 \qquad \text{and} \qquad
\phi'\big(\pi - c\, |\Omega|^{-\frac{1}{2}} \big) <0 \:. \]
Moreover, the potential is positive on the interval~$(c\, |\Omega|^{-\frac{1}{2}}, \pi - c\, |\Omega|^{-\frac{1}{2}})$,
implying that~$\phi$ is convex on this interval (see for example~\cite[Section~5]{angular}).
This is a contradiction.
\QED

\begin{Prp} \label{prpangular}
There is a constant~$c$ such that the eigenvalues~$\lambda_n$ satisfy the inequalities
\beq \label{lames}
\frac{(n+1)^2}{c} \leq \lambda_n \leq c\,|\Omega|\, (n+1)^2 \qquad \text{for all~$n \in
\N \cup \{0\}$ and~$\Omega \in \R$}\:.
\eeq
\end{Prp}
\Proof We shall apply~\cite[Corollary~7.6]{tspectral}, which states that if for given~$\lambda_n$,
we choose two intervals~$I_L, I_R \subset (0, \pi)$ such that potential~$V$ is non-negative on
the complement of these intervals, and if we choose any two solutions~$\acute{y}$
and~$\grave{y}$ of the Riccati equation on the intervals~$I_L$ respectively~$I_R$
which lie in the upper half plane~$\{ \im y > 0\}$, then
\[ \pi \,( n-5 ) \:<\: \int_{I_L} \im \acute{y} + \int_{I_R} \im \grave{y} \:\leq\: \pi \,(n+2)  \:. \]
We choose the intervals as~$I_L = \big( 0, \min(u^L_+, \umax) \big)$ and~$I_R = \big( \max(u^R_+, \umax), \pi \big)$
(with~$u^{L\!/\!R}_+$ and~$\umax$ as introduced in~\cite[Section~10.1]{tspectral}).
Then obviously~$\re V \geq 0$ on the complement of these intervals.

Thus our task is to estimate the integral of~$\im \acute{y}$ over the interval~$I_L$
(and similarly the integral of~$\im \grave{y}$ over~$I_R$).
Here we want to use the results of the detailed estimates of the 
Riccati solutions near the poles and in the WKB and Airy regions 
carried out in~\cite{special, tinvariant, tspectral}. These estimates apply in the parameter range
(see~\cite[Section~10.1]{tspectral}, keeping in mind that now the potential is real)
\begin{align}
|\Omega| &\geq \Const_4 \label{Omegalarge} \\
\lambda &\geq \Const_5\, |\Omega|\:. \label{lamlarge}
\end{align}
Let us argue why we may restrict attention to this parameter range.
First, for~$\Omega$ in a compact set, the inequalities~\eqref{lames}
follow immediately from a continuity argument and Weyl's asymptotics
(see~\cite[Section~7.3]{tspectral}). Therefore, we may restrict
attention to large~$|\Omega|$, \eqref{Omegalarge}.
Next, for proving the upper bound in~\eqref{lames}, it is clearly no restriction
to assume that~\eqref{lamlarge} holds.
For the lower bound, we can argue as follows: Given~$N \in \N$,
for the first~$N$ eigenvalues, the lower bound in~\eqref{lames} follows
immediately from Lemma~\ref{lemmaA1}.
On the other hand, by choosing~$N$ sufficiently large,
we can apply~\cite[Proposition~7.7]{tspectral} to conclude that
\[ \lambda_n \geq \Const_5\, |\Omega| \qquad \text{for all~$n \geq N$}\:. \]
Therefore, we may indeed restrict attention to the parameter range~\eqref{Omegalarge}
and~\eqref{lamlarge}.

The detailed estimates of the Riccati solutions near the poles and in the WKB and Airy regions 
carried out in~\cite{special, tinvariant, tspectral} show that
\[ \frac{1}{\Const} \int_{\ul^L}^{\ur^L} \sqrt{-V} \leq \int_{I_L} \im \acute{y} \leq  
\Const \int_{\ul^L}^{\ur^L} \sqrt{-V} \:, \]
where~$\ul^L$ and~$\ur^L$ are the boundaries of the WKB region as
introduced in~\cite[Section~10.1]{tspectral}
(and similarly for the integral of~$\im \grave{y}$ over~$I_R$).
Thus it remains to estimate the integral of~$\sqrt{-V}$. In order to get upper bounds,
we restrict attention to the interval~$(\ul, u_0)$ near the pole,
where (see~\cite[eqns~(10.8), (11.12) and~(11.19)]{tspectral})
\[ \ul = \frac{\Const_1}{\sqrt{\re \lambda}} \qquad \text{and} \qquad
u_0 = \Lambda^{\frac{1}{4}}\, |\Omega|^{-\frac{1}{2}} + \O \big( |\Omega|^{-\frac{3}{2}} \big) \:. \]
On this interval, the potential can be estimated by (see~\cite[eqns~(11.14) and~(11.18)]{tspectral})
\[ V(u) \leq \frac{\const}{u^2} - \frac{3 \lambda}{4} \leq -\frac{\lambda}{2}\:, \]
where in the last step we increased the constant~$\Const_1$. Hence
\[ n \gtrsim \int_{\ul}^{\ur} \sqrt{-V} \gtrsim \sqrt{\lambda} \: \big( u^0 - \ul\big) \gtrsim \left(
\frac{\lambda_n}{|\Omega|} \right)^\frac{1}{2} \:. \]
We thus obtain the estimate
\[ \lambda_n \lesssim |\Omega|\, n^2 \:, \]
proving the upper bound in~\eqref{lames}.

In order to obtain simple lower bounds for the eigenvalues, we make use of the fact
that away from the pole region, the potential is bounded by
\[ V\big|_{(u_0, \pi)} \gtrsim -\lambda \:. \]
Hence
\[ n \lesssim \int_{\ul}^{\ur} \sqrt{-V}
= \int_{\ul}^{u_0} \sqrt{-V} + \int_{u_0}^{\ur} \sqrt{-V}
\lesssim \left( \frac{\lambda_n}{|\Omega|} \right)^\frac{1}{2}  + \sqrt{\lambda_n} \:, \]
giving rise to the estimate
\[ \lambda_n \gtrsim n^2 \:. \]
This concludes the proof.
\QED

\begin{Lemma} \label{lemmaangular1}
For any given~$c_1>0$, we  let~$U \subset \C$ be the region
\beq \label{ocond}
|{\mbox{\rm{Im}}}\, \Omega| < c_1 \:, \qquad |\Omega| \geq \Const_4 \:.
\eeq
Then for any~$n \in \N \cup \{0\}$, there is a constant~$c=c(n,c_1, \Const_4)>0$ such
all spectral points~$\lambda$ of the angular operator~$\A_\omega$
restricted to the image of the operator~$Q^\omega_n$ are in the range
\beq \label{lambes}
\frac{|\Omega|}{c} \leq |\lambda| \leq c\, |\Omega|
\qquad \text{for all~$\Omega \in U$}\:.
\eeq
\end{Lemma}
\Proof If~$\Omega$ is real, the inequalities~\eqref{lambes} were already derived
in Lemma~\ref{lemmaA1} and Proposition~\ref{prpangular}.
In order to extend these results to complex~$\Omega$,
similar as in~\cite[Section~16]{tspectral} we consider the homotopy
\beq \label{Wtaudef}
W_\tau = \tau \,W[\Omega] + (1-\tau) \,W[\re \Omega]
\eeq
with~$\tau \in [0,1]$ (where the argument in the square brackets is the
respective value for the parameter~$\Omega$ in~\eqref{Wpot})
Then for~$\tau=0$, the potential~$W_\tau$ is real, and the results
of Lemma~\ref{lemmaA1} and Proposition~\ref{prpangular} apply.

In~\cite[Section~16]{tspectral} a similar homotopy was considered,
and the eigenvalues were traced in detail. However, as we will become clear below,
these estimates are not good enough for getting the lower bound in~\eqref{ocond}
for the first~$N$ spectral points, making it necessary to slightly refine the method.
In preparation, we now explain an alternative method for tracking the eigenvalues.
Apart from giving a different point of view, this method has the advantage that it can
be refined to get the required control of the first~$N$ spectral points.

For families of self-adjoint operators, the change of the
eigenvalues can be estimated in terms of the sup-norm of the perturbation, i.e.\
(see for example~\cite{kato})
\beq \label{katoes}
\big| \Delta \lambda_n \big| \leq \| \Delta W\|\:.
\eeq
This inequality is not necessarily true for non-selfadjoint operators. But it holds in our
setting up to a uniform constant, if we make use of the fact that the spectral operators~$Q^\omega_n$
are uniformly bounded~\eqref{Qnb}. In order to make the argument precise, we first consider
a non-degenerate eigenspace, in which case the eigenvalue~$\lambda_n(\tau)$ depends smoothly
on~$\tau$. Differentiating the eigenvalue equation
\[ \big( H(\tau) - \lambda_n(\tau)  \big)\: \phi_n(\tau) = 0 \]
with respect to~$\tau$, we obtain
\[ \big( H - \lambda_n \big)\: \dot{\phi}_n = 
\big( \dot{H} - \dot{\lambda}_n \big)\: \phi_n \]
(where we omitted the argument~$\tau$, and the dot denotes the $\tau$-derivative).
Multiplying by~$Q^\omega_n$, the left side vanishes, and thus
\beq \label{flow}
0 = Q^\omega_n \,\big( \dot{H} - \dot{\lambda}_n \big)\: \phi_n 
= Q^\omega_n \,\dot{H} \phi_n - \dot{\lambda}_n\: \phi_n \:.
\eeq
Taking the norm and using~\eqref{Qnb}, we obtain the estimate
\beq \label{lambdadot}
\big| \dot{\lambda}_n \big| \leq c_2\: \big\| \dot{H} \big\| = c_2\:\big\| \dot{W}_\tau \big\| \:.
\eeq
Integrating this inequality, we obtain~\eqref{katoes}, up to the constant~$c_2$,
\beq \label{katonew}
\big| \Delta \lambda_n \big| \leq c_2\: \| \Delta W\| \:.
\eeq
In the general case with degeneracies, the situation is more involved, because the
eigenvalues no longer depend smoothly on~$\tau$.
But the spectrum is still continuous in~$\tau$. Moreover, the eigenvalues depend
smoothly on~$\tau$ except at the points where degeneracies form
and the dimensions of the invariant subspaces change.
Therefore, the inequality~\eqref{lambdadot} shows that the total variation
of the change of the spectral points can be estimated by a constant times~$\|\Delta W\|$.

Using the assumption~\eqref{ocond} in~\eqref{Wpot}, one sees that for our homotopy~\eqref{Wtaudef},
\[ \| \Delta W\| \equiv \|W_1 - W_0 \| \lesssim |\Omega| \:. \]
Using this inequality in~\eqref{katonew}, we obtain
\beq \label{coarse}
\big| \lambda_n[\Omega] - \lambda_n[\re \Omega] \big| \lesssim |\Omega| \:.
\eeq
Combining this estimate with the upper bound in~\eqref{lames}, we obtain the upper
bound in~\eqref{lambes}. Moreover, the lower bound in~\eqref{lames} gives
the lower bound in~\eqref{lambes}, but only if~$n$ is sufficiently large and~$|\Omega|$ is not too large.

In order to derive the lower bound in~\eqref{lames}, we make use of the fact that for large~$|\Omega|$,
the eigenfunction is localized mainly near the poles, where the imaginary part of~$W$ is bounded
uniformly in~$|\Omega|$. This is made precise by the following estimate:
We choose a test function~$\eta \in C^\infty((0, \pi))$ taking values in the interval~$[0,1]$ with
the properties
\beq \label{etaprop}
\text{supp}\, \eta \subset \Big[ |\Omega|^{-\frac{1}{2}}, \pi - |\Omega|^{-\frac{1}{2}} \Big] \qquad \text{and} \qquad
\eta \big|_{\big[ 2 \,|\Omega|^{-\frac{1}{2}}, \;\pi - 2 \,|\Omega|^{-\frac{1}{2}} \big]} \equiv 1 \:.
\eeq
Clearly, $\eta$ can be chosen such that~$\sup_{[0,\pi]} |\eta''| \lesssim |\Omega|$.
Integrating by parts, we obtain
\[ 2 \int_0^\pi \eta\, \re V\: |\phi|^2\: du = \int_0^\pi \eta\,\Big( \overline{\phi} \phi'' + \overline{\phi''} \phi \Big) \: du
= \int_0^\pi \Big( \eta'' \,|\phi|^2 - \eta\, |\phi'|^2 \Big) \: du \:, \]
giving rise to the inequality
\[ \int_0^\pi \eta\, \re V\: |\phi|^2\: du \lesssim |\Omega|\: \|\phi\|^2_{L^2}\:.  \]
Using~\eqref{Wpot} as well as the upper bound in~\eqref{lambes}, we conclude that
\beq \label{witheta}
\re \big(\Omega^2 \big) \int_0^\pi \eta \:|\phi|^2\: \sin^2 u\:du \lesssim |\Omega|\: \|\phi\|^2_{L^2}
\eeq
(note that, in view of~\eqref{etaprop}, the summands with poles in~\eqref{Wpot} are pointwise bounded
on the support of~$\eta$ by a constant times~$|\Omega|$). Moreover, using that the function~$\sin u$
has zeros at~$u=0$ and~$u=\pi$, we also have
\[ \re \big(\Omega^2 \big) \int_0^\pi \big(1 - \eta) \:|\phi|^2\: \sin^2 u\:du \lesssim |\Omega|\: \|\phi\|^2_{L^2} \:. \]
Adding this estimate to~\eqref{witheta}, we obtain the inequality
\beq \label{sinint}
\int_0^\pi |\phi|^2\: \sin^2 u\:du \lesssim \frac{\|\phi\|^2_{L^2}}{|\Omega|} \:.
\eeq
We now multiply our flow equation~\eqref{flow} by~$\overline{\phi}_n$ and integrate. Omitting the index~$n$
of the wave function~$\phi_n$, this gives the estimate
\beq \label{dotlamimprove}
\big|\dot{\lambda}_n\big| \|\phi\|_{L^2} \leq \big\| Q^\omega_n \big\|\: \big\| \dot{W}_\tau \phi \big\|_{L^2} \:.
\eeq
Next, we estimate the last factor as follows,
\begin{align*}
\big\| \dot{W}_\tau \phi \big\|_{L^2}^2 &= \int_0^\pi \big| \dot{W}_\tau \big|^2\: |\phi|^2\: du
\leq \big\| \dot{W}_\tau \big\|_\infty \int_0^\pi \big| \dot{W}_\tau \big|\: |\phi|^2\: du
\lesssim |\Omega|  \int_0^\pi \big| \dot{W}_\tau \big|\: |\phi|^2\: du \:.
\end{align*}
Again using the explicit form of the potential~\eqref{Wpot}, one sees that
\[  \big| \dot{W}_\tau - 2 i \Omega\: \im \Omega\: \sin^2 u \big| \lesssim 1 \:. \]
We thus obtain
\[ \big\| \dot{W}_\tau \phi \big\|_{L^2}^2 \lesssim  |\Omega| \:\|\phi\|_{L^2}^2
+ |\Omega|^2  \int_0^\pi \sin^2 u\: |\phi|^2\: du 
\overset{\eqref{sinint}}{\lesssim} |\Omega| \:\|\phi\|_{L^2}^2 \:. \]
Using this inequality in~\eqref{dotlamimprove}, we get
\[ \big|\dot{\lambda}_n\big| \lesssim \sqrt{|\Omega|}\: \big\| Q^\omega_n \big\| \lesssim \sqrt{|\Omega|} \:, \]
where in the last step we again used~\eqref{Qnb}.
We thus obtain the following improvement of~\eqref{coarse},
\[ \big| \lambda_n[\Omega] - \lambda_n[\re \Omega] \big| \lesssim \sqrt{|\Omega|} \:. \]
Combining this estimate with the result of Lemma~\ref{lemmaA1} gives the lower
bound in~\eqref{lambes}. This concludes the proof.
\QED

\Thanks {{\em{Acknowledgments:}}
We would like to thank Agnes B\"auml, Yunlong Zang and the referee for helpful comments on the manuscript.
F.F.\ is grateful to the Center of Mathematical Sciences and Applications at
Harvard University for hospitality and support. 

\providecommand{\bysame}{\leavevmode\hbox to3em{\hrulefill}\thinspace}
\providecommand{\MR}{\relax\ifhmode\unskip\space\fi MR }
\providecommand{\MRhref}[2]{%
  \href{http://www.ams.org/mathscinet-getitem?mr=#1}{#2}
}
\providecommand{\href}[2]{#2}

\end{document}